\documentclass[aps,prc,showpacs,showkeys,nofootinbib]{revtex4}

\usepackage{color} 

\usepackage{graphicx}
\usepackage{dcolumn}
\usepackage{amsthm}
\usepackage{bm}

\usepackage{isotope}
\usepackage{braket}
\usepackage{physics}

\newcommand{\sigmabf}{\mbox{\boldmath $\sigma$}}
\newcommand{\xibf}{\mbox{\boldmath $\xi$}}

\begin{document}

\title{Nucleon microscopy in proton-nucleus scattering via analysis of bremsstrahlung emission
}


\author{Sergei~P.~Maydanyuk}%
\email{maidan@kinr.kiev.ua}%
\affiliation{Institute of Modern Physics, Chinese Academy of Sciences, Lanzhou, 730000, China}
\affiliation{Institute for Nuclear Research, National Academy of Sciences of Ukraine, Kiev, 03680, Ukraine}
\author{Peng-Ming~Zhang}%
\email{zhpm@impcas.ac.cn} %
\affiliation{Institute of Modern Physics, Chinese Academy of Sciences, Lanzhou, 730000, China}
\author{Li-Ping~Zou}%
\email{zoulp@impcas.ac.cn} %
\affiliation{Institute of Modern Physics, Chinese Academy of Sciences, Lanzhou, 730000, China}

\date{\small\today}

\begin{abstract}
We investigate an idea, how to use analysis of the bremsstrahlung photons to study
the internal structure of proton under nuclear reaction with nucleus.
A new model is constructed to describe bremsstrahlung emission of photons which accompanies the scattering of protons off nuclei.
Our bremsstrahlung formalism uses many-nucleon basis that allows to analyze coherent and incoherent bremsstrahlung emissions.
As scattered proton can be under the influence of strong forces and produces the largest bremsstrahlung contribution to full spectrum,
we focus on accurate determination of its quantum evolution concerning nucleus
basing on quantum mechanics. 
For such a motivation,
we at first time generalize Pauli equation with interacting potential describing evolution of fermion inside strong field,
with including the electromagnetic form-factors of nucleon basing on DIS theory.
Anomalous magnetic momenta of nucleons reinforce our motivation to develop such a formalism.
The full bremsstrahlung spectrum in our model (after renormalization) is dependent on form-factors of the scattered proton.

In our calculations we choose the scattering of $p + ^{197}{\rm Au}$ at proton beam energy of 190~MeV,
where experimental bremsstrahlung data were obtained with high accuracy.
We show that the full bremsstrahlung spectrum is sensitive to the form-factors of the scattered proton.
In the limit without such form-factors, we reconstruct our previous result (where internal structure of the scattered proton was not studied).
\end{abstract}




\maketitle

\section{Introduction
\label{sec.introduction}}

Understanding the internal structure of nucleons has attracted researchers for a long time.
The first experimental investigations of structure of protons were performed in inelastic electron--proton scattering at high energies in SLAC in 1968
(see Ref.~\cite{Greiner.Chromodynamics.2002}, p.~96).
Now high energy lepton-nucleon scattering (deep inelastic scattering, DIS) plays a key role in determining internal (partonic) structure of protons.
Relevant information is summarized in reports [see Review PDG~\cite{Review_particle_phys.2018},
also reviews~\cite{DeRoeck.2011.PPNP,Forte.2013.ARNPS,Blumlein.2013.PPNP,Perez.2013.RPP,Ball.2013.JHEP}].
Investigations of nucleon-nucleon collisions at TEVATRON (Fermilab), RHIC (Brookhaven), LHC (CERN) show us new important information~\cite{Review_particle_phys.2018}.


Moreover, we know that full information of nuclear forces (strong interactions) cannot be obtained on the basis of analysis of any type of reactions between two nucleons (clear example is fusion in nuclei, see Ref.~\cite{Maydanyuk_Zhang_Zou.2017.PRC}, also
reviews~\cite{Back.2014.RMP,Birkelund.1979.PRep,Vaz.1981.PRep,Birkelund.1983.ARNPS,Beckerman.1985.PRep,Steadman.1986.ARNPS,
Beckerman.1988.RPP,Rowley.1991.PLB,Vandenbosch.1992.ARNPS,Reisdorf.1994.JPG,Dasgupta.1998.ARNPS,Balantekin.1998.RMP,Liang.2005.IJMPE,
Canto.2006.PRep,Keeley.2007.PPNP,Hagino.2012.PTP}).
So, analysis of interactions between two nucleons is not enough
(for example, Nijmegen data set~\cite{Stoks.1993.PRC}),
and we have to include to analysis systems with more nucleons (i.e. nuclei).
From microscopic point of view, nuclear interactions should be completely formed with internal structure of nucleons.
This situation leads naturally to new investigations of interactions (reactions) between nucleons and nuclei, in order to obtain more complete information about internal structure of nucleons.

The bremsstrahlung emission of photons accompanying nuclear reactions is a traditional sector in nuclear physics,
which has been causing much interest for a long time
(see reviews~\cite{Pluiko.1987.PEPAN,Kamanin.1989.PEPAN}). 
This is because of such photons provide rich independent information about the studied nuclear process.
Dynamics of the nuclear process, interactions between nucleons, types of nuclear forces, structure of nuclei, quantum effects, anisotropy (deformations) can be included in the model describing the bremsstrahlung emission.
At the same time, measurements of such photons and their analysis provide information about all these aspects, and verify suitability of the models.
So, bremsstrahlung photons is the tool to obtain experimental information for all above questions.

Anomalous magnetic momenta for neutrons (and protons) is another strong motivation,
why there is sense to include internal structure of nucleons to the bremsstrahlung formalism in nuclear reactions (even for low energies of beams).
As we shown in Ref.~\cite{Maydanyuk.2012.PRC}, magnetic momenta of nucleons play an important role in formation of the magnetic emission in proton-nucleus scattering.
According to our estimations~\cite{Maydanyuk.2012.PRC}, the electric and magnetic bremsstrahlung emissions have similar magnitudes in such a reaction.
It was shown in Ref.~\cite{Maydanyuk_Zhang.2015.PRC}, that incoherent emission in experimental bremsstrahlung data \cite{Goethem.2002.PRL}
for scattering $p+^{208}{\rm Pb}$ at 145~MeV of the proton beam energy
is not small at low energies (at 20--120~MeV of photons emitted, see Fig.~5 for details in that paper).
Such a type of emission is formed due to individual interactions between the scattered proton and nucleons of nucleus.
However, we did not take into account anomalous magnetic momenta of nucleons in our model, calculations and analysis of bremsstrahlung experimental information.
As anomalous magnetic moment for neutron is essentially different from its Dirac magnetic moment, one can suppose that after inclusion of anomalous momenta to the model changes of results~\cite{Maydanyuk_Zhang.2015.PRC} can be not small
(for example, this can give essentially different estimation of role of incoherent emission on background of the full bremsstrahlung spectrum).
So, inclusion of internal structure of nucleons to the bremsstrahlung model for nuclear reactions for energies from low to high is motivated task.

In this paper, we focus on solution how to realize ideas described above.
We construct a new bremsstrahlung model for proton nucleus scattering, where we include internal structure of the scattered proton.
As starting basis for such developments, we use our previous formalism~\cite{Maydanyuk.2012.PRC,Maydanyuk_Zhang.2015.PRC} applied for proton nucleus scattering.
%
We use the first approximation of generalisation of Dirac equation to describe system of nucleons with interacting potential.
Of course, such an equation should describe spinor properties of fermions, and should have fully quantum description of system of nucleons,
in full correspondence with quantum mechanics.
As we found, this is many-nucleon generalisation of Pauli equation with interacting potential investigated in Refs.~\cite{Maydanyuk.2012.PRC,Maydanyuk_Zhang.2015.PRC,Maydanyuk_Zhang_Zou.2016.PRC,Maydanyuk_Zhang_Zou.2018.PRC,Liu_Maydanyuk_Zhang_Liu.2018.arxiv}.
This way allows us to construct formalism with some connection with
many-nucleon bremsstrahlung developments of other researchers
~\cite{Liu.1981.PRC.v23,Baye.1985.NPA,Liu.1990.PRC.v41,Liu.1990.PRC.v42,
Baye.1991.NPA,Liu.1992.FBS,Dohet-Eraly.2011.JPCS,Dohet-Eraly.2011.PRC,Dohet-Eraly.2013.PhD,
Dohet-Eraly.2013.JPCS,Dohet-Eraly.2013.PRC,Dohet-Eraly.2014.PRC.v89,Dohet-Eraly.2014.PRC.v90}
(here, evolution of nucleons is investigated as many body problem of quantum mechanics that allows to save quantum properties maximally completely).
%
%
%
%
On the other side, our formalism is started from approximation of Dirac equation, so it allows to include next relativistic corrections in quantum way.
DIS theory provides accurate description of internal structure of nucleons via form-factors, so we implement this formalism to our bremsstrahlung theory.
We estimate new bremsstrahlung contributions of emitted photons, caused by such new addition to the model.
The obtained bremsstrahlung probabilities already dependent on internal structure of the scattered proton.
We analyze and estimate such a dependence.


The paper is organized by the following way.
In Sec.~\ref{sec.model} we present our new model of the bremsstrahlung photons emitted during proton nucleus scattering.
In Sec.~\ref{sec.results} we give the results of study for the scattering of $p + ^{197}{\rm Au}$ at proton beam energy of 190~MeV.
We summarize conclusions in Sect.~\ref{sec.conclusions}.
We present main part of calculations in Appendixes (see Sect.~\ref{sec.app.1}--\ref{sec.app.9}).

\section{Model
\label{sec.model}}

\subsection{Generalized equation of Pauli for spinor particle with mass $m$ in field $V(\mathbf{r})$ with electromagnetic form factors
\label{sec.1}}

Let us consider Dirac equation for nucleon with mass $m$ inside field $V(\mathbf{r})$ (see \cite{Ahiezer.1981}, p.~21 (1.2.3), p.~32):
%
\begin{equation}
\begin{array}{lcl}
  i \hbar \displaystyle\frac{\partial \psi}{\partial t} =
  \biggl\{
      c\, \alpha\:
      \Bigl( \mathbf{p} - \displaystyle\frac{ze}{c} \mathbf{A} \Bigr) +
      \beta m c^{2} + ze\, A_{0} +
    V(\mathbf{r})
  \biggr\}\; \psi.
\end{array}
\label{eq.2.1}
\end{equation}
This equation is written in coordinates of Euclidian space (in frameworks of formalism in~\cite{Ahiezer.1981}, which includes potentials in Dirac equation).
In particular, we have the following relations between coordinates and corresponding momenta of Euclidian and pseudo-Euclidian spaces:
%
\begin{equation}
\begin{array}{lllllll}
  x_{1,2,3}^{\rm (ev)} = \mathbf{r}, &
  x_{4}^{\rm (ev)} = it, &
  p_{1,2,3}^{\rm (ev)} = -i\hbar\, \displaystyle\frac{d}{dx_{1,2,3}}, &
  p_{4}^{\rm (ev)} = i\,p_{0}^{\rm (ps)}.
\end{array}
\label{eq.2.1.addev}
\end{equation}
We change the wave function as
\begin{equation}
\begin{array}{lll}
  \Psi = \psi\; e^{i\, m c^{2}t/\hbar}, &

  \displaystyle\frac{\partial \Psi}{\partial t} =

  e^{i\, m c^{2}t / \hbar}\, \Bigl( \displaystyle\frac{\partial \psi }{\partial t} + \displaystyle\frac{i\, m c^{2}}{\hbar}\,\psi\, \Bigr), &

  \displaystyle\frac{\partial \psi }{\partial t} =
    e^{- i\, m c^{2}t / \hbar}\, \displaystyle\frac{\partial \Psi}{\partial t} - \displaystyle\frac{i\, m c^{2}}{\hbar}\,\psi
\end{array}
\label{eq.2.2}
\end{equation}
and equation is transformed to the following:
%
%
%
\begin{equation}
\begin{array}{lcl}
\vspace{1mm}
  -\, \beta \hbar \displaystyle\frac{\partial \Psi}{\partial t}
  & = &
%
%
  \biggl\{
    i\, c\, \beta \alpha\:
    \Bigl( \mathbf{p} - \displaystyle\frac{ze}{c} \mathbf{A} \Bigr) +
    i\, m c^{2} +
    i \beta \bigl[ ze\, A_{0} + V(\mathbf{r}) - m c^{2} \bigr]
  \biggr\}\; \Psi.
\end{array}
\label{eq.2.4}
\end{equation}
We rewrite this equation via matrixes $\gamma_{\mu}$ в эвклидовой метрике
(we define them, according to Ref.~\cite{Ahiezer.1981}):
\begin{equation}
\begin{array}{lcl}
\vspace{1mm}
  \gamma_{4} = \beta =
  \left(
  \begin{array}{lcl}
    1 & 0 \\
    0 & -1
  \end{array}
  \right), &

  \gamma_{k} = -i\, \beta \alpha_{k} =
  \left(
  \begin{array}{lcl}
    0 & -i \sigma_{k} \\
    i \sigma_{k} & 0
  \end{array}
  \right)
\end{array}
\label{eq.2.5}
\end{equation}
and obtain:
\begin{equation}
\begin{array}{lcl}
  - \hbar\, \gamma_{4}\, \displaystyle\frac{\partial \Psi}{\partial t}
  & = &
  \biggl\{
    - c\, \gamma_{k} \:
    \Bigl( \mathbf{p} - \displaystyle\frac{ze}{c} \mathbf{A} \Bigr) +
    i\, m c^{2} +
    i \gamma_{4} \bigl[ ze\, A_{0} + V(\mathbf{r}) - m c^{2} \bigr]
  \biggr\}\; \Psi.
\end{array}
\label{eq.2.6}
\end{equation}
In order to describe internal structure of the scattered proton, we introduce matrixes $\Gamma_{\mu}$ of DIS theory
[
we define them according to formalism in Ref.~\cite{Greiner.Chromodynamics.2002}, see (3.6) p.~78;
we apply approximation where we neglect structure of nucleons of nucleus%
] instead of Dirac's matrixes $\gamma_{\mu}$:
%
%
\begin{equation}
  \gamma_{\mu} \to \Gamma_{\mu} =
    A\, \gamma_{\mu} + B p_{\mu}^{\prime} + C p_{\mu} + i D p^{\prime\, \nu} \sigma_{\mu\nu} + i\, E p^{\nu} \sigma_{\mu\nu}.
\label{eq.2.7}
\end{equation}
$A$, $B$, $C$, $D$, $E$ are functions depended on the transferred momentum $q^{2}$ between the scattered proton and nucleon of nucleus.
They characterize internal structure of the scattered proton.
$p$ and $p^{\prime}$ are momenta of the scattered proton before its interaction with virtual photon (emitted by nucleon of the nucleus) and after it.
%
%
One of motivations to use transition (\ref{eq.2.7}) in the formalism is the following.
At $q^{2} \to 0$ we should obtain $A(q^{2}=0) = 1$, and
components $B$, $C$, $D$, $E$ should describe (anomalous) magnetic moment of the scattered proton.

Now equation (\ref{eq.2.6}) is rewritten as
\begin{equation}
\begin{array}{lcl}
\vspace{1mm}
  - \hbar\, \Gamma_{4}\, \displaystyle\frac{\partial \Psi}{\partial t}
  & = &
  \biggl\{
    - c\, \Gamma_{k} \:
    \Bigl( \mathbf{p} - \displaystyle\frac{ze}{c} \mathbf{A} \Bigr) +
    i\, m c^{2} +
    i \Gamma_{4} \bigl[ ze\, A_{0} + V(\mathbf{r}) - m c^{2} \bigr]
  \biggr\}\; \Psi.
\end{array}
\label{eq.2.8}
\end{equation}
Substituting explicit form (\ref{eq.2.7}) of matrixes $\Gamma_{\mu}$, this equation is transformed as
\begin{equation}
\begin{array}{lcl}
\vspace{1mm}
  & & -\, \hbar\,
  \bigl( A\, \gamma_{4} + B p_{4}^{\prime} + C p_{4} + i D p^{\prime\, \nu} \sigma_{4\nu} + i\, E p^{\nu} \sigma_{4\nu} \bigr)\:
    \displaystyle\frac{\partial \Psi}{\partial t} = \\
  & = &
  \biggl\{
    - c\,
    \bigl( A\, \gamma_{k} + B p_{k}^{\prime} + C p_{k} + i D p^{\prime\, \nu} \sigma_{k\nu} + i\, E p^{\nu} \sigma_{k\nu} \bigr)\:
      \Bigl( \mathbf{p} - \displaystyle\frac{ze}{c} \mathbf{A} \Bigr) +
    i\, m c^{2} + \\
\vspace{1mm}
  & + &
    i\, \bigl( A\, \gamma_{4} + B p_{4}^{\prime} + C p_{4} + i D p^{\prime\, \nu} \sigma_{4\nu} + i\, E p^{\nu} \sigma_{4\nu} \bigr)\,
      \bigl[ ze\, A_{0} + V(\mathbf{r}) - m c^{2} \bigr]
  \biggr\}\: \Psi \\
\end{array}
\label{eq.2.9}
\end{equation}

Using Gordon transformation and conditions $B = C$, $E = - D$ from DIS theory (see \cite{Greiner.Chromodynamics.2002}, p.~79),
equation (\ref{eq.2.9}) is transformed to
\begin{equation}
\begin{array}{lcl}
\vspace{2mm}
  -\, \hbar\,
  \bigl( A\, \gamma_{4} + i B q^{\nu}\, \sigma_{4\nu} \bigr)\:
    \displaystyle\frac{\partial \Psi}{\partial t} & = &
  \Bigl\{
    - c\, \bigl( A\, \gamma_{k} + i B q^{\nu}\, \sigma_{k\nu} \bigr)\:
      \Bigl( \mathbf{p} - \displaystyle\frac{ze}{c} \mathbf{A} \Bigr) +
    i\, m c^{2}\, + \\
  & + &
  i\, \bigl( A\, \gamma_{4} + i B q^{\nu}\,\sigma_{4\nu} \bigr)\,
    \bigl[ ze\, A_{0} + V(\mathbf{r}) - m c^{2} \bigr]
  \Bigr\}\: \Psi,
\end{array}
\label{eq.2.13}
\end{equation}
where $q^{\mu} = p^{\prime\,\mu} - p^{\mu}$,
$q^{2} = \mathbf{q}^{2} + q_{4}^{2}$.
In DIS theory $A = F_{1}$ and $B = F_{2}$ are Dirac and Pauli form-factors of nucleon.
%
According to Ref.~\cite{Schindler.2005.EPJA}, $F_{1}$ and $F_{2}$ represent electric charge and (anomalous) magnetic moment of nucleon,
they are
%
\begin{equation}
\begin{array}{lcll}
  F_{1}^{p} (Q^{2} = -q^{2} = 0) = 1, &
  F_{1}^{n} (0) = 0, &
  F_{2}^{p} (0) = 1.793, &
  F_{2}^{n} (0) = -1.913.
\end{array}
\label{eq.2.14}
\end{equation}

We rewrite bi-spinor wave function via spinor components as
\begin{equation}
  \Psi =
  \left(
  \begin{array}{c}
    \varphi \\
    \chi
  \end{array}
  \right).
\label{eq.2.15}
\end{equation}
Taking explicit form of matrixes $\gamma_{\mu}$ (\ref{eq.2.5}) into account, we rewrite equation (\ref{eq.2.13}) by components as
\begin{equation}
\begin{array}{lcl}
\vspace{2mm}
  -\, \hbar\,
  \biggl\{
  F_{1}
  \left(
  \begin{array}{cc}
    1 & 0 \\
    0 & -1
  \end{array}
  \right) +

  i F_{2} q^{\nu} \sigma_{4\nu} \biggr\}\:

  \left(
  \begin{array}{c}
    \displaystyle\frac{\partial \varphi}{\partial t} \\
    \displaystyle\frac{\partial \chi}{\partial t}
  \end{array}
  \right) & = &

  \biggl\{
    - c\, \Bigl[ F_{1}
  \left(
  \begin{array}{cc}
    0 & - i \sigma_{k} \\
    i \sigma_{k} & 0
  \end{array}
  \right) +

  i F_{2} q^{\nu} \sigma_{k\nu} \Bigr]\:
      \Bigl( \mathbf{p} - \displaystyle\frac{ze}{c} \mathbf{A} \Bigr) +

  i\, m c^{2}
  \left(
  \begin{array}{cc}
    1 & 0 \\
    0 & 1
  \end{array}
  \right) + \\

\vspace{1mm}
  & + &
  i\, \Bigl[ F_{1}
  \left(
  \begin{array}{cc}
    1 & 0 \\
    0 & -1
  \end{array}
  \right) +

  i F_{2} q^{\nu} \sigma_{4\nu} \Bigr]\,
      \bigl[ ze\, A_{0} + V(\mathbf{r}) - m c^{2} \bigr]
  \biggr\}\:
    \left(
  \begin{array}{c}
    \varphi \\
    \chi
  \end{array}
  \right).
\end{array}
\label{eq.2.16}
\end{equation}
According to definition of $\sigma_{\mu\nu}$ (see \cite{Ahiezer.1981}, p.~23) we have:
\begin{equation}
  \sigma_{\mu\nu} =
  \displaystyle\frac{1}{2i}
  (\gamma_{\mu}\gamma_{\nu} - \gamma_{\nu}\gamma_{\mu})
\label{eq.2.17}
\end{equation}
and
\begin{equation}
\begin{array}{lcl}
  \sigma_{44} = 0, &
  \sigma_{k4} =
  \sigma_{k}
  \left(
  \begin{array}{cc}
    0 & 1 \\
    1 & 0
  \end{array}
  \right), &
  \sigma_{km} =
  \displaystyle\frac{\sigma_{k}\sigma_{m} - \sigma_{m}\sigma_{k}}{2i}
  \left(
  \begin{array}{cc}
    1 & 0 \\
    0 & 1
  \end{array}
  \right) =
  \varepsilon_{kmj}\, \sigma_{j}
  \left(
  \begin{array}{cc}
    1 & 0 \\
    0 & 1
  \end{array}
  \right),
\end{array}
\label{eq.2.21}
\end{equation}
where we use property of commutator of Pauli matrixes (see Ref.~\cite{Okun.1966}, p.~32, $i,j,k = 1,2,3$):
\begin{equation}
  \sigma_{i}\sigma_{j} = \delta_{ij} I + i\, \varepsilon_{ijk} \sigma_{k},
\label{eq.2.19}
\end{equation}
and $\varepsilon_{ijk} $ is unit antisymmetric tensor, $\varepsilon_{123} = 1$.
%
%
Substituting these components to (\ref{eq.2.16}), we obtain:
\begin{equation}
\begin{array}{lll}
\vspace{1mm}
  i\, \hbar F_{1}\, \displaystyle\frac{\partial \varphi}{\partial t} -
  \hbar F_{2} q^{k} \sigma_{k}\, \displaystyle\frac{\partial \chi}{\partial t} & = &

  \Bigl\{
    - c\, \varepsilon_{kmj}\, F_{2} q^{k}\, \sigma_{j}\:
    \Bigl( \mathbf{p} - \displaystyle\frac{ze}{c} \mathbf{A} \Bigr) +
    m c^{2} +
    F_{1}\, \bigl[ ze\, A_{0} + V(\mathbf{r}) - m c^{2} \bigr]
  \Bigr\}\:
  \varphi\; + \\

\vspace{4mm}
  & + & \;
  \Bigl\{
    c\, \bigl[ F_{1} \sigma_{m} - F_{2} q^{4} \sigma_{m} \bigr]\:
      \Bigl( \mathbf{p} - \displaystyle\frac{ze}{c} \mathbf{A} \Bigr) -
    i\, F_{2} q^{k} \sigma_{k}\, \bigl[ ze\, A_{0} + V(\mathbf{r}) - m c^{2} \bigr]
  \Bigr\}\: \chi, \\

\vspace{1mm}
  -i\, \hbar F_{1}\, \displaystyle\frac{\partial \chi}{\partial t} -
  \hbar F_{2} q^{k} \sigma_{k}\, \displaystyle\frac{\partial \varphi}{\partial t} & = &

  \Bigl\{
    - c\, \bigl[ F_{1} \sigma_{m} + F_{2} q^{4} \sigma_{m} \bigr]\:
      \Bigl( \mathbf{p} - \displaystyle\frac{ze}{c} \mathbf{A} \Bigr) -
    i\, F_{2} q^{k} \sigma_{k}\, \bigl[ ze\, A_{0} + V(\mathbf{r}) - m c^{2} \bigr]
  \Bigr\}\:
  \varphi\; + \\

  & + &
  \Bigl\{
    - c\, \varepsilon_{kmj}\, F_{2} q^{k}\, \sigma_{j}\:
      \Bigl( \mathbf{p} - \displaystyle\frac{ze}{c} \mathbf{A} \Bigr) +
    m c^{2} -
    F_{1}\, \bigl[ ze\, A_{0} + V(\mathbf{r}) - m c^{2} \bigr]
  \Bigr\}\, \chi.
\end{array}
\label{eq.2.23}
\end{equation}

We shall solve system of equations (\ref{eq.2.23}). We shall find equations in dependence on $\frac{\partial \varphi}{\partial t}$ and $\frac{\partial \chi}{\partial t}$ отдельно.
The first equation is obtained, after summarizing the second equation with the first one (\ref{eq.2.23})
with multiplication on corresponding factors:
\begin{equation}
\begin{array}{lll}
\vspace{0.9mm}
  & i\hbar\,
    \Bigl\{ F_{1}^{2} - F_{2}^{2} \bigl(q^{k} \sigma_{k} \bigr)^{2} \Bigr\}\,
    \displaystyle\frac{\partial \varphi}{\partial t}\; - \\

\vspace{0.1mm}
  - &
  \biggl\{
    c\, F_{2}\,
    \Bigl[
      - \varepsilon_{kmj}\, F_{1}\, q^{k}\, \sigma_{j} -
      i \bigl( F_{1} + F_{2} q^{4} \bigr)\, q^{k} \sigma_{k}\, \sigma_{m}
    \Bigr]
    \Bigl( \mathbf{p} - \displaystyle\frac{ze}{c} \mathbf{A} \Bigr)\; + \\

\vspace{1.7mm}
  + & F_{1}\, m c^{2} +
    \bigl[F_{1}^{2}\, + F_{2}^{2} (q^{k} \sigma_{k})^{2} \bigr]\,
      \bigl[ ze\, A_{0} + V(\mathbf{r}) - m c^{2} \bigr]
  \biggr\}\, \varphi\; = \\

  = &
  \biggl\{
    c
    \Bigl[
      F_{1} \bigl( F_{1} - F_{2} q^{4} \bigr)\, \sigma_{m} -
      i F_{2}^{2}\: q^{k} \sigma_{k}\, \varepsilon_{nmj}\, q^{n}\, \sigma_{j}
    \Bigr]
      \Bigl( \mathbf{p} - \displaystyle\frac{ze}{c} \mathbf{A} \Bigr) +
    i\, m c^{2}\, F_{2} q^{k} \sigma_{k} -
    2i\,F_{1}F_{2} q^{k} \sigma_{k} \bigl[ ze\, A_{0} + V(\mathbf{r}) - m c^{2} \bigr]
  \biggr\} \chi.
\end{array}
\label{eq.2.27}
\end{equation}

The second new equation is obtained, when we remove the second equation of system (\ref{eq.2.23}) from the first one with multiplication on
corresponding coefficients:
\begin{equation}
\begin{array}{lll}
\vspace{2.0mm}
  & i\hbar\,
    \Bigl\{ F_{1}^{2} - F_{2}^{2} \bigl(q^{k} \sigma_{k} \bigr)^{2} \Bigr\}\,
    \displaystyle\frac{\partial \chi}{\partial t}\; = \\

  = &
  \biggl\{
  c\,
  \Bigl[
    - i F_{2}^{2} q^{k} \sigma_{k}\, \varepsilon_{nmj}\, q^{n}\, \sigma_{j}\, +
    F_{1}\, \bigl( F_{1} + F_{2} q^{4} \bigr)\, \sigma_{m}\,
  \Bigr]\,
    \Bigl( \mathbf{p} - \displaystyle\frac{ze}{c} \mathbf{A} \Bigr) +
    i F_{2} q^{k} \sigma_{k}\, m c^{2}\; + \\

\vspace{2.0mm}
  + &
  2\,i\, F_{1}\, F_{2} q^{k} \sigma_{k}\, \bigl[ ze\, A_{0} + V(\mathbf{r}) - m c^{2} \bigr]
  \biggr\}\, \varphi\;+ \\

  + &
  \biggl\{
    c\,
    \Bigl[
      i F_{2} q^{k} \sigma_{k}\,
      \bigl( F_{1} - F_{2} q^{4} \bigr)\, \sigma_{m} +
      F_{1}\, \varepsilon_{kmj}\, F_{2} q^{k}\, \sigma_{j}
    \Bigr]\,
      \Bigl( \mathbf{p} - \displaystyle\frac{ze}{c} \mathbf{A} \Bigr)\; - \\

   - &
   F_{1}\, m c^{2} +
    \bigl[ F_{1}^{2} + F_{2}^{2} (q^{k} \sigma_{k})^{2} \bigr]\,
      \bigl[ ze\, A_{0} + V(\mathbf{r}) - m c^{2} \bigr]
  \biggr\}\, \chi.
\end{array}
\label{eq.2.28}
\end{equation}
One can see that at limit of
\begin{equation}
\begin{array}{lcl}
  F_{1} = 1 &
  F_{2} = 0,
\end{array}
\label{eq.2.29}
\end{equation}
the obtained equations (\ref{eq.2.27}) and (\ref{eq.2.28}) are transformed to known system of equations (1.3.3) in \cite{Ahiezer.1981}, p.~32]
(for one particle with addition of the potential $V$):
\begin{equation}
\begin{array}{lcl}
  i \hbar \displaystyle\frac{\partial \varphi}{\partial t} =
  c\, \sigmabf\: \Bigl( \mathbf{p} - \displaystyle\frac{ze}{c} \mathbf{A} \Bigr)\, \chi +
  ze\, A_{0}\, \varphi + V(\mathbf{r})\, \varphi, \\
  i \hbar \displaystyle\frac{\partial \chi}{\partial t} =
  c\, \sigmabf\: \Bigl( \mathbf{p} - \displaystyle\frac{ze}{c} \mathbf{A} \Bigr)\, \varphi -
  2\,m c^{2}\:\chi + ze\, A_{0}\:\chi + V(\mathbf{r})\; \chi.
\end{array}
\label{eq.2.30}
\end{equation}

Now we shall apply expansion over powers of $1/c$. Here, we follow idea given in \cite{Ahiezer.1981} (see p.~32--33 in that book).
Let us assume that $\chi$ has similar magnitude as $\varphi/c$.
In obtaining new equation in the first approximation, one can omit all terms with $\chi$
(with exception of $2\,m_{i} c^{2}\:\chi$ which includes $c^{2}$; but we keep terms with $\frac{ze}{c} \mathbf{A}$) in the second equation (\ref{eq.2.28}).
We obtain:
\begin{equation}
\begin{array}{lll}
\vspace{1.5mm}
  & \bigl[ F_{1} + F_{1}^{2} + F_{2}^{2} (q^{k} \sigma_{k})^{2} \bigr]\, \chi = \\

  = &
  \biggl\{
    \displaystyle\frac{1}{mc}\,
    \Bigl[
      - i F_{2}^{2} q^{k} \sigma_{k}\, \varepsilon_{nmj}\, q^{n}\, \sigma_{j}\, +
      F_{1} \bigl( F_{1} + F_{2} q^{4} \bigr)\, \sigma_{m}
    \Bigr]\,
      \Bigl( \mathbf{p} - \displaystyle\frac{ze}{c} \mathbf{A} \Bigr) +

    i\, (1 - 2\,F_{1})\, F_{2}\, q^{k} \sigma_{k}
  \biggr\}\, \varphi.
\end{array}
\label{eq.2.32}
\end{equation}

We take into account properties of Pauli matrixes:
\begin{equation}
\begin{array}{ll}
  (\sigma_{1})^{2} = (\sigma_{2})^{2} = (\sigma_{3})^{2}  = 1, &
  \sigma_{i}\sigma_{j} = \delta_{ij} I + i\, \varepsilon_{ijk} \sigma_{k},
\end{array}
\label{eq.2.33}
\end{equation}
where $\varepsilon_{ijk}$ is unit antisymmetric tensor, $\varepsilon_{123} = 1$.
We find:
\begin{equation}
\begin{array}{lll}
  \bigl( q^{k} \sigma_{k} \bigr)^{2} & = &
%
%
  I\, \mathbf{q}^{2},
\end{array}
\label{eq.2.36}
\end{equation}
where $I$ is unit matrix.
From (\ref{eq.2.32}) we obtain:
\begin{equation}
\begin{array}{lll}
  f(|\mathbf{q}|)\, \chi =
  \biggl\{
    \displaystyle\frac{1}{mc}\,
    \Bigl[
      - i F_{2}^{2} q^{k} \sigma_{k}\, \varepsilon_{nmj}\, q^{n}\, \sigma_{j}\, +
      F_{1} \bigl( F_{1} + F_{2} q^{4} \bigr)\, \sigma_{m}
    \Bigr]\,
      \Bigl( \mathbf{p} - \displaystyle\frac{ze}{c} \mathbf{A} \Bigr) +

    i\, (1 - 2\,F_{1})\, F_{2}\, q^{k} \sigma_{k}
  \biggr\}\, \varphi,
\end{array}
\label{eq.2.38}
\end{equation}
where we introduced a new notation:
\begin{equation}
  f(|\mathbf{q}|) =
  F_{1} + F_{1}^{2} + F_{2}^{2} (q^{k} \sigma_{k})^{2} =
  F_{1} + F_{1}^{2} + F_{2}^{2}\, \mathbf{q}^{2}.
\label{eq.2.37}
\end{equation}

As next step, we have to substitute this equation to equation (\ref{eq.2.27}).
Taking (\ref{eq.2.37}) into account, we obtain:
\begin{equation}
\begin{array}{lll}
\vspace{0.9mm}
  & i\hbar\,
    \Bigl\{ F_{1}^{2} - F_{2}^{2}\, \mathbf{q}^{2} \Bigr\}\,
    f(|\mathbf{q}|)
    \displaystyle\frac{\partial \varphi}{\partial t}\; - 

    \biggl\{
    c\, F_{2}\,
    \Bigl[
      - \varepsilon_{kmj}\, F_{1}\, q^{k}\, \sigma_{j} -
      i \bigl( F_{1} + F_{2} q^{4} \bigr)\, q^{k} \sigma_{k}\, \sigma_{m}
    \Bigr]
    \Bigl( \mathbf{p} - \displaystyle\frac{ze}{c} \mathbf{A} \Bigr)\; + \\

\vspace{1.7mm}
  + & F_{1}\, m c^{2} +
    \bigl(F_{1}^{2} + F_{2}^{2}\, \mathbf{q}^{2} \bigr)\,
    \bigl( ze\, A_{0} + V(\mathbf{r}) - m c^{2} \bigr)
  \biggr\} \cdot f(|\mathbf{q}|) \varphi\; = \\

  = &
  \biggl\{
    c\,
    \Bigl[
      F_{1} \bigl( F_{1} - F_{2} q^{4} \bigr)\, \sigma_{m} -
      i F_{2}^{2}\: q^{k} \sigma_{k}\, \varepsilon_{nmj}\, q^{n}\, \sigma_{j}
    \Bigr]\,
      \Bigl( \mathbf{p} - \displaystyle\frac{ze}{c} \mathbf{A} \Bigr) +
    i\, m c^{2}\, F_{2} q^{k} \sigma_{k} -
    2i\,F_{1}F_{2} q^{k} \sigma_{k} \bigl[ ze\, A_{0} + V(\mathbf{r}) - m c^{2} \bigr]
  \biggr\}\, \times \\

  \times &
  \biggl\{
    \displaystyle\frac{1}{mc}\,
    \Bigl[
      - i F_{2}^{2} q^{k} \sigma_{k}\, \varepsilon_{nmj}\, q^{n}\, \sigma_{j}\, +
      F_{1} \bigl( F_{1} + F_{2} q^{4} \bigr)\, \sigma_{m}
    \Bigr]\,
      \Bigl( \mathbf{p} - \displaystyle\frac{ze}{c} \mathbf{A} \Bigr) +
    i\, (1 - 2\,F_{1})\, F_{2}\, q^{k} \sigma_{k}
  \biggr\}\, \varphi.
\end{array}
\label{eq.2.41}
\end{equation}
So, we obtain a new equation which depends on one spinor function $\varphi$ only.
This equation is generalization of Pauli equation [see Eqs.~(1.3.5)--(1.3.7) in Ref.~\cite{Ahiezer.1981}, p.~33], but with included electromagnetic form-factors of nucleon and the interacting potential $V(\mathbf{r})$ [along to formalism in Ref.~\cite{Ahiezer.1981}, see p.~48--60].
It is convenient to rewrite this equation in compact form:
\begin{equation}
  i\hbar\,
    \Bigl\{ F_{1}^{2} - F_{2}^{2}\, \mathbf{q}^{2} \Bigr\}\,
    f(|\mathbf{q}|)
    \displaystyle\frac{\partial \varphi}{\partial t} =
  A \cdot f(|\mathbf{q}|) \cdot \varphi + B \cdot \varphi,
\label{eq.4.1.1}
\end{equation}
where
\begin{equation}
\begin{array}{lll}
\vspace{1.1mm}
  A & = &
  c\, F_{2}\,
  \Bigl[
    - \varepsilon_{kmj}\, F_{1}\, q^{k}\, \sigma_{j} -
    i \bigl( F_{1} + F_{2} q^{4} \bigr)\, q^{k} \sigma_{k}\, \sigma_{m}
  \Bigr]
  \Bigl( \mathbf{p} - \displaystyle\frac{ze}{c} \mathbf{A} \Bigr)\; + \\

  & + &
  F_{1}\, m c^{2} +
    \bigl(F_{1}^{2} + F_{2}^{2}\, \mathbf{q}^{2} \bigr)\,
    \bigl( ze\, A_{0} + V(\mathbf{r}) - m c^{2} \bigr),
\end{array}
\label{eq.2.43}
\end{equation}
\begin{equation}
\begin{array}{lll}
\vspace{0.9mm}
  B & = &
  \biggl\{
  c
  \Bigl[
    F_{1} \bigl( F_{1} - F_{2} q^{4} \bigr)\, \sigma_{m} -
    i F_{2}^{2}\: q^{k} \sigma_{k}\, \varepsilon_{nmj}\, q^{n}\, \sigma_{j}
  \Bigr]
    \Bigl( \mathbf{p} - \displaystyle\frac{ze}{c} \mathbf{A} \Bigr) +
    i\, m c^{2}\, F_{2} q^{k} \sigma_{k} -
    2i\,F_{1}F_{2} q^{k} \sigma_{k} \bigl[ ze\, A_{0} + V(\mathbf{r}) - m c^{2} \bigr]
  \biggr\}\, \times \\

  & \times &
  \biggl\{
    \displaystyle\frac{1}{mc}\,
    \Bigl[
      - i F_{2}^{2} q^{k} \sigma_{k}\, \varepsilon_{nmj}\, q^{n}\, \sigma_{j}\, +
      F_{1} \bigl( F_{1} + F_{2} q^{4} \bigr)\, \sigma_{m}
    \Bigr]\,
      \Bigl( \mathbf{p} - \displaystyle\frac{ze}{c} \mathbf{A} \Bigr) +
    i\, (1 - 2\,F_{1})\, F_{2}\, q^{k} \sigma_{k}
  \biggr\}.
\end{array}
\label{eq.2.44}
\end{equation}

After calculations, we obtain the following solutions for functions $A$ and $B$ (see Appendix A):
\begin{equation}
\begin{array}{lll}
  B - B_{1} + A \cdot f(|\mathbf{q}|) =
  i\, cF_{2}\,
  \Bigl\{ b_{1}\, q^{m} + b_{2}\, \varepsilon_{mjl}\, q^{j}\, \sigma_{l} \Bigr\}
  \Bigl( \mathbf{p} - \displaystyle\frac{ze}{c} \mathbf{A} \Bigr)_{m}\; +
  mc^{2}\,b_{3},
\end{array}
\label{eq.4.1.2}
\end{equation}
where
\begin{equation}
\begin{array}{lll}
  m B_{1} & = &
    a_{1}\;
    \Bigl[
      \Bigl( \mathbf{p}_{i} - \displaystyle\frac{z_{i}e}{c} \mathbf{A}_{i} \Bigr)^{2} -
      \displaystyle\frac{z_{i}e}{c}\, \sigmabf \mathbf{H}
    \Bigr]\; +
    \bigl(F_{2}^{4} \mathbf{q}^{2} + a_{2} + a_{3}\bigr)\, \Bigl( \mathbf{qp} - \displaystyle\frac{ze}{c}\, \mathbf{qA} \Bigr)^{2}\; + m\, \bar{B}_{10},
\end{array}
\label{eq.4.1.4}
\end{equation}
\begin{equation}
\begin{array}{lll}
  m \bar{B}_{10} & = &
  i\, \varepsilon_{lm^{\prime}k}\, \sigma_{k}\: q^{l} q^{m}\,
  \Bigl\{
    (a_{2}-a_{3}) \Bigl(p_{m} p_{m'} - \displaystyle\frac{ze}{c} ( A_{m'} p_{m} + A_{m} p_{m'} ) + \displaystyle\frac{z^{2}e^{2}}{c^{2}}\, A_{m} A_{m'} \Bigr) +
    \displaystyle\frac{i\hbar ze}{c} \Bigl[ a_{2}\, \displaystyle\frac{dA_{m'}}{dx_{m}} - a_{3}\, \displaystyle\frac{dA_{m}}{dx_{m'}} \Bigr]
  \Bigr\},
\end{array}
\label{eq.4.1.5}
\end{equation}
and
\begin{equation}
\begin{array}{lll}
\vspace{1.1mm}
  b_{1} & = &
  F_{1}^{2} (1 - F_{1}) + F_{1} F_{2} (3F_{1} - 1)\, q^{4} -
  F_{2}^{2} \bigl( F_{1} + F_{2} q^{4} \bigr)\, \mathbf{q}^{2} -
  \displaystyle\frac{2\,F_{1}^{2}}{mc^{2}} \bigl( F_{1} + F_{2} q^{4} \bigr) \bigl[ ze\, A_{0} + V(\mathbf{r}) \bigr], \\

\vspace{1.1mm}
  b_{2} & = &
    i\, \Bigl[
      2\, F_{1}^{2} +
      F_{1} F_{2} \bigl( 3 F_{1} - 1 \bigr)\, q^{4} -
      F_{2}^{2} \bigl( 2 + F_{2} q^{4} \bigr)\, \mathbf{q}^{2} -
      \displaystyle\frac{2\,F_{1}}{mc^{2}} \bigl( F_{1}^{2} + F_{1} F_{2} q^{4} - F_{2}^{2}\, \mathbf{q}^{2} \bigr) \bigl( ze\, A_{0} + V(\mathbf{r}) \bigr)
    \Bigr], \\

  b_{3} & = &
  \Bigl\{
    F_{1}^{2}\, (1 - F_{1}^{2}) -
    (1 - 2\,F_{1}^{2})\, F_{2}^{2}\, \mathbf{q}^{2} -
    F_{2}^{4}\, \mathbf{q}^{4}
  \Bigr\} +
  \displaystyle\frac{1}{mc^{2}}\,
  \Bigl\{
    F_{1}^{3}\, (1 + F_{1}) +
    F_{1} F_{2}^{2}\, (3 - 2 F_{1})\, \mathbf{q}^{2} +
    F_{2}^{4}\, \mathbf{q}^{4}
  \Bigr\} \bigl[ ze\, A_{0} + V(\mathbf{r}) \bigr].
\end{array}
\label{eq.4.1.3}
\end{equation}
Coefficients $a_{1}$, $a_{2}$ and $a_{3}$ are defined in Appendix~\ref{sec.2.3} [see Eqs.~(\ref{eq.app.2.3.3}) and (\ref{eq.app.2.3.8})].


\subsection{Operator of emission of bremsstrahlung photons
\label{sec.4}}

In this paper we use approximation $A_{0} = 0$.
%
%
On such a basis, from Eqs.~(\ref{eq.2.37}), (\ref{eq.4.1.2})--(\ref{eq.4.1.5})
%
%
we obtain (see Appendix~\ref{sec.app.4}, for details)
\begin{equation}
  i\hbar\, \Bigl\{ F_{1}^{3} (1 + F_{1}) - F_{1} F_{2}^{2}\, \mathbf{q}^{2} - F_{4}^{2}\, \mathbf{q}^{4} \Bigr\}\, \displaystyle\frac{\partial \varphi}{\partial t} =
  \bigl( h_{0} + h_{\gamma 0} + h_{\gamma 1} \bigr) \cdot \varphi,
\label{eq.4.1.8}
\end{equation}
where
\begin{equation}
\begin{array}{lll}
\vspace{0.1mm}
  h_{0} & = &
  \displaystyle\frac{a_{1}\, \mathbf{p}_{i}^{2}}{m} +
  \Bigl(
    F_{1}^{3}\, (1 + F_{1}) +
    F_{1} F_{2}^{2}\, (3 - 2 F_{1})\, \mathbf{q}^{2} +
    F_{2}^{4}\, \mathbf{q}^{4}
  \Bigr)\, V(\mathbf{r})\; + \\
\vspace{1.5mm}
  & + &
  mc^{2}
  \Bigl[
    F_{1}^{2}\, (1 - F_{1}^{2}) -
    (1 - 2\,F_{1}^{2})\, F_{2}^{2}\, \mathbf{q}^{2} -
    F_{2}^{4}\, \mathbf{q}^{4}
  \Bigr]\; - \\

\vspace{0.1mm}
  & - &
  i\, \displaystyle\frac{2F_{1}F_{2}}{mc}\,
  \Bigl\{
    F_{1} \bigl( F_{1} + F_{2} q^{4} \bigr)\, q^{m} +
    i\, \bigl( F_{1}^{2} + F_{1} F_{2} q^{4} - F_{2}^{2}\, \mathbf{q}^{2} \bigr) \varepsilon_{mjl}\, q^{j}\, \sigma_{l}
  \Bigr\}\, V(\mathbf{r})\, \mathbf{p}_{m}\; + \\
\vspace{0.1mm}
  & + &
  \displaystyle\frac{1}{m}\,
    \Bigl\{
      \bigl(F_{2}^{4} \mathbf{q}^{2} + a_{2} + a_{3}\bigr)\, \bigl( \mathbf{qp} \bigr)^{2} +
      i\, \varepsilon_{lm^{\prime}k}\, \sigma_{k}\: q^{l} q^{m}\, (a_{2}-a_{3}) p_{m} p_{m'}
    \Bigr\}\; + \\
\vspace{0.1mm}
  & + &
  i\, cF_{2}\,
  \Bigl\{
  \Bigl[
    F_{1}^{2} (1 - F_{1}) + F_{1} F_{2} (3F_{1} - 1)\, q^{4} -
    F_{2}^{2} \bigl( F_{1} + F_{2} q^{4} \bigr)\, \mathbf{q}^{2}
  \Bigr]\, q^{m}\; + \\
  & + &
  i\, \Bigl[
    2\, F_{1}^{2} +
    F_{1} F_{2} \bigl( 3 F_{1} - 1 \bigr)\, q^{4} -
    F_{2}^{2} \bigl( 2 + F_{2} q^{4} \bigr)\, \mathbf{q}^{2}
  \Bigr]\, \varepsilon_{mjl}\, q^{j}\, \sigma_{l} \Bigr\}\, \mathbf{p}_{m},
\end{array}
\label{eq.4.2.2}
\end{equation}
\begin{equation}
\begin{array}{lll}
\vspace{1.1mm}
  h_{\gamma 0} & = &
  \displaystyle\frac{a_{1}}{m}\,
    \Bigl[
      - \displaystyle\frac{z_{i}e}{c} (-i\hbar\, \mathbf{div A} + 2\,\mathbf{Ap}) +
      \displaystyle\frac{z_{i}^{2}e^{2}}{c^{2}}\, \mathbf{A}^{2} -
      \displaystyle\frac{z_{i}e}{c}\, \sigmabf \mathbf{H}
    \Bigr], \\

\vspace{0.5mm}
  h_{\gamma 1} & = &
  -\, i\, cF_{2}\,
    \Bigl\{ b_{1}\, q^{m} + b_{2}\, \varepsilon_{mjl}\, q^{j}\, \sigma_{l} \Bigr\}\,
    \displaystyle\frac{ze}{c} \mathbf{A}_{m} +
  \displaystyle\frac{1}{m}\,
  \Bigl\{
    \bigl(F_{2}^{4} \mathbf{q}^{2} + a_{2} + a_{3}\bigr)\,
      \Bigl[ -2 \mathbf{(qp)} \displaystyle\frac{ze}{c}\, \mathbf{(qA)} + \displaystyle\frac{z^{2}e^{2}}{c^{2}}\, \mathbf{(qA)}^{2} \Bigr]\; + \\
  & + &
  i\, \varepsilon_{lm^{\prime}k}\, \sigma_{k}\: q^{l} q^{m}\,
  \Bigl[
    (a_{2}-a_{3}) \Bigl(- \displaystyle\frac{ze}{c} ( A_{m'} p_{m} + A_{m} p_{m'} ) + \displaystyle\frac{z^{2}e^{2}}{c^{2}}\, A_{m} A_{m'} \Bigr) +
    \displaystyle\frac{i\hbar ze}{c} \Bigl( a_{2}\, \displaystyle\frac{dA_{m'}}{dx_{m}} - a_{3}\, \displaystyle\frac{dA_{m}}{dx_{m'}} \Bigr) \Bigr]
  \Bigr\}.
\end{array}
\label{eq.4.3.2}
\end{equation}
Here, the first term $h_{\gamma 0}$ is operator, describing electric and magnetic emissions of the bremsstrahlung photons without characteristics of the internal structure of the scattered proton. Peculiarities of such types of the (coherent and incoherent) bremsstrahlung emission in the proton-nucleus scattering were studied in details in Refs.~\cite{Maydanyuk.2012.PRC,Maydanyuk_Zhang.2015.PRC}.
The second term $h_{\gamma 1}$ is operator of emission, describing contribution to the full bremsstrahlung spectrum caused taking the internal structure of the scattered proton into account.


\subsection{Elastic scattering of virtual photon on proton 
\label{sec.5}}

For the first estimations of emission of the bremsstrahlung photons on the basis of the developed formalism,
we shall analyze elastic scattering of virtual photon on proton (scattered off the nucleus).
So, energy of this proton in the scattering is conserved and we have:
%
%
\begin{equation}
\begin{array}{lll}
  q_{4} = 0, &
  \mathbf{q}^{2} = - q^{2} = Q^{2}.
\end{array}
\label{eq.5.1.1}
\end{equation}
From Eqs.~(\ref{eq.app.2.3.3}) and (\ref{eq.app.2.3.8}) we calculate coefficients:
\begin{equation}
\begin{array}{lll}
  a_{1} = (F_{1}^{2} - F_{2}^{2}\, Q^{2})^{2} = \displaystyle\frac{a_{2}^{2}}{F_{2}^{4}}, &
  a_{2} = a_{3} = F_{2}^{2} \bigl[ F_{1}^{2} - F_{2}^{2}\, Q^{2} \bigr].
\end{array}
\label{eq.5.1.2}
\end{equation}
\begin{equation}
\begin{array}{lll}
\vspace{0.9mm}
  b_{1} & = &
  F_{1}^{2} (1 - F_{1}) -
  F_{1} F_{2}^{2}\, Q^{2} -
  \displaystyle\frac{2\,F_{1}^{3}}{mc^{2}}\, V(\mathbf{r}), \\

\vspace{0.9mm}
  b_{2} & = &
    i\, \Bigl[
      2\, F_{1}^{2} -
      2 F_{2}^{2}\, Q^{2} -
      \displaystyle\frac{2\,F_{1}}{mc^{2}} \bigl( F_{1}^{2} - F_{2}^{2} Q^{2} \bigr)\, V(\mathbf{r})
    \Bigr], \\

  b_{3} & = &
  \Bigl\{
    F_{1}^{2}\, (1 - F_{1}^{2}) -
    (1 - 2\,F_{1}^{2})\, F_{2}^{2}\, Q^{2} -
    F_{2}^{4}\, Q^{4}
  \Bigr\} +
  \displaystyle\frac{1}{mc^{2}}\,
  \Bigl\{
    F_{1}^{3}\, (1 + F_{1}) +
    F_{1} F_{2}^{2}\, (3 - 2 F_{1})\, Q^{2} +
    F_{2}^{4}\, Q^{4}
  \Bigr\}\, V(\mathbf{r}).
\end{array}
\label{eq.5.1.3}
\end{equation}

DIS theory provides kinematic relation for virtual photon and proton (see Eq.~(3.7) in Ref.~\cite{Greiner.Chromodynamics.2002}, p.~79; we do not take the internal structure of nucleon of the nucleus into account)%
\footnote{In our formalism developed in the space representation, application of DIS kinematic relations for virtual photons is approximation.}.
We obtain:
%
%
\begin{equation}
  \mathbf{qp} = -\displaystyle\frac{1}{2}\, Q^{2}.
\label{eq.5.2.1}
\end{equation}
%
%
We shall use QED representation for the vector potential of the bremsstrahlung emission:
\begin{equation}
\begin{array}{lcl}
  \mathbf{A} & = &
  \displaystyle\sum\limits_{\alpha=1,2}
    \sqrt{\displaystyle\frac{2\pi\hbar c^{2}}{w_{\rm ph}}}\; \mathbf{e}^{(\alpha),\,*}
    e^{-i\, \mathbf{k_{\rm ph}r}}.
\end{array}
\label{eq.5.3.1}
\end{equation}
Here, $\mathbf{e}^{(\alpha)}$ are unit vectors of polarization of the photon emitted ($\mathbf{e}^{(\alpha), *} = \mathbf{e}^{(\alpha)}$), $\mathbf{k}_{\rm ph}$ is wave vector of the photon and $w_{\rm ph} = k_{\rm ph} c = \bigl| \mathbf{k}_{\rm ph}\bigr|c$. Vectors $\mathbf{e}^{(\alpha)}$ are perpendicular to $\mathbf{k}_{\rm ph}$ in Coulomb calibration.
We have two independent polarizations $\mathbf{e}^{(1)}$ and $\mathbf{e}^{(2)}$ for the photon with impulse $\mathbf{k}_{\rm ph}$ ($\alpha=1,2$).
We have properties:
\begin{equation}
\begin{array}{lclc}
  \Bigl[ \mathbf{k}_{\rm ph} \times \mathbf{e}^{(1)} \Bigr] = k_{\rm ph}\, \mathbf{e}^{(2)}, &
  \Bigl[ \mathbf{k}_{\rm ph} \times \mathbf{e}^{(2)} \Bigr] = -\, k_{\rm ph}\, \mathbf{e}^{(1)}, &
  \Bigl[ \mathbf{k}_{\rm ph} \times \mathbf{e}^{(3)} \Bigr] = 0, &
  \displaystyle\sum\limits_{\alpha=1,2,3}
  \Bigl[ \mathbf{k}_{\rm ph} \times \mathbf{e}^{(\alpha)} \Bigr] =
  k_{\rm ph}\, (\mathbf{e}^{(2)} - \mathbf{e}^{(1)}).
\end{array}
\label{eq.5.3.2}
\end{equation}
We also need in determination of the scalar multiplication of the vectors $\mathbf{q}$ and $\mathbf{A}$.
As the first approximation, we shall introduce a new angle $\varphi_{ph}$ between these vectors:
\begin{equation}
\begin{array}{lcl}
  \mathbf{qA} & = & qA \cdot \sin \varphi_{ph}.
\end{array}
\label{eq.5.3.3}
\end{equation}
After calculations we obtain the following terms for the hamiltonian (see Appendix~\ref{sec.app.5}, for details):
\begin{equation}
\begin{array}{lll}
\vspace{0.1mm}
  h_{0} & = &
  \displaystyle\frac{a_{1}\, \mathbf{p}^{2}}{m} +
  \Bigl( F_{1}^{3}\, (1 + F_{1}) + F_{1} F_{2}^{2}\, (3 - 2 F_{1})\, Q^{2} + F_{2}^{4}\, Q^{4} + i\, \displaystyle\frac{F_{1}^{3}F_{2}\, Q^{2}}{mc} \Bigr) \bigl[ ze\, A_{0} + V(\mathbf{r}) \bigr]\; + \\
\vspace{1.1mm}
  & + &
  mc^{2} \Bigl[  F_{1}^{2}\, (1 - F_{1}^{2}) - (1 - 2\,F_{1}^{2})\, F_{2}^{2}\, Q^{2} - F_{2}^{4}\, Q^{4} \Bigr] +
  \displaystyle\frac{Q^{4}}{4m}\, \bigl(F_{2}^{4} Q^{2} + 2a_{2}\bigr) -
  \displaystyle\frac{i\, cF_{2}\, Q^{2}}{2}\, \Bigl[ F_{1}^{2} (1 - F_{1}) - F_{2}^{2} F_{1} \, Q^{2} \Bigr]\; + \\
  & + &
  \Bigl\{ \displaystyle\frac{2F_{1}F_{2}}{mc}\, \bigl[ ze\, A_{0} + V(\mathbf{r}) \bigr]\, - 2cF_{2} \Bigr\}
    \bigl( F_{1}^{2} - F_{2}^{2}\, Q^{2} \bigr) \varepsilon_{mjl}\, q^{j}\, \sigma_{l}\, \mathbf{p}_{m}.
\end{array}
\label{eq.5.3.4}
\end{equation}
\begin{equation}
\begin{array}{lll}
  h_{\gamma 1} & = &
  ze\, F_{2}\, \sqrt{\displaystyle\frac{\pi\hbar c^{2}}{w_{\rm ph}}}\; e^{-i\, \mathbf{k_{\rm ph}r}}\;
  \Bigl\{
    2 Q\, \sin \varphi_{ph}\,
      \Bigl[ -\, i\, b_{1} + \displaystyle\frac{F_{2}\, Q^{2}}{mc}\, \bigl( 2F_{1}^{2} - F_{2}^{2}\, Q^{2} \bigr) \Bigr]\; - \\
  & - &
    i\, \sqrt{2}\, b_{2}\, \varepsilon_{mjl}\, q^{j}\, \sigma_{l} \sum\limits_{\alpha=1,2} \mathbf{e}_{m}^{(\alpha),\,*} +
    \displaystyle\frac{4ze}{mc}\, \sqrt{\displaystyle\frac{\pi\hbar}{w_{\rm ph}}}\;
      F_{2}Q^{2}\, \bigl( 2F_{1}^{2} - F_{2}^{2}\, Q^{2} \bigr)\;
      e^{-i\, \mathbf{k_{\rm ph}r}}\, \sin^{2} \varphi_{ph}
  \Bigr\}.
\end{array}
\label{eq.5.3.5}
\end{equation}
$h_{\gamma 0}$ is not changed.

\vspace{0.7mm}
\subsection{Matrix element of emission of bremsstrahlung photons
\label{sec.6}}

Now we calculate the full matrix element of the bremsstrahlung emission.
We define it, using as basis our previous formalism~\cite{Maydanyuk.2012.PRC,Maydanyuk_Zhang.2015.PRC,Maydanyuk_Zhang_Zou.2016.PRC,Maydanyuk_Zhang_Zou.2018.PRC,Liu_Maydanyuk_Zhang_Liu.2018.arxiv}
(see also Refs.~\cite{Maydanyuk.2003.PTP,Maydanyuk.2006.EPJA,Maydanyuk.2008.EPJA,Maydanyuk.2008.MPLA,Maydanyuk.2009.NPA,Maydanyuk.2009.TONPPJ,%
Maydanyuk.2009.JPS,Maydanyuk.2010.PRC,Maydanyuk.2011.JPCS,Maydanyuk.2011.JPG}) as
\begin{equation}
\begin{array}{lcl}
  F_{fi} & \equiv &
  F_{fi,0} + F_{fi,1}  =
    \Bigl< k_{f} \Bigl|\, h_{\gamma 0} + h_{\gamma 1}\, \Bigr| \,k_{i} \Bigr> =
    \displaystyle\int
      \psi^{*}_{f}(\mathbf{r})\:
      (h_{\gamma 0} + h_{\gamma 1})\:
      \psi_{i}(\mathbf{r})\;
      \mathbf{dr},
\end{array}
\label{eq.6.1.1}
\end{equation}
where $\psi_{i}(\mathbf{r}) = |k_{i}\bigr>$ and $\psi_{f}(\mathbf{r}) = |k_{f}\bigr>$ are the stationary wave functions of the proton-nucleus system in the initial $i$-state (i.e. state before emission of the bremsstrahlung photon) and final $f$-state (i.e. state after emission of this photon) which do not contain number of photons emitted.

In further development of our formalism, we shall assume that it is impossible to fix polarization of virtual photon concerning polarization of the bremsstrahlung photon.
So, we have to average the matrix elements of emission over angle $\varphi_{ph}$ and
obtain (see Appendix~\ref{sec.app.6} for details):
\begin{equation}
\begin{array}{lll}
\vspace{0.5mm}
  F_{fi,0} & = &
  \bigl< k_{f} \bigl|\,  h_{\gamma 0}\, \bigr| \,k_{i} \bigr> \quad = \quad
  Z_{\rm eff}\, \displaystyle\frac{e}{mc}\,
    \sqrt{\displaystyle\frac{2\pi\hbar c^{2}}{w}}\;
    \Bigl\{ p_{\rm el} + p_{\rm mag, 1} + p_{\rm mag, 2} \Bigr\}, \\

  F_{fi,1} & = &
  \bigl< k_{f} \bigl|\,  h_{\gamma 0}\, \bigr| \,k_{i} \bigr> \quad = \quad
    Z_{\rm eff}\; e\, F_{2}\, \sqrt{\displaystyle\frac{\pi\hbar c^{2}}{w_{\rm ph}}}\;
    \Bigl\{ p_{\rm q,1} + p_{\rm q, 2} + p_{\rm q, 3} \Bigr\},
\end{array}
\label{eq.6.1.2}
\end{equation}
where
\begin{equation}
\begin{array}{lcl}
  \vspace{2mm}
  p_{\rm el} & = &
  i \displaystyle\sum\limits_{\alpha=1,2}
    \mathbf{e}^{(\alpha)}\,
    \Bigl< k_{f} \Bigl|\, e^{-i\,\mathbf{kr}}\: \nabla\, \Bigr| \,k_{i} \Bigr>, \\

  \vspace{2mm}
  p_{\rm mag, 1} & = &
  \displaystyle\frac{1}{2}\:
  \displaystyle\sum\limits_{\alpha=1,2}
  \Bigl< k_{f} \Bigl|\,
    e^{-i\,\mathbf{kr}}\;
    \sigmabf\cdot \Bigl[ \mathbf{e}^{(\alpha)} \times \nabla \Bigr]\,
  \Bigr| \,k_{i} \Bigr>, \\

  p_{\rm mag, 2} & = & -
  i\,\displaystyle\frac{1}{2}\:
  \displaystyle\sum\limits_{\alpha=1,2}
  \Bigl[ \mathbf{k} \times \mathbf{e}^{(\alpha)} \Bigr]\:
  \Bigl< k_{f} \Bigl|\, e^{-i\,\mathbf{kr}}\; \sigmabf\, \Bigr| \,k_{i} \Bigr>,
\end{array}
\label{eq.6.1.3}
\end{equation}
\begin{equation}
\begin{array}{lcl}
\vspace{1.3mm}
  \tilde{p}_{\rm q,1} & = &
    i\, A_{1} (Q, F_{1}, F_{2})\, \Bigl< k_{f} \Bigl|\, e^{-i\, \mathbf{k_{\rm ph}r}}\; \Bigr| \,k_{i} \Bigr> +
    i\, B_{1} (Q, F_{1}, F_{2})\, \Bigl< k_{f} \Bigl|\, e^{-i\, \mathbf{k_{\rm ph}r}}\; V(\mathbf{r}) \Bigr| \,k_{i} \Bigr>, \\

\vspace{1.3mm}
  \tilde{p}_{\rm q,2} & = &
    i\, A_{2} (Q, F_{1}, F_{2})\, \Bigl< k_{f} \Bigl|\, e^{-i\, \mathbf{k_{\rm ph}r}}\; \Bigr| \,k_{i} \Bigr> +
    i\, B_{2} (Q, F_{1}, F_{2})\, \Bigl< k_{f} \Bigl|\, e^{-i\, \mathbf{k_{\rm ph}r}}\; V(\mathbf{r}) \Bigr| \,k_{i} \Bigr>, \\

  \tilde{p}_{\rm q,3} & = &
    i\, A_{3} (Q, F_{1}, F_{2})\, \Bigl< k_{f} \Bigl|\, e^{-i\, \mathbf{2k_{\rm ph}r}}\; \Bigr| \,k_{i} \Bigr>
\end{array}
\label{eq.7.1.4}
\end{equation}
and
\begin{equation}
\begin{array}{lcl}
\vspace{1.5mm}
  A_{1} (Q, F_{1}, F_{2}) & = &
    -\, \displaystyle\frac{4 Q}{\pi}
    \Bigl\{ \Bigl[ F_{1}^{2} (1 - F_{1}) - F_{1} F_{2}^{2}\, Q^{2} \Bigr] +
       i\, \displaystyle\frac{F_{2}\, Q^{2}}{\pi\, mc}\, \bigl( 2F_{1}^{2} - F_{2}^{2}\, Q^{2} \bigr) \Bigr\}, \\
\vspace{1.5mm}
  B_{1} (Q, F_{1}, F_{2}) & = & 8 Q\, \displaystyle\frac{F_{1}^{3}}{\pi\, mc^{2}}, \\

\vspace{1.5mm}
  A_{2} (Q, F_{1}, F_{2}) & = &
    - i\, 2\, \bigl( F_{1}^{2} - F_{2}^{2}\, Q^{2} \bigr)\,
    \sqrt{2}\, \varepsilon_{mjl}\, q^{j}\, \sigma_{l} \sum\limits_{\alpha=1,2} \mathbf{e}_{m}^{(\alpha),\,*}, \\

\vspace{1.2mm}
  B_{2} (Q, F_{1}, F_{2}) & = &
    i\, 2\, \bigl( F_{1}^{2} - F_{2}^{2} Q^{2} \bigr)\,
    \displaystyle\frac{\sqrt{2}\, F_{1}}{mc^{2}}\,
    \varepsilon_{mjl}\, q^{j}\, \sigma_{l} \sum\limits_{\alpha=1,2} \mathbf{e}_{m}^{(\alpha),\,*}, \\

  A_{3} (Q, F_{1}, F_{2}) & = &
  - i\, \displaystyle\frac{2 ze}{mc}\, \sqrt{\displaystyle\frac{\pi\hbar}{w_{\rm ph}}}\;
  F_{2}Q^{2}\, \bigl( 2F_{1}^{2} - F_{2}^{2}\, Q^{2} \bigr).
\end{array}
\label{eq.7.1.5}
\end{equation}
One can see that the magnetic moment of the scattered proton gives own correction to the full magnetic bremsstrahlung emission via components $p_{\rm q,1}$, $p_{\rm q,2}$ and $p_{\rm q,3}$ (i.e. the magnetic field of the full nuclear system is changed).
Now a new physical question has been appeared about magnitude of such a magnetic emission.
Also a new type of space distribution of the emitted photons is appeared via term $\Bigl< k_{f} \Bigl|\, e^{-i\, \mathbf{k_{\rm ph}r}}\; V(\mathbf{r}) \Bigr| \,k_{i} \Bigr>$.


\vspace{1.5mm}
\subsection{Wave function of nuclear system and summation over spinor states
\label{sec.8}}

We define the wave function of the proton in field of the nucleus, according to formalism in Ref.~\cite{Maydanyuk.2012.PRC} (see Sect.~C, Eqs.~(12)--(13) in that paper).
We construct it in form of bilinear combination of eigenfunctions of orbital and spinor subsystems
(as Eq.~(1.4.2) in \cite{Ahiezer.1981}, p.~42).
%
%
However, we shall assume that it is not possible to fix experimentally states for selected $M$ (eigenvalue of momentum operator $\hat{J}_{z}$).
So, we shall be interesting in superposition over all states with different $M$ and define the wave function so%
\footnote{Here, the function (\ref{eq.8.1.1}) is spinor (i.e. two component) solution of equation (\ref{eq.4.1.8}) (which is generalization of the Pauli equation).
At the same time, wave function (\ref{eq.2.15}) is bi-spinor (i.e. four component) solution of the Dirac equation.
Along to QED formalism (see Ref.~\cite{Ahiezer.1981}, p.~42--44), the wave function (\ref{eq.4.1.8}) is fully characterized by quantum number $l$
(while two components of unite solution (\ref{eq.2.15}) of the Dirac equation have different values of $l$, and so different radial components
in the spherically symmetric consideration).
So, representation (\ref{eq.8.1.1}) in determination of solution of equation (\ref{eq.4.1.8}) is correct.
%
}:
\begin{equation}
  \varphi_{jl} (\mathbf{r}, s) =
  R_{l}\,(r)\:
  \displaystyle\sum\limits_{m=-l}^{l}
  \displaystyle\sum\limits_{\mu = \pm 1/2}
    C_{lm 1/2 \mu}^{j,M=m+\mu}\, Y_{lm}(\mathbf{n}_{\rm r})\, v_{\mu} (s),
\label{eq.8.1.1}
\end{equation}
where $R\,(r)$ is radial scalar function (not dependent on $m$ at the same $l$), $\mathbf{n}_{\rm r} = \mathbf{r} / r$ is unit vector directed along $\mathbf{r}$, $Y_{lm}(\mathbf{n}_{\rm r})$ are spherical functions (we use definition (28,7)--(28,8), p.~119 in~\cite{Landau.v3.1989}), $C_{lm 1/2 \mu}^{jM}$ are Clebsh-Gordon coefficients, $s$ is variable of spin, $M = m + \mu$ and $l = j \pm 1/2$.
For convenience of calculations we shall use spacial wave function as
\begin{equation}
  \varphi_{lm} (\mathbf{r}) =
  R_{l}\,(r)\: Y_{lm}(\mathbf{n}_{\rm r}).
\label{eq.8.1.2}
\end{equation}


Using representation (\ref{eq.8.1.1}) for the wave functions, we calculate the matrix elements (\ref{eq.6.1.3}) and (\ref{eq.7.1.4}),
according to formalism in Ref.~\cite{Maydanyuk.2012.PRC} [see Sect.~C, Eqs.~(14)--(23) in that paper]:
\begin{equation}
\begin{array}{lcl}
  \vspace{2mm}
  p_{\rm el} & = &
  i\, \displaystyle\sum\limits_{m_{f}, m_{i}}
  \displaystyle\sum\limits_{\mu_{i},\, \mu_{f} = \pm 1/2}
    C_{l_{f}m_{f} 1/2 \mu_{f}}^{j_{f}M_{f},\,*}\,
    C_{l_{i}m_{i} 1/2 \mu_{i}}^{j_{i}M_{i}} \cdot
    (\mathbf{e}^{(1)} + \mathbf{e}^{(2)})\,
    \Bigl< k_{f} \Bigl|\, e^{-i\,\mathbf{kr}}\: \nabla\, \Bigr| \,k_{i} \Bigr>_\mathbf{r}, \\

  \vspace{1mm}
  p_{\rm mag,\,1} & = &
  \displaystyle\frac{1}{2}
  \displaystyle\sum\limits_{m_{f}, m_{i}}
  \displaystyle\sum\limits_{\mu_{i},\, \mu_{f} = \pm 1/2}
    C_{l_{f}m_{f} 1/2 \mu_{f}}^{j_{f}M_{f},\,*}\,
    C_{l_{i}m_{i} 1/2 \mu_{i}}^{j_{i}M_{i}} \cdot
    \Bigl[
      \mathbf{e}_{\rm x} +
      \mathbf{e}_{\rm y}\, i\, \Bigl\{ \delta_{\mu_{i}, +1/2}\; -\; \delta_{\mu_{i}, -1/2} \Bigr\} +
      \mathbf{e}_{\rm z} \Bigr] \times \\
  \vspace{2mm}
  & \times &
    \biggl[
      \displaystyle\sum\limits_{\alpha=1,2} \mathbf{\rm e}^{(\alpha)} \times
      \Bigl< k_{f} \Bigl|\, e^{-i\,\mathbf{kr}}\, \nabla \Bigr| \,k_{i} \Bigr>_\mathbf{r} \biggr], \\

  p_{\rm mag,\,2} & = &
  \displaystyle\frac{-i\,k}{2}\,
  \displaystyle\sum\limits_{m_{f}, m_{i}}
  \displaystyle\sum\limits_{\mu_{i},\, \mu_{f} = \pm 1/2}
    C_{l_{f}m_{f} 1/2 \mu_{f}}^{j_{f}M_{f},\,*}\,
    C_{l_{i}m_{i} 1/2 \mu_{i}}^{j_{i}M_{i}} \cdot
  \Bigl[ -1 + i\, \Bigl\{ \delta_{\mu_{i}, +1/2}\; -\; \delta_{\mu_{i}, -1/2} \Bigr\} \Bigr]\:
  \Bigl< k_{f} \Bigl|\, e^{-i\,\mathbf{kr}}\, \Bigr| \,k_{i} \Bigr>_\mathbf{r}
\end{array}
\label{eq.8.1.12}
\end{equation}
and we obtain formulas for new matrix elements:
\begin{equation}
\begin{array}{lcl}
\vspace{1.3mm}
  \tilde{p}_{\rm q,1} & = &
  i \displaystyle\sum\limits_{m_{f}, m_{i}}
  \displaystyle\sum\limits_{\mu_{i},\, \mu_{f} = \pm 1/2}
    C_{l_{f}m_{f} 1/2 \mu_{f}}^{j_{f}M_{f},\,*}\,
    C_{l_{i}m_{i} 1/2 \mu_{i}}^{j_{i}M_{i}}\; \times \\

\vspace{1.3mm}
  & \times &
  \Bigl\{
    A_{1} (Q, F_{1}, F_{2})\, \Bigl< k_{f} \Bigl|\, e^{-i\, \mathbf{k_{\rm ph}r}}\; \Bigr| \,k_{i} \Bigr>_\mathbf{r} +
    B_{1} (Q, F_{1}, F_{2})\, \Bigl< k_{f} \Bigl|\, e^{-i\, \mathbf{k_{\rm ph}r}}\; V(\mathbf{r}) \Bigr| \,k_{i} \Bigr>_\mathbf{r}
  \Bigr\}, \\

  \tilde{p}_{\rm q,3} & = &
  i \displaystyle\sum\limits_{m_{f}, m_{i}}
  \displaystyle\sum\limits_{\mu_{i},\, \mu_{f} = \pm 1/2}
    C_{l_{f}m_{f} 1/2 \mu_{f}}^{j_{f}M_{f},\,*}\,
    C_{l_{i}m_{i} 1/2 \mu_{i}}^{j_{i}M_{i}} \cdot
    A_{3} (Q, F_{1}, F_{2})\, \Bigl< k_{f} \Bigl|\, e^{-2i\, \mathbf{k_{\rm ph}r}}\; \Bigr| \,k_{i} \Bigr>_\mathbf{r}.
\end{array}
\label{eq.8.2.2}
\end{equation}
Here, $\bigl< k_{f} \bigl|\, \ldots \bigr| \,k_{i} \bigr>_\mathbf{r}$ is one-component matrix element
\begin{equation}
\begin{array}{lcl}
  \Bigl< k_{f} \Bigl|\, \hat{f}  \Bigr| \,k_{i} \Bigr>_\mathbf{r} & \equiv &
  \displaystyle\int
    R_{f}^{*}\,(r)\:
    Y_{l_{f}m_{f}}(\mathbf{n}_{\rm r})^{*}\;
    \hat{f} \:
    R_{i}\,(r)\:
    Y_{l_{i}m_{i}}(\mathbf{n}_{\rm r})\; \mathbf{dr},
\end{array}
\label{eq.8.1.10}
\end{equation}
where integration should be performed over space coordinates only.
Here, we orient frame vectors $\mathbf{e}_{\rm x}$, $\mathbf{e}_{\rm y}$ and $\mathbf{e}_{\rm z}$ so, that $\mathbf{e}_{\rm z}$ be directed along to $\mathbf{k}$. Then, vectors $\mathbf{e}_{\rm x}$ and $\mathbf{e}_{\rm y}$ can be directed along $\mathbf{e}^{(1)}$ and $\mathbf{e}^{(2)}$, correspondingly.
In Coulomb gauge we have:
\begin{equation}
\begin{array}{lcccc}
  \mathbf{e}_{\rm x} = \mathbf{e}^{(1)}, &
  \mathbf{e}_{\rm y} = \mathbf{e}^{(2)}, &
  |\mathbf{e}_{\rm x}| = |\mathbf{e}_{\rm y}| = |\mathbf{e}_{\rm z}| = 1, &
  |\mathbf{e}^{(3)}| = 0.
\end{array}
\label{eq.8.1.11}
\end{equation}

\vspace{1.2mm}
\subsection{Calculations of matrix elements of emission in multipolar expansion
\label{sec.9}}

As nest step, we apply multipolar expansion in order to calculate the matrix elements of bremsstrahlung emission.
Such calculations are straightforward and they are presented in Appendix~\ref{sec.app.9}.
Calculation of radial integrals in the obtained solutions (\ref{eq.app.9.3.8}), (\ref{eq.app.9.4.1}), (\ref{eq.app.9.7.3}), (\ref{eq.app.9.7.7}) and (\ref{eq.app.9.8.4}) for the matrix elements in that Appendix is the most difficult numeric part in this research. But, they do not depend on $\mu$, and also $m_{i}$, $m_{f}$.
So, one can rewrite the matrix elements, performing summation over such quantum numbers.

\subsubsection{Representation for matrix element $p_{\rm el}$
\label{sec.9.9.1}}

For $p_{\rm el}$ from (\ref{eq.app.9.3.8}) and (\ref{eq.app.9.7.4}) we obtain:
\begin{equation}
\begin{array}{lcl}
\vspace{1.5mm}
  p_{\rm el}^{M} & = &
  \sqrt{\displaystyle\frac{\pi}{2}}\:
  \displaystyle\sum\limits_{l_{\rm ph}=1}\,
    (-i)^{l_{\rm ph}}\, \sqrt{2l_{\rm ph}+1}\; \cdot
    \biggl\{
      \sqrt{\displaystyle\frac{l_{i}}{2l_{i}+1}}\,
      K_{\rm el}^{M} (l_{i},l_{f}, l_{\rm ph}, l_{i}-1) \cdot
    \Bigl[ J_{1}(l_{i},l_{f},l_{\rm ph}) + (l_{i}+1) \cdot J_{2}(l_{i},l_{f},l_{\rm ph}) \Bigr]\; - \\

\vspace{1.5mm}
  & - &
  \sqrt{\displaystyle\frac{l_{i}+1}{2l_{i}+1}}\,
  K_{\rm el}^{M} (l_{i},l_{f}, l_{\rm ph}, l_{i}+1) \cdot
  \Bigl[ J_{1}(l_{i},l_{f},l_{\rm ph}) - l_{i} \cdot J_{2}(l_{i},l_{f},l_{\rm ph}) \Bigr]
  \biggr\}, \\

\vspace{1.5mm}
  p_{\rm el}^{E} & = &
  \sqrt{\displaystyle\frac{\pi}{2}}\:
  \displaystyle\sum\limits_{l_{\rm ph}=1}\,
    (-i)^{l_{\rm ph}}\, \sqrt{2l_{\rm ph}+1}\; \times \\
\vspace{1mm}
  & \times &
    \biggl\{
      \sqrt{\displaystyle\frac{l_{i}\,(l_{\rm ph}+1)}{(2l_{i}+1)(2l_{\rm ph}+1)}} \cdot
      K_{E}(l_{i},l_{f}, l_{\rm ph}, l_{i}-1, l_{\rm ph}-1) \cdot
      \Bigl[
        J_{1}(l_{i},l_{f},l_{\rm ph}-1)\; +
        (l_{i}+1) \cdot J_{2}(l_{i},l_{f},l_{\rm ph}-1)
      \Bigr]\; - \\
\vspace{1mm}
    & - &
    \sqrt{\displaystyle\frac{l_{i}\,l_{\rm ph}}{(2l_{i}+1)(2l_{\rm ph}+1)}} \cdot
      K_{E} (l_{i},l_{f}, l_{\rm ph}, l_{i}-1, l_{\rm ph}+1) \cdot
      \Bigl[ J_{1}(l_{i},l_{f},l_{\rm ph}+1)\; + (l_{i}+1) \cdot J_{2}(l_{i},l_{f},l_{\rm ph}+1) \Bigr]\; + \\
\vspace{1mm}
  & + &
    \sqrt{\displaystyle\frac{(l_{i}+1)(l_{\rm ph}+1)}{(2l_{i}+1)(2l_{\rm ph}+1)}} \cdot
      K_{E} (l_{i},l_{f},l_{\rm ph}, l_{i}+1, l_{\rm ph}-1) \cdot
      \Bigl[ J_{1}(l_{i},l_{f},l_{\rm ph}-1) - l_{i} \cdot J_{2}(l_{i},l_{f},l_{\rm ph}-1) \Bigr]\; - \\
  & - &
    \sqrt{\displaystyle\frac{(l_{i}+1)\,l_{\rm ph}}{(2l_{i}+1)(2l_{\rm ph}+1)}} \cdot
      K_{E} (l_{i},l_{f}, l_{\rm ph}, l_{i}+1, l_{\rm ph}+1) \cdot
      \Bigl[ J_{1}(l_{i},l_{f},l_{\rm ph}+1) - l_{i} \cdot J_{2}(l_{i},l_{f},l_{\rm ph}+1) \Bigr]
  \biggr\},
\end{array}
\label{eq.9.9.1.1}
\end{equation}
where
\begin{equation}
\begin{array}{lcl}
\vspace{1.5mm}
  K_{\rm el}^{M} (l_{i},l_{f}, l_{\rm ph}, l_{i}-1) & = &
  \displaystyle\sum\limits_{\mu=\pm 1} h_{\mu} \cdot i\,\mu\;
  \displaystyle\sum\limits_{m_{i}, m_{f}}
  \displaystyle\sum\limits_{\mu_{i},\, \mu_{f} = \pm 1/2}
    C_{l_{f}m_{f} 1/2 \mu_{f}}^{j_{f}M_{f},\,*}\,
    C_{l_{i}m_{i} 1/2 \mu_{i}}^{j_{i}M_{i}}\;
    I_{M}(l_{i},l_{f}, l_{\rm ph}, l_{i}-1, \mu), \\

  K_{\rm el}^{M} (l_{i},l_{f}, l_{\rm ph}, l_{i}+1) & = &
  \displaystyle\sum\limits_{\mu=\pm 1} h_{\mu} \cdot i\,\mu\;
  \displaystyle\sum\limits_{m_{i}, m_{f}}
  \displaystyle\sum\limits_{\mu_{i},\, \mu_{f} = \pm 1/2}
    C_{l_{f}m_{f} 1/2 \mu_{f}}^{j_{f}M_{f},\,*}\,
    C_{l_{i}m_{i} 1/2 \mu_{i}}^{j_{i}M_{i}}\;
    I_{M}(l_{i},l_{f}, l_{\rm ph}, l_{i}+1, \mu).
\end{array}
\label{eq.9.9.1.2}
\end{equation}
\begin{equation}
\begin{array}{lcl}
\vspace{1.5mm}
\vspace{1mm}
  K_{E}(l_{i},l_{f}, l_{\rm ph}, l_{i}-1, l_{\rm ph}-1) & = &
  \displaystyle\sum\limits_{\mu=\pm 1} h_{\mu}
  \displaystyle\sum\limits_{m_{i}, m_{f}}
  \displaystyle\sum\limits_{\mu_{i},\, \mu_{f} = \pm 1/2}
    C_{l_{f}m_{f} 1/2 \mu_{f}}^{j_{f}M_{f},\,*}\,
    C_{l_{i}m_{i} 1/2 \mu_{i}}^{j_{i}M_{i}}\;
    I_{E}(l_{i},l_{f}, l_{\rm ph}, l_{i}-1, l_{\rm ph}-1, \mu), \\

\vspace{1mm}
  K_{E} (l_{i},l_{f}, l_{\rm ph}, l_{i}-1, l_{\rm ph}+1) & = &
  \displaystyle\sum\limits_{\mu=\pm 1} h_{\mu}
  \displaystyle\sum\limits_{m_{i}, m_{f}}
  \displaystyle\sum\limits_{\mu_{i},\, \mu_{f} = \pm 1/2}
    C_{l_{f}m_{f} 1/2 \mu_{f}}^{j_{f}M_{f},\,*}\,
    C_{l_{i}m_{i} 1/2 \mu_{i}}^{j_{i}M_{i}}\;
    I_{E} (l_{i},l_{f}, l_{\rm ph}, l_{i}-1, l_{\rm ph}+1), \mu), \\

\vspace{1mm}
  K_{E}(l_{i},l_{f}, l_{\rm ph}, l_{i}+1, l_{\rm ph}-1) & = &
  \displaystyle\sum\limits_{\mu=\pm 1} h_{\mu}
  \displaystyle\sum\limits_{m_{i}, m_{f}}
  \displaystyle\sum\limits_{\mu_{i},\, \mu_{f} = \pm 1/2}
    C_{l_{f}m_{f} 1/2 \mu_{f}}^{j_{f}M_{f},\,*}\,
    C_{l_{i}m_{i} 1/2 \mu_{i}}^{j_{i}M_{i}}\;
    I_{E}(l_{i},l_{f}, l_{\rm ph}, l_{i}+1, l_{\rm ph}-1, \mu), \\

  K_{E}(l_{i},l_{f}, l_{\rm ph}, l_{i}+1, l_{\rm ph}+1) & = &
  \displaystyle\sum\limits_{\mu=\pm 1} h_{\mu}
  \displaystyle\sum\limits_{m_{i}, m_{f}}
  \displaystyle\sum\limits_{\mu_{i},\, \mu_{f} = \pm 1/2}
    C_{l_{f}m_{f} 1/2 \mu_{f}}^{j_{f}M_{f},\,*}\,
    C_{l_{i}m_{i} 1/2 \mu_{i}}^{j_{i}M_{i}}\;
    I_{E}(l_{i},l_{f}, l_{\rm ph}, l_{i}+1, l_{\rm ph}+1, \mu).
\end{array}
\label{eq.9.9.1.4}
\end{equation}

\subsubsection{Representation for matrix element $p_{\rm mag,1}$
\label{sec.9.9.2}}

For electric and magnetic components of $p_{\rm mag,1}$ from (\ref{eq.app.9.3.8}) and (\ref{eq.app.9.7.7}) we have:
\begin{equation}
\begin{array}{lcl}
\vspace{1.5mm}
  p_{\rm mag,1}^{M} & = &
  \displaystyle\frac{1}{2}
  \sqrt{\displaystyle\frac{\pi}{2}}\:
  \displaystyle\sum\limits_{l_{\rm ph}=1}\,
    (-i)^{l_{\rm ph}}\, \sqrt{2l_{\rm ph}+1}\; \cdot
    \biggl\{
      \sqrt{\displaystyle\frac{l_{i}}{2l_{i}+1}}\,
      K_{\rm mag,1}^{M} (l_{i},l_{f}, l_{\rm ph}, l_{i}-1) \cdot
    \Bigl[ J_{1}(l_{i},l_{f},l_{\rm ph}) + (l_{i}+1) \cdot J_{2}(l_{i},l_{f},l_{\rm ph}) \Bigr]\; - \\

\vspace{1.5mm}
  & - &
  \sqrt{\displaystyle\frac{l_{i}+1}{2l_{i}+1}}\,
  K_{\rm mag,1}^{M} (l_{i},l_{f}, l_{\rm ph}, l_{i}+1) \cdot
  \Bigl[ J_{1}(l_{i},l_{f},l_{\rm ph}) - l_{i} \cdot J_{2}(l_{i},l_{f},l_{\rm ph}) \Bigr]
  \biggr\}, \\

\vspace{1.5mm}
  p_{\rm mag,1}^{E} & = &
  \displaystyle\frac{1}{2}\,
  \sqrt{\displaystyle\frac{\pi}{2}}\:
  \displaystyle\sum\limits_{l_{\rm ph}=1}\,
    (-i)^{l_{\rm ph}}\, \sqrt{2l_{\rm ph}+1}\; \times \\
\vspace{1mm}
  & \times &
    \biggl\{
      \sqrt{\displaystyle\frac{l_{i}\,(l_{\rm ph}+1)}{(2l_{i}+1)(2l_{\rm ph}+1)}} \cdot
      K_{E}(l_{i},l_{f}, l_{\rm ph}, l_{i}-1, l_{\rm ph}-1) \cdot
      \Bigl[
        J_{1}(l_{i},l_{f},l_{\rm ph}-1)\; +
        (l_{i}+1) \cdot J_{2}(l_{i},l_{f},l_{\rm ph}-1)
      \Bigr]\; - \\
\vspace{1mm}
    & - &
    \sqrt{\displaystyle\frac{l_{i}\,l_{\rm ph}}{(2l_{i}+1)(2l_{\rm ph}+1)}} \cdot
      K_{E} (l_{i},l_{f}, l_{\rm ph}, l_{i}-1, l_{\rm ph}+1) \cdot
      \Bigl[ J_{1}(l_{i},l_{f},l_{\rm ph}+1)\; + (l_{i}+1) \cdot J_{2}(l_{i},l_{f},l_{\rm ph}+1) \Bigr]\; + \\
\vspace{1mm}
  & + &
    \sqrt{\displaystyle\frac{(l_{i}+1)(l_{\rm ph}+1)}{(2l_{i}+1)(2l_{\rm ph}+1)}} \cdot
      K_{E} (l_{i},l_{f},l_{\rm ph}, l_{i}+1, l_{\rm ph}-1) \cdot
      \Bigl[ J_{1}(l_{i},l_{f},l_{\rm ph}-1) - l_{i} \cdot J_{2}(l_{i},l_{f},l_{\rm ph}-1) \Bigr]\; - \\
  & - &
    \sqrt{\displaystyle\frac{(l_{i}+1)\,l_{\rm ph}}{(2l_{i}+1)(2l_{\rm ph}+1)}} \cdot
      K_{E} (l_{i},l_{f}, l_{\rm ph}, l_{i}+1, l_{\rm ph}+1) \cdot
      \Bigl[ J_{1}(l_{i},l_{f},l_{\rm ph}+1) - l_{i} \cdot J_{2}(l_{i},l_{f},l_{\rm ph}+1) \Bigr]
  \biggr\},
\end{array}
\label{eq.9.9.2.1}
\end{equation}
where
\begin{equation}
\begin{array}{lcl}
\vspace{1.5mm}
  K_{\rm mag,1}^{M} (l_{i},l_{f}, l_{\rm ph}, l_{i}-1) & = &
  \displaystyle\sum\limits_{\mu=\pm 1} h_{\mu} \cdot i
  \displaystyle\sum\limits_{m_{i}, m_{f}}
  \displaystyle\sum\limits_{\mu_{i},\, \mu_{f} = \pm 1/2}
    C_{l_{f}m_{f} 1/2 \mu_{f}}^{j_{f}M_{f},\,*}\,
    C_{l_{i}m_{i} 1/2 \mu_{i}}^{j_{i}M_{i}}\;
    I_{M}(l_{i},l_{f}, l_{\rm ph}, l_{i}-1, \mu), \\

  K_{\rm mag,1}^{M} (l_{i},l_{f}, l_{\rm ph}, l_{i}+1) & = &
  \displaystyle\sum\limits_{\mu=\pm 1} h_{\mu} \cdot i
  \displaystyle\sum\limits_{m_{i}, m_{f}}
  \displaystyle\sum\limits_{\mu_{i},\, \mu_{f} = \pm 1/2}
    C_{l_{f}m_{f} 1/2 \mu_{f}}^{j_{f}M_{f},\,*}\,
    C_{l_{i}m_{i} 1/2 \mu_{i}}^{j_{i}M_{i}}\;
    I_{M}(l_{i},l_{f}, l_{\rm ph}, l_{i}+1, \mu).
\end{array}
\label{eq.9.9.2.2}
\end{equation}
\begin{equation}
\begin{array}{lcl}
\vspace{1.5mm}
  K_{\rm mag,1}^{E}(l_{i},l_{f}, l_{\rm ph}, l_{i}-1, l_{\rm ph}-1) & = &
  \displaystyle\sum\limits_{\mu=\pm 1} h_{\mu}\, \mu
  \displaystyle\sum\limits_{m_{i}, m_{f}}
  \displaystyle\sum\limits_{\mu_{i},\, \mu_{f} = \pm 1/2}
    C_{l_{f}m_{f} 1/2 \mu_{f}}^{j_{f}M_{f},\,*}\,
    C_{l_{i}m_{i} 1/2 \mu_{i}}^{j_{i}M_{i}}\;
    I_{E}(l_{i},l_{f}, l_{\rm ph}, l_{i}-1, l_{\rm ph}-1, \mu), \\

\vspace{1mm}
  K_{\rm mag,1}^{E} (l_{i},l_{f}, l_{\rm ph}, l_{i}-1, l_{\rm ph}+1) & = &
  \displaystyle\sum\limits_{\mu=\pm 1} h_{\mu}\, \mu
  \displaystyle\sum\limits_{m_{i}, m_{f}}
  \displaystyle\sum\limits_{\mu_{i},\, \mu_{f} = \pm 1/2}
    C_{l_{f}m_{f} 1/2 \mu_{f}}^{j_{f}M_{f},\,*}\,
    C_{l_{i}m_{i} 1/2 \mu_{i}}^{j_{i}M_{i}}\;
    I_{E} (l_{i},l_{f}, l_{\rm ph}, l_{i}-1, l_{\rm ph}+1), \mu), \\

\vspace{1mm}
  K_{\rm mag,1}^{E}(l_{i},l_{f}, l_{\rm ph}, l_{i}+1, l_{\rm ph}-1) & = &
  \displaystyle\sum\limits_{\mu=\pm 1} h_{\mu}\, \mu
  \displaystyle\sum\limits_{m_{i}, m_{f}}
  \displaystyle\sum\limits_{\mu_{i},\, \mu_{f} = \pm 1/2}
    C_{l_{f}m_{f} 1/2 \mu_{f}}^{j_{f}M_{f},\,*}\,
    C_{l_{i}m_{i} 1/2 \mu_{i}}^{j_{i}M_{i}}\;
    I_{E}(l_{i},l_{f}, l_{\rm ph}, l_{i}+1, l_{\rm ph}-1, \mu), \\

  K_{\rm mag,1}^{E}(l_{i},l_{f}, l_{\rm ph}, l_{i}+1, l_{\rm ph}+1) & = &
  \displaystyle\sum\limits_{\mu=\pm 1} h_{\mu}\, \mu
  \displaystyle\sum\limits_{m_{i}, m_{f}}
  \displaystyle\sum\limits_{\mu_{i},\, \mu_{f} = \pm 1/2}
    C_{l_{f}m_{f} 1/2 \mu_{f}}^{j_{f}M_{f},\,*}\,
    C_{l_{i}m_{i} 1/2 \mu_{i}}^{j_{i}M_{i}}\;
    I_{E}(l_{i},l_{f}, l_{\rm ph}, l_{i}+1, l_{\rm ph}+1, \mu).
\end{array}
\label{eq.9.9.2.4}
\end{equation}

\subsubsection{Representation for matrix element $p_{\rm mag,2}$
\label{sec.9.9.3}}

For electric and magnetic components of $p_{\rm mag,2}$ from (\ref{eq.app.9.3.8}) and (\ref{eq.app.9.7.7}) we have:
\begin{equation}
\begin{array}{lcl}
\vspace{1.5mm}
  p_{\rm mag,2}^{M} & = &
  \displaystyle\frac{1}{2}\,
  \sqrt{\displaystyle\frac{\pi}{2}}\, k\,
  \displaystyle\sum\limits_{l_{\rm ph}=1}\,
    (-i)^{l_{\rm ph}}\, \sqrt{2l_{\rm ph}+1} \cdot
    K_{\rm mag,2}^{M} (l_{i},l_{f}, l_{\rm ph}, l_{\rm ph}) \cdot
    \tilde{J} (l_{i},l_{f},l_{\rm ph}), \\

\vspace{1.5mm}
  p_{\rm mag,2}^{E} & = &
  \displaystyle\frac{1}{2}\,
  \sqrt{\displaystyle\frac{\pi}{2}}\, k
  \displaystyle\sum\limits_{l_{\rm ph}=1}\,
    (-i)^{l_{\rm ph}}\, \sqrt{2l_{\rm ph}+1} \cdot
    \biggl\{
      \sqrt{\displaystyle\frac{l_{\rm ph}+1}{2l_{\rm ph}+1}} \cdot
      K_{\rm mag,2}^{E} (l_{i},l_{f},l_{\rm ph},l_{\rm ph}-1) \cdot
      \tilde{J}\,(l_{i},l_{f},l_{\rm ph}-1)\; - \\
  & - &
      \sqrt{\displaystyle\frac{l_{\rm ph}}{2l_{\rm ph}+1}} \cdot
      K_{\rm mag,2}^{E} (l_{i},l_{f},l_{\rm ph},l_{\rm ph}+1) \cdot
      \tilde{J}\,(l_{i},l_{f},l_{\rm ph}+1)
  \biggr\},
\end{array}
\label{eq.9.9.3.1}
\end{equation}
where
\begin{equation}
\begin{array}{lcl}
\vspace{1.5mm}
  K_{\rm mag,2}^{M} (l_{i},l_{f}, l_{\rm ph}, l_{\rm ph}) & = &
  \displaystyle\sum\limits_{\mu=\pm 1} i\mu
  \displaystyle\sum\limits_{m_{i}, m_{f}}
  \displaystyle\sum\limits_{\mu_{i},\, \mu_{f} = \pm 1/2}
    C_{l_{f}m_{f} 1/2 \mu_{f}}^{j_{f}M_{f},\,*}\,
    C_{l_{i}m_{i} 1/2 \mu_{i}}^{j_{i}M_{i}}\; \times \\
\vspace{1.5mm}
  & \times &
    \Bigl[ -1 + i\, \Bigl\{ \delta_{\mu_{i}, +1/2}\; -\; \delta_{\mu_{i}, -1/2} \Bigr\} \Bigr]\,
    \tilde{I}\,(l_{i},l_{f},l_{\rm ph}, l_{\rm ph}, \mu), \\

\vspace{1.5mm}
  K_{\rm mag,2}^{E}(l_{i},l_{f}, l_{\rm ph}, l_{\rm ph}-1) & = &
  \displaystyle\sum\limits_{\mu=\pm 1}
  \displaystyle\sum\limits_{m_{i}, m_{f}}
  \displaystyle\sum\limits_{\mu_{i},\, \mu_{f} = \pm 1/2}
    C_{l_{f}m_{f} 1/2 \mu_{f}}^{j_{f}M_{f},\,*}\,
    C_{l_{i}m_{i} 1/2 \mu_{i}}^{j_{i}M_{i}}\;
    \Bigl[ -1 + i\, \Bigl\{ \delta_{\mu_{i}, +1/2}\; -\; \delta_{\mu_{i}, -1/2} \Bigr\} \Bigr]\; \times \\
\vspace{2.5mm}
  & \times &
    \tilde{I}\,(l_{i},l_{f},l_{\rm ph},l_{\rm ph}-1,\mu), \\

\vspace{1.5mm}
  K_{\rm mag,2}^{E} (l_{i},l_{f}, l_{\rm ph}, l_{\rm ph}+1) & = &
  \displaystyle\sum\limits_{\mu=\pm 1}
  \displaystyle\sum\limits_{m_{i}, m_{f}}
  \displaystyle\sum\limits_{\mu_{i},\, \mu_{f} = \pm 1/2}
    C_{l_{f}m_{f} 1/2 \mu_{f}}^{j_{f}M_{f},\,*}\,
    C_{l_{i}m_{i} 1/2 \mu_{i}}^{j_{i}M_{i}}\;
    \Bigl[ -1 + i\, \Bigl\{ \delta_{\mu_{i}, +1/2}\; -\; \delta_{\mu_{i}, -1/2} \Bigr\} \Bigr]\; \times \\
  & \times &
    \tilde{I}\,(l_{i},l_{f},l_{\rm ph},l_{\rm ph}+1,\mu).
\end{array}
\label{eq.9.9.3.2}
\end{equation}

\subsubsection{Representation for matrix element $\tilde{p}_{\rm q,1}$
\label{sec.9.9.4}}

For electric and magnetic components of $\tilde{p}_{\rm q,1}$ from (\ref{eq.app.9.4.1}), (\ref{eq.app.9.7.7}) and (\ref{eq.app.9.8.4}) we have:
\begin{equation}
\begin{array}{lcl}
\vspace{1.3mm}
  \tilde{p}_{\rm q,1}^{M} & = &
  \sqrt{\displaystyle\frac{\pi}{2}}\:
  \displaystyle\sum\limits_{l_{\rm ph}=1}\,
    (-i)^{l_{\rm ph}}\, \sqrt{2l_{\rm ph}+1} \cdot
    K_{\rm q,1}^{M} (l_{i},l_{f}, l_{\rm ph}, l_{\rm ph})\;
  \Bigl\{
    A_{1} (Q, F_{1}, F_{2}) \cdot \tilde{J}\, (l_{i},l_{f},l_{\rm ph}) +
    B_{1} (Q, F_{1}, F_{2}) \cdot \breve{J}\, (l_{i},l_{f},l_{\rm ph})
  \Bigr\}, \\

\vspace{0.7mm}
  \tilde{p}_{\rm q,1}^{E} & = &
  \sqrt{\displaystyle\frac{\pi}{2}}\:
  \displaystyle\sum\limits_{l_{\rm ph}=1}\,
    (-i)^{l_{\rm ph}}\, \sqrt{2l_{\rm ph}+1}\; \times \\
\vspace{0.9mm}
  & \times &
  \biggl\{
    K_{\rm q,1}^{E} (l_{i},l_{f}, l_{\rm ph}, l_{\rm ph}-1) \cdot
    \sqrt{\displaystyle\frac{l_{\rm ph}+1}{2l_{\rm ph}+1}} \cdot
    \Bigl[
      A_{1} (Q, F_{1}, F_{2}) \cdot \tilde{J}\,(l_{i},l_{f},l_{\rm ph}-1) +
      B_{1} (Q, F_{1}, F_{2}) \cdot \breve{J}\,(l_{i},l_{f},l_{\rm ph}-1)
    \Bigr]\; - \\
  & - &
    K_{\rm q,1}^{E} (l_{i},l_{f}, l_{\rm ph}, l_{\rm ph}+1) \cdot
    \sqrt{\displaystyle\frac{l_{\rm ph}}{2l_{\rm ph}+1}} \cdot
    \Bigl[
      A_{1} (Q, F_{1}, F_{2}) \cdot \tilde{J}\,(l_{i},l_{f},l_{\rm ph}+1) +
      B_{1} (Q, F_{1}, F_{2}) \cdot \breve{J}\,(l_{i},l_{f},l_{\rm ph}+1)
    \Bigr]
  \biggr\},
\end{array}
\label{eq.9.9.4.1}
\end{equation}
where
\begin{equation}
\begin{array}{lcl}
  K_{\rm q,1}^{M} (l_{i},l_{f}, l_{\rm ph}, l_{\rm ph}) & = &
  \displaystyle\sum\limits_{\mu=\pm 1} i\,\mu
  \displaystyle\sum\limits_{m_{i}, m_{f}}
  \displaystyle\sum\limits_{\mu_{i},\, \mu_{f} = \pm 1/2}
    C_{l_{f}m_{f} 1/2 \mu_{f}}^{j_{f}M_{f},\,*}\,
    C_{l_{i}m_{i} 1/2 \mu_{i}}^{j_{i}M_{i}} \cdot
    \tilde{I}\,(l_{i},l_{f},l_{\rm ph}, l_{\rm ph}, \mu), \\

  K_{\rm q,1}^{E} (l_{i},l_{f}, l_{\rm ph}, l_{\rm ph}) & = &
  \displaystyle\sum\limits_{\mu=\pm 1}
  \displaystyle\sum\limits_{m_{i}, m_{f}}
  \displaystyle\sum\limits_{\mu_{i},\, \mu_{f} = \pm 1/2}
    C_{l_{f}m_{f} 1/2 \mu_{f}}^{j_{f}M_{f},\,*}\,
    C_{l_{i}m_{i} 1/2 \mu_{i}}^{j_{i}M_{i}} \cdot
    \tilde{I}\,(l_{i},l_{f},l_{\rm ph}, l_{\rm ph}, \mu).
\end{array}
\label{eq.9.9.4.2}
\end{equation}

\subsection{Probability of emission of the bremsstrahlung photon
\label{sec.10}}

We define the probability of the emitted bremsstrahlung photons on the bass of the full matrix element $p_{fi}$ 
in frameworks of formalism given in \cite{Maydanyuk_Zhang_Zou.2016.PRC,Maydanyuk.2012.PRC,Maydanyuk_Zhang.2015.PRC} and we do not repeat it in this paper.
In result, we obtain the bremsstrahlung probability as%
%
%
\begin{equation}
\begin{array}{ccl}
  \displaystyle\frac{d\,P }{dw_{\rm ph}} & = &
  \displaystyle\frac{e^{2}}{2\pi\,c^{5}}\: \displaystyle\frac{w_{\rm ph}\,E_{i}}{m_{\rm p}^{2}\,k_{i}}\: \bigl| p_{fi} \bigr|^{2}.
\end{array}
\label{eq.10.1}
\end{equation}
%
%
In further analysis we will calculate the different contributions of the emitted photons to the full bremsstrahlung spectrum.
For estimation of the interesting contribution, we just use the corresponding term $p_{\rm el}$, $p_{\rm mag,1}$, $p_{\rm mag,2}$ or $p_{\rm q,1}$.


\section{Analysis, discussions
\label{sec.results}}

Let us estimate the bremsstrahlung probability accompanying the scattering of protons off nuclei, using formalism above.
For calculations and analysis we choose the reaction of $p + ^{197}{\rm Au}$ at proton beam energy of 190~MeV,
where experimental bremsstrahlung data \cite{Goethem.2002.PRL} were obtained with high accuracy.
Wave function of relative motion between proton and center-of-mass of nucleus is determined concerning to the proton-nucleus potential in form of $V (r) = v_{c}(r) + v_{N}(r) + v_{\rm so}(r) + v_{l} (r)$,
where $v_{c}(r)$, $v_{N}(r)$, $v_{\rm so}(r)$, and $v_{l} (r)$ are Coulomb, nuclear, spin-orbital, and centrifugal components defined with parameters in
Eqs.~(46)--(47) in Ref.~\cite{Maydanyuk_Zhang.2015.PRC}.


We analyzed these data in our previous paper~\cite{Maydanyuk_Zhang.2015.PRC} in details (without consideration of the internal structure of nucleons).
In particular, we constructed formalism describing the coherent and incoherent emissions of the bremsstrahlung photons.
We found that inclusion of the incoherent emission to the model allows to improve essentially agreement between calculations and experimental data~\cite{Goethem.2002.PRL}.
So, in the current research we focus on estimation of role of the internal structure of the scattered proton in forming the bremsstrahlung spectrum.
In order to preform such an analysis clearly and obtain the first estimations, we shall neglect by incoherent emission (which will make analysis and formalism to be essentially more complicated) at current step.
But, it turns out that this is enough to obtain the first conclusions about our approach from such an analysis.

\begin{figure}[htbp]
\centerline{\includegraphics[width=85mm]{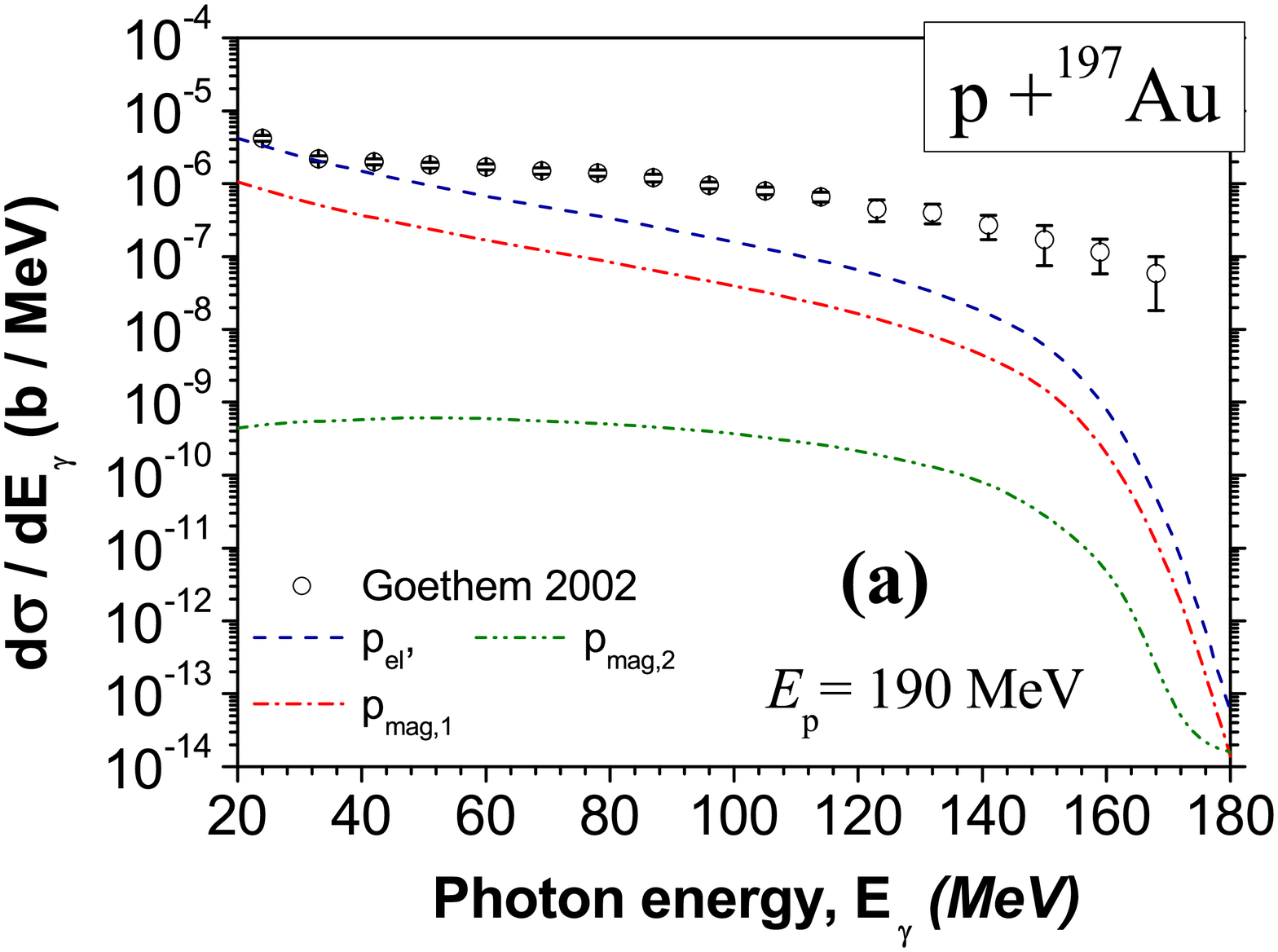}
\hspace{-1mm}\includegraphics[width=85mm]{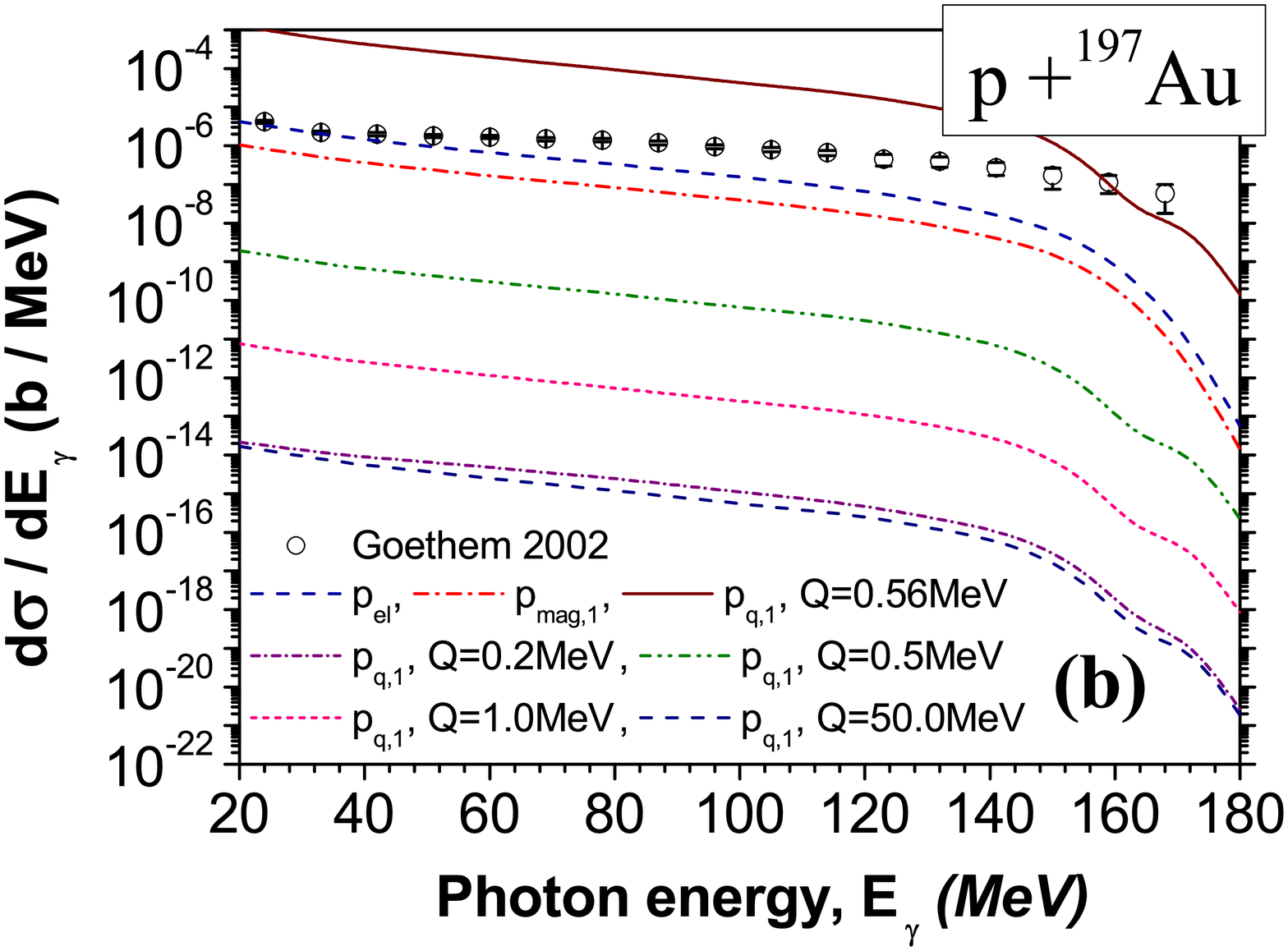}
}
\vspace{-4mm}
\caption{\small (Color online)
The calculated electric and magnetic bremsstrahlung emissions in the scattering of protons off the $^{197}{\rm Au}$ nuclei at energy of proton beam of $E_{\rm p}=190$~MeV
in comparison with experimental data~\cite{Goethem.2002.PRL}
[components are defined in Eqs.~(\ref{eq.8.1.2}), (\ref{eq.9.9.1.1})--(\ref{eq.9.9.3.2})]:
(a) electric emission defined by $p_{\rm el}$ (blue dashed line),
magnetic emission defined by $p_{\rm mag, 1}$ (red dash-dotted line) and
magnetic emission defined by $p_{\rm mag, 2}$ (green dash-double dotted line).
(b) Contributions of the bremsstrahlung emission given by term $p_{\rm q,1}$ with different $Q$ in comparison with electric and magnetic emissions.
\label{fig.1}}
\end{figure}

At first, we analyze contributions of the electric and magnetic emissions, given by terms $p_{\rm el}$, $p_{\rm mag, 1}$ and $p_{\rm mag, 2}$, to the full bremsstrahlung spectrum.
Results of such calculations are presented in Fig.~\ref{fig.1}~(a).
%
%
We see that the contributions from electrical and first magnetic terms $p_{\rm el}$ and $p_{\rm mag, 1}$ are similar.
But, such two contributions are essentially larger than contribution for the second magnetic term $p_{\rm mag, 2}$.
This result is in complete agreement with analysis given in Ref.~\cite{Maydanyuk.2012.PRC}.
After reconstruction of the logic given in Ref.~\cite{Maydanyuk.2012.PRC}, now we include the internal structure of the scattered proton to calculations.
Such calculations for contribution of the emitted photons caused by internal structure of the scattered proton are presented in next Fig.~\ref{fig.1}~(b), where we provide the bremsstrahlung contributions from $p_{\rm q,1}$ in dependence on different values for $Q$.
From this figure one can see, that these spectra are essentially different.
Moreover, at some values of $Q$ this contribution of the emitted photons is even larger than the electric and magnetic contributions presented in Fig.~\ref{fig.1}~(a).

\begin{figure}[htbp]
\centerline{\includegraphics[width=85mm]{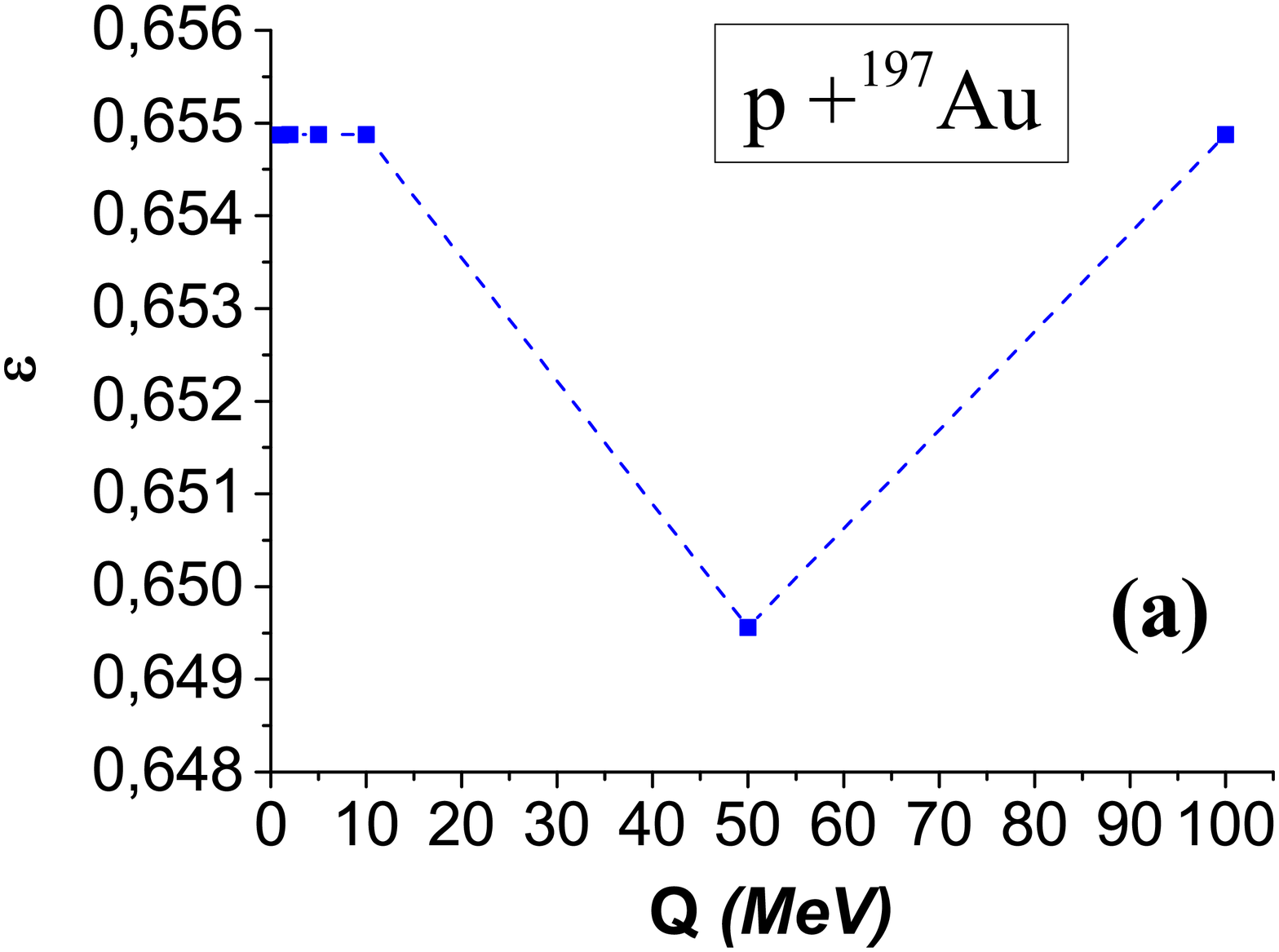}
}
\vspace{-4mm}
\caption{\small (Color online)
Function of errors defined in Eq.~(\ref{eq.result.1}) in dependence on $Q$ for
bremsstrahlung emissions in the scattering of protons off the $^{197}{\rm Au}$ nuclei at energy of proton beam of $E_{\rm p}=190$~MeV
[we calculate the full matrix element as summation of terms $p_{\rm el}$, $p_{\rm mag,1}$, $p_{\rm mag,2}$ and $p_{\rm q,1}(Q)$;
experimental data are used from Ref.~~\cite{Goethem.2002.PRL}].
\label{fig.2}}
\end{figure}

Comparing Eqs.~(\ref{eq.6.1.3}) and (\ref{eq.7.1.4}), one can find that matrix element $p_{\rm q,1}$ has different dependence on energy of the emitted photons
in comparison with $p_{\rm el}$, $p_{\rm mag,1}$ and $p_{\rm mag, 2}$.
Moreover, dependence of $p_{\rm q,1}$ on energy of photons is changed in variations of $Q$.
So, the summarized bremsstrahlung spectrum with inclusion of term $p_{\rm q,1}$ will be changed even after renormalization of calculations at different $Q$ on the same experimental data.
We use such an idea in order to find value for $Q$, which gives the most close agrement between calculations and experimental data.
In order to realize such an idea, we use our functions of errors previously introduced to the nuclear bremsstrahlung theory
(see Eqs.~(23)--(24) in Ref.~\cite{Maydanyuk_Zhang_Zou.2016.PRC}, Eqs.~(20) in Ref.~\cite{Maydanyuk.2015.NPA}, also Ref.~\cite{Maydanyuk_Zhang_Zou.2017.PRC}),
which we reformulate as
\begin{equation}
\begin{array}{lcl}
  \varepsilon (Q) & = &
  \displaystyle\frac{1}{N}
  \displaystyle\sum\limits_{k=1}^{N}
    \displaystyle\frac{\Bigl|\sigma^{\rm (theor)} (E_{k}, Q) - \sigma^{\rm (exp)} (E_{k}) \Bigr|}{\sigma^{\rm (exp)} (E_{k})}.
\end{array}
\label{eq.result.1}
\end{equation}
Here, $\sigma^{\rm (theor)} (E_{k})$ and $\sigma^{\rm (exp)} (E_{k})$ are theoretical and experimental bremsstrahlung cross-sections at energy $E_{k}$ of the emitted photon,
the summation is performed over experimental data ($N=17$ for data in Ref.~\cite{Goethem.2002.PRL}).
Such calculations are presented in Fig.~\ref{fig.2}.
Here, we find $Q=50$~MeV corresponding to the minimal function of errors (\ref{eq.result.1}).

Note that at limit of neglecting of the internal structure of the scattering proton via limit of $F_{1}(Q) \to 1$, $F_{2}(Q) \to 0$,
we reconstruct our previous results in Ref.~\cite{Maydanyuk.2012.PRC} completely.

\section{Conclusions
\label{sec.conclusions}}

In this paper we 
investigate an idea, how to use analysis of the bremsstrahlung photons to study the internal structure of proton, which is under nuclear reaction with nucleus
(which can be light, middle or superheavy).
We construct a new model describing bremsstrahlung emission of photons which accompanies the scattering of protons off nuclei.

From physical grounds, emission of photons is formed as result of relative motions (accelerations) of nucleons of the nucleus-target and the scattering proton.
By such a reason, we construct the bremsstrahlung formalism from many-nucleon basis \cite{Maydanyuk_Zhang.2015.PRC,Maydanyuk_Zhang_Zou.2016.PRC}.
This allows to analyze different contributions of coherent and incoherent bremsstrahlung emissions.

In the model, we focus on the new description of internal structure of the scattered proton%
\footnote{At current formalism, we neglect structure of nucleons of nucleus-target.}.
We are interested if such a structure can be visible in the full bremsstrahlung spectrum.
To realize this aim, we implement electromagnetic form-factors for nucleons on the basis of DIS theory to our model.
As a result, in our formalism the full bremsstrahlung spectrum is dependent on such form-factors of the scattered proton.
In the limit without such form-factors, we reconstruct our previous results in Ref.~\cite{Maydanyuk.2012.PRC} completely.
As the scattered proton can be under influence of strong forces and gives the largest bremsstrahlung contribution to full spectrum,
we focus on maximally accurate description of its evolution concerning nucleons of nucleus.
%
Quantum effects and evolution of such a complicated nuclear system are well described by the scattering theory
that has been deeply studied and tested experimentally well.
From such motivations, (for the first time) we generalize Pauli equation with interacting potential (describing quantum evolution of fermion inside strong field),
with including the formalism of electromagnetic form-factors of nucleon.
%
Note that the idea of generalizations of Pauli equation has been successfully applied in studying coherent bremsstrahlung for the proton-nucleus scattering \cite{Maydanyuk.2012.PRC},
and allows to add incoherent processes from individual nucleon-nucleon interactions \cite{Maydanyuk_Zhang.2015.PRC}.


In order to analyze and test our approach, for calculations we choose the reaction of $p + ^{197}{\rm Au}$ at proton beam energy of 190~MeV,
where experimental bremsstrahlung data \cite{Goethem.2002.PRL} were obtained with high accuracy.
Anomalous magnetic momenta of nucleons (important in estimations of bremsstrahlung) reinforce our motivation to develop formalism at such energies of protons.
Conclusions from analysis of this model are the following:

\begin{enumerate}
\item
At first, 
we analyze contributions of the electric and magnetic emissions, defined by terms $p_{\rm el}$, $p_{\rm mag, 1}$ and $p_{\rm mag, 2}$
, to the full bremsstrahlung spectrum (see Fig.~\ref{fig.1}~(a)).
We find that the electrical and first magnetic contributions from terms $p_{\rm el}$ and $p_{\rm mag, 1}$ are similar.
But, such two contributions are essentially larger than contribution from the second magnetic term $p_{\rm mag, 2}$.
So, we reconstruct completely our old result~\cite{Maydanyuk.2012.PRC} (where the electric and magnetic coherent emissions were studied in details in such a reaction)
without inclusion of the internal structure of the scattered proton into analysis and calculations.

\vspace{3mm}
\item
In next step, we analyze and estimate the contribution of bremsstrahlung emission after we include the internal structure of scattered proton to the model and calculations.
This is a new type of bremsstrahlung emission defined by term $p_{\rm q,1}$, which we introduce to the bremsstrahlung theory in nuclear physics (see Fig.~\ref{fig.1}~(b)).
This emission depends on the form-factors of scattered proton.
We find that at some value of $Q$ such an emission can be larger in comparison with the electrical and magnetic emissions (presented in Fig.~\ref{fig.1}~(a) and studied in Ref.~\cite{Maydanyuk.2012.PRC}).


\vspace{3mm}
\item
Important advance of our approach is in that such a dependence of the bremsstrahlung spectra on form-factors exists also after renormalization of calculations on experimental data.
Using idea of function of errors (\ref{eq.result.1})
(see also Eqs.~(23)--(24) in Ref.~\cite{Maydanyuk_Zhang_Zou.2016.PRC}, Eqs.~(20) in Ref.~\cite{Maydanyuk.2015.NPA}, Ref.~\cite{Maydanyuk_Zhang_Zou.2017.PRC}),
we extract proper value for $Q$ (we obtain $Q=50$~MeV),
which corresponds to the most close agreement between calculated full spectrum and experimental data~\cite{Goethem.2002.PRL} (see Fig.~\ref{fig.2}).

\vspace{3mm}
\item
In the limit of neglecting of the internal structure of the scattering proton (at $F_{1}(Q) \to 1$, $F_{2}(Q) \to 0$),
we reconstruct results in Ref.~\cite{Maydanyuk.2012.PRC} completely.

\end{enumerate}
Our results confirm that the full bremsstrahlung spectrum is sensitive to the form-factors of scattered proton (characterizing its internal structure).
This is the first indication that this is possible to construct a new type of microscopy (with higher resolution in comparison with existed one), to study internal structure of nucleons
in experimental way by means of bremsstrahlung analysis.


\section*{Acknowledgements
\label{sec.acknowledgements}}

S.~P.~M. thanks the Institute of Modern Physics of Chinese Academy of Sciences for warm hospitality and support.
This work was supported by the Major State Basic Research Development Program in China (No. 2015CB856903),
the National Natural Science Foundation of China (Grant Nos. 11575254, 11447105 and 11175215),
the Chinese Academy of Sciences fellowships for researchers from developing countries (No. 2014FFJA0003).

\appendix

\section{Solution of equation (\ref{eq.4.1.1})
\label{sec.app.1}}

\subsection{Calculations of parameters for Eq.~(\ref{eq.4.1.1})
\label{sec.2.2}}

In this Appendix we solve equation (\ref{eq.4.1.1}).
At first, we shall transform function $A$ in Eq.~(\ref{eq.2.43}).
Taking the following property into account
\begin{equation}
\begin{array}{ll}
  \sigma_{i}\sigma_{j} = \delta_{ij} I + i\, \varepsilon_{ijk} \sigma_{k}, &
  \varepsilon_{ijk}\, \varepsilon_{ilm} = \delta_{jl}\delta_{km} - \delta_{jk}\delta_{lm},
\end{array}
\label{eq.app.2.2.1}
\end{equation}
we obtain:
\begin{equation}
\begin{array}{lll}
  - \varepsilon_{kmj}\, F_{1}\, q^{k}\, \sigma_{j} -
  i \bigl( F_{1} + F_{2} q^{4} \bigr)\, q^{k} \sigma_{k}\, \sigma_{m} = 

%
%
  - i \bigl( F_{1} + F_{2} q^{4} \bigr)\, q^{m} + F_{2}\, q^{4} q^{k}\, \varepsilon_{kmj} \sigma_{j}.
\end{array}
\label{eq.app.2.2.2}
\end{equation}
Taking this expression into account, now we simplify $A$ in Eq.~(\ref{eq.2.43}):
%
\begin{equation}
\begin{array}{lll}
\vspace{0.7mm}
  A & = &
  c\, F_{2}\,
  \Bigl[
    - i \bigl( F_{1} + F_{2} q^{4} \bigr)\, q^{m} +
    F_{2}\, q^{4} q^{k}\, \varepsilon_{kmj} \sigma_{j}
  \Bigr]
  \Bigl( \mathbf{p} - \displaystyle\frac{ze}{c} \mathbf{A} \Bigr)_{m}\; + \\

  & + &
    F_{1}\, m c^{2} +
    \bigl(F_{1}^{2} + F_{2}^{2}\, \mathbf{q}^{2} \bigr)\,
    \bigl( ze\, A_{0} + V(\mathbf{r}) - m c^{2} \bigr).
\end{array}
\label{eq.app.2.2.3}
\end{equation}
We transform function $B$ in Eq.~(\ref{eq.2.44}) as
%
\begin{equation}
\begin{array}{lll}
\vspace{0.4mm}
  B & = &
%
  B_{1} +
  i\,c\, F_{2}\,
  \biggl\{
    \Bigl[
      F_{1} \bigl( F_{1} - F_{2} q^{4} \bigr)\, \sigma_{m} -
      i F_{2}^{2}\: q^{k} \sigma_{k}\, \varepsilon_{nmj}\, q^{n}\, \sigma_{j}
    \Bigr]\,
    (1 - 2\,F_{1})\, q^{k} \sigma_{k}\; + \\

\vspace{0.5mm}
  & + &
    q^{k} \sigma_{k}\:
    \Bigl[
      - i F_{2}^{2} q^{k} \sigma_{k}\, \varepsilon_{nmj}\, q^{n}\, \sigma_{j}\, +
      F_{1} \bigl( F_{1} + F_{2} q^{4} \bigr)\, \sigma_{m}
    \Bigr]\; - \\

\vspace{3.5mm}
  & - &
    2\,F_{1} q^{k} \sigma_{k} \bigl[ ze\, A_{0} + V(\mathbf{r}) - m c^{2} \bigr]\,
    \displaystyle\frac{1}{mc^{2}}\,
    \Bigl[
      - i F_{2}^{2} q^{k} \sigma_{k}\, \varepsilon_{nmj}\, q^{n}\, \sigma_{j}\, +
      F_{1} \bigl( F_{1} + F_{2} q^{4} \bigr)\, \sigma_{m}
    \Bigr]\,
  \biggr\}\, \Bigl( \mathbf{p} - \displaystyle\frac{ze}{c} \mathbf{A} \Bigr)\; - \\

  & - &
  m c^{2}\, (1 - 2\,F_{1})\, F_{2}^{2}\, \mathbf{q}^{2} +
  2\,(1 - 2\,F_{1})\, F_{1}F_{2}^{2}\, \bigl[ ze\, A_{0} + V(\mathbf{r}) - m c^{2} \bigr]\, \mathbf{q}^{2},
\end{array}
\label{eq.app.2.2.4}
\end{equation}
where %
\begin{equation}
\begin{array}{lll}
\vspace{0.8mm}
  B_{1} & = &
  \displaystyle\frac{1}{m}\,
    \Bigl[
      \bigl[ F_{1}^{2} - F_{1} F_{2} q^{4} - F_{2}^{2}\, \mathbf{q}^{2} \bigr]\, \sigma_{m} + F_{2}^{2}\, q^{m} q^{l}\, \sigma_{l}
    \Bigr]
      \Bigl( p_{m} - \displaystyle\frac{ze}{c} A_{m} \Bigr)\; \times \\
  & \times &
    \Bigl[
      \bigl[ F_{1}^{2} + F_{1} F_{2} q^{4} - F_{2}^{2}\, \mathbf{q}^{2} \bigr]\, \sigma_{m^{\prime}} + F_{2}^{2}\: q^{m^{\prime}} q^{l}\, \sigma_{l}
    \Bigr]
      \Bigl( p_{m'} - \displaystyle\frac{ze}{c} A_{m'} \Bigr).
\end{array}
\label{eq.app.2.2.5}
\end{equation}
Here, we taken property~(\ref{eq.app.2.2.8}) into account. It is obtained so. Taking properties~(\ref{eq.app.2.2.1}) into account, we simplify term ($i,j,k=1,2,3$)
%
\begin{equation}
\begin{array}{lll}
\vspace{2.0mm}
  i F_{2}^{2}\: q^{k} \sigma_{k}\, \varepsilon_{nmj}\, q^{n}\, \sigma_{j} =
  i F_{2}^{2}\: q^{k} q^{n}\, \varepsilon_{nmj}\; (\sigma_{k} \sigma_{j}) =
  i F_{2}^{2}\: q^{k} q^{n}\, \varepsilon_{nmj}\; (\delta_{kj} I + i\, \varepsilon_{kjl} \sigma_{l}) =
  i F_{2}^{2}\: q^{k} q^{n}\, \varepsilon_{nmj}\, \delta_{kj} +
  i F_{2}^{2}\: q^{k} q^{n}\, \varepsilon_{nmj}\; i\, \varepsilon_{kjl} \sigma_{l}\; = \\


  = i F_{2}^{2}\: q^{k} q^{n}\, \varepsilon_{nmk} + F_{2}^{2}\: q^{k} q^{k}\, \sigma_{m} - F_{2}^{2}\: q^{m} q^{l}\, \sigma_{l} =
  i F_{2}^{2}\: q^{k} q^{n}\, \varepsilon_{nmk} + F_{2}^{2}\, \mathbf{q}^{2}\, \sigma_{m} - F_{2}^{2}\: q^{m} q^{l}\, \sigma_{l}.
\end{array}
\label{eq.app.2.2.6}
\end{equation}
Here, the first term equas to zero, as summation is performed over two indexes of antisymmetric $\varepsilon_{nmk}$:
%
\begin{equation}
\begin{array}{lll}
  q^{k} q^{n}\, \varepsilon_{nmk} =
  - q^{k} q^{n}\, \varepsilon_{nkm} =
  - \displaystyle\frac{1}{2}\, \Bigl( q^{k} q^{n}\, \varepsilon_{nkm} + q^{n} q^{k}\, \varepsilon_{knm} \Bigr) =
  - \displaystyle\frac{1}{2}\, q^{k} q^{n}\, \Bigl( \varepsilon_{nkm} - \varepsilon_{nkm} \Bigr) = 0.
\end{array}
\label{eq.app.2.2.7}
\end{equation}
So, we write:
%
\begin{equation}
  i F_{2}^{2}\: q^{k} \sigma_{k}\, \varepsilon_{nmj}\, q^{n}\, \sigma_{j} =
  F_{2}^{2}\, \mathbf{q}^{2}\, \sigma_{m} - F_{2}^{2}\: q^{m} q^{l}\, \sigma_{l}.
\label{eq.app.2.2.8}
\end{equation}
Now we simplify the first term in final expression in Eq.~(\ref{eq.app.2.2.4}):
%
\begin{equation}
\begin{array}{lll}
\vspace{2.0mm}
  & F_{1} \bigl( F_{1} - F_{2} q^{4} \bigr)\, \sigma_{m} -
  i F_{2}^{2}\: q^{k} \sigma_{k}\, \varepsilon_{nmj}\, q^{n}\, \sigma_{j} =

  F_{1} \bigl( F_{1} - F_{2} q^{4} \bigr)\, \sigma_{m} -
  F_{2}^{2}\, \mathbf{q}^{2}\, \sigma_{m} + F_{2}^{2}\: q^{m} q^{l}\, \sigma_{l}\; = \\

  = & \bigl[ F_{1}^{2} - F_{1} F_{2} q^{4} - F_{2}^{2}\, \mathbf{q}^{2} \bigr]\, \sigma_{m} + F_{2}^{2}\: q^{m} q^{l}\, \sigma_{l}.
\end{array}
\label{eq.app.2.2.9}
\end{equation}
Basing on Eqs.~(\ref{eq.app.2.2.9}) and (\ref{eq.2.36}), we find:
%
\begin{equation}
\begin{array}{lll}
  \Bigl[ F_{1} \bigl( F_{1} - F_{2} q^{4} \bigr)\, \sigma_{m} - i F_{2}^{2}\: q^{k} \sigma_{k}\, \varepsilon_{nmj}\, q^{n}\, \sigma_{j} \Bigr]\, q^{k} \sigma_{k} =


%
%
%

  \bigl[ F_{1}^{2} - F_{1} F_{2} q^{4} \bigr]\, q^{m} +
  i\, \bigl[ F_{1}^{2} - F_{1} F_{2} q^{4} - F_{2}^{2}\, \mathbf{q}^{2} \bigr]\, \varepsilon_{mjl}\, q^{j}\, \sigma_{l}.
\end{array}
\label{eq.app.2.2.10}
\end{equation}
For the second term in Eq.~(\ref{eq.app.2.2.4}) we find:
%
\begin{equation}
\begin{array}{lll}
  -\, i F_{2}^{2} q^{k} \sigma_{k}\, \varepsilon_{nm'j}\, q^{n}\, \sigma_{j} + F_{1} \bigl( F_{1} + F_{2} q^{4} \bigr)\, \sigma_{m'} =

%
  \bigl[ F_{1}^{2} + F_{1} F_{2} q^{4} - F_{2}^{2}\, \mathbf{q}^{2} \bigr]\, \sigma_{m^{\prime}} + F_{2}^{2}\: q^{m^{\prime}} q^{l}\, \sigma_{l}.
\end{array}
\label{eq.app.2.2.11}
\end{equation}
and, using logic of transformations (\ref{eq.app.2.2.10}), we obtain:
%
\begin{equation}
\begin{array}{lll}
  \Bigl[ - i F_{2}^{2}\: q^{k} \sigma_{k}\, \varepsilon_{nmj}\, q^{n}\, \sigma_{j} + F_{1} \bigl( F_{1} + F_{2} q^{4} \bigr)\, \sigma_{m} \Bigr]\, q^{k} \sigma_{k}\; = 
  \bigl[ F_{1}^{2} + F_{1} F_{2} q^{4} \bigr]\, q^{m} +
  i\, \bigl[ F_{1}^{2} + F_{1} F_{2} q^{4} - F_{2}^{2}\, \mathbf{q}^{2} \bigr]\, \varepsilon_{mjl}\, q^{j}\, \sigma_{l}.
\end{array}
\label{eq.app.2.2.12}
\end{equation}
For the third term in Eq.~(\ref{eq.app.2.2.4}) [taking formula~(\ref{eq.app.2.2.12}) into account] we have
%
\begin{equation}
\begin{array}{lll}
\vspace{1.5mm}
  & -
    2\,F_{1} q^{k} \sigma_{k} \bigl[ ze\, A_{0} + V(\mathbf{r}) - m c^{2} \bigr]\,
    \displaystyle\frac{1}{mc^{2}}\,
    \Bigl[
      - i F_{2}^{2} q^{k} \sigma_{k}\, \varepsilon_{nmj}\, q^{n}\, \sigma_{j}\, +
      F_{1} \bigl( F_{1} + F_{2} q^{4} \bigr)\, \sigma_{m}
    \Bigr]\, = \\

%

\vspace{1.0mm}
  = & -
    \displaystyle\frac{2\,F_{1} }{mc^{2}}\, \bigl[ ze\, A_{0} + V(\mathbf{r}) \bigr]\,
    \cdot
    \Bigl\{
      \bigl[ F_{1}^{2} + F_{1} F_{2} q^{4} \bigr]\, q^{m} +
      i\, \bigl[ F_{1}^{2} + F_{1} F_{2} q^{4} - F_{2}^{2}\, \mathbf{q}^{2} \bigr]\, \varepsilon_{mjl}\, q^{j}\, \sigma_{l}
    \Bigr\}\; + \\

  + &
    2\,F_{1}\,
    \cdot
    \Bigl\{
      \bigl[ F_{1}^{2} + F_{1} F_{2} q^{4} \bigr]\, q^{m} +
      i\, \bigl[ F_{1}^{2} + F_{1} F_{2} q^{4} - F_{2}^{2}\, \mathbf{q}^{2} \bigr]\, \varepsilon_{mjl}\, q^{j}\, \sigma_{l}
    \Bigr\}.
\end{array}
\label{eq.app.2.2.13}
\end{equation}
Summation of this expression and term~(\ref{eq.app.2.2.12}) equals to
%
\begin{equation}
\begin{array}{lll}
\vspace{1.3mm}
  &
  \Bigl[ - i F_{2}^{2}\: q^{k} \sigma_{k}\, \varepsilon_{nmj}\, q^{n}\, \sigma_{j} + F_{1} \bigl( F_{1} + F_{2} q^{4} \bigr)\, \sigma_{m} \Bigr]\,
    q^{k} \sigma_{k}\; - \\
\vspace{2.0mm}
  - &
    2\,F_{1} q^{k} \sigma_{k} \bigl[ ze\, A_{0} + V(\mathbf{r}) - m c^{2} \bigr]\,
    \displaystyle\frac{1}{mc^{2}}\,
    \Bigl[
      - i F_{2}^{2} q^{k} \sigma_{k}\, \varepsilon_{nmj}\, q^{n}\, \sigma_{j}\, +
      F_{1} \bigl( F_{1} + F_{2} q^{4} \bigr)\, \sigma_{m}
    \Bigr]\, = \\

\vspace{1.0mm}
  = & -
    \displaystyle\frac{2\,F_{1} }{mc^{2}}\, \bigl[ ze\, A_{0} + V(\mathbf{r}) \bigr]\,
    \cdot
    \Bigl\{
      \bigl[ F_{1}^{2} + F_{1} F_{2} q^{4} \bigr]\, q^{m} +
      i\, \bigl[ F_{1}^{2} + F_{1} F_{2} q^{4} - F_{2}^{2}\, \mathbf{q}^{2} \bigr]\, \varepsilon_{mjl}\, q^{j}\, \sigma_{l}
    \Bigr\}\; + \\
  + &
    (2\,F_{1} + 1)\,
    \cdot
    \Bigl\{
      \bigl[ F_{1}^{2} + F_{1} F_{2} q^{4} \bigr]\, q^{m} +
      i\, \bigl[ F_{1}^{2} + F_{1} F_{2} q^{4} - F_{2}^{2}\, \mathbf{q}^{2} \bigr]\, \varepsilon_{mjl}\, q^{j}\, \sigma_{l}
    \Bigr\}.
\end{array}
\label{eq.app.2.2.14}
\end{equation}

Now we simplify $B$ in Eq.~(\ref{eq.app.2.2.4}) further:
%
\begin{equation}
\begin{array}{lll}
\vspace{0.3mm}
  B & = &
  B_{1} +
  i\,c\, F_{2}\,
  \biggl\{
    (1 - 2\,F_{1})\,
    \Bigl[
      \bigl( F_{1}^{2} - F_{1} F_{2} q^{4} \bigr)\, q^{m} +
      i\, \bigl( F_{1}^{2} - F_{1} F_{2} q^{4} - F_{2}^{2}\, \mathbf{q}^{2} \bigr)\, \varepsilon_{mjl}\, q^{j}\, \sigma_{l}
    \Bigr]\; - \\

\vspace{1.0mm}
  & - &
    \displaystyle\frac{2\,F_{1} }{mc^{2}}\, \bigl[ ze\, A_{0} + V(\mathbf{r}) \bigr]\,
    \cdot
    \Bigl\{
      \bigl[ F_{1}^{2} + F_{1} F_{2} q^{4} \bigr]\, q^{m} +
      i\, \bigl[ F_{1}^{2} + F_{1} F_{2} q^{4} - F_{2}^{2}\, \mathbf{q}^{2} \bigr]\, \varepsilon_{mjl}\, q^{j}\, \sigma_{l}
    \Bigr\}\; + \\

\vspace{0.9mm}
  & + &
    (2\,F_{1} + 1)\,
    \cdot
    \Bigl\{
      \bigl[ F_{1}^{2} + F_{1} F_{2} q^{4} \bigr]\, q^{m} +
      i\, \bigl[ F_{1}^{2} + F_{1} F_{2} q^{4} - F_{2}^{2}\, \mathbf{q}^{2} \bigr]\, \varepsilon_{mjl}\, q^{j}\, \sigma_{l}
    \Bigr\}
  \biggr\}\,
  \Bigl( \mathbf{p} - \displaystyle\frac{ze}{c} \mathbf{A} \Bigr)_{m}\; - \\

  & - &
  m c^{2}\, (1 - 2\,F_{1})\, F_{2}^{2}\, \mathbf{q}^{2} +
  2\,(1 - 2\,F_{1})\, F_{1}F_{2}^{2}\, \bigl[ ze\, A_{0} + V(\mathbf{r}) - m c^{2} \bigr]\, \mathbf{q}^{2}.
\end{array}
\label{eq.app.2.2.15}
\end{equation}

Let us simplify term at $\bigl( \mathbf{p} - \frac{ze}{c} \mathbf{A} \bigr)$. We have
\begin{equation}
\begin{array}{lll}
\vspace{0.1mm}
  &
  (1 - 2\,F_{1})\,
  \Bigl[
    \bigl( F_{1}^{2} - F_{1} F_{2} q^{4} \bigr)\, q^{m} +
    i\, \bigl( F_{1}^{2} - F_{1} F_{2} q^{4} - F_{2}^{2}\, \mathbf{q}^{2} \bigr)\, \varepsilon_{mjl}\, q^{j}\, \sigma_{l}
  \Bigr]\; - \\
\vspace{0.3mm}
  - &
    \displaystyle\frac{2\,F_{1} }{mc^{2}}\, \bigl[ ze\, A_{0} + V(\mathbf{r}) \bigr]\,
    \cdot
    \Bigl\{
      \bigl[ F_{1}^{2} + F_{1} F_{2} q^{4} \bigr]\, q^{m} +
      i\, \bigl[ F_{1}^{2} + F_{1} F_{2} q^{4} - F_{2}^{2}\, \mathbf{q}^{2} \bigr]\, \varepsilon_{mjl}\, q^{j}\, \sigma_{l}
    \Bigr\}\; + \\
\vspace{2.8mm}
  + &
    (2\,F_{1} + 1)\,
    \cdot
    \Bigl\{
      \bigl[ F_{1}^{2} + F_{1} F_{2} q^{4} \bigr]\, q^{m} +
      i\, \bigl[ F_{1}^{2} + F_{1} F_{2} q^{4} - F_{2}^{2}\, \mathbf{q}^{2} \bigr]\, \varepsilon_{mjl}\, q^{j}\, \sigma_{l}
    \Bigr\}\; = \\


\vspace{0.1mm}
  = &
  \Bigl[
    (1 - 2\,F_{1})\, \bigl( F_{1}^{2} - F_{1} F_{2} q^{4} \bigr) +
    (2\,F_{1} + 1) \bigl( F_{1}^{2} + F_{1} F_{2} q^{4} \bigr) -
    \displaystyle\frac{2\,F_{1} }{mc^{2}}\, \bigl( ze\, A_{0} + V(\mathbf{r}) \bigr) \bigl( F_{1}^{2} + F_{1} F_{2} q^{4} \bigr)
 \Bigr]\, q^{m}\; + \\
\vspace{0.1mm}
  + &
  i\, \Bigl[
    (1 - 2\,F_{1}) \bigl( F_{1}^{2} - F_{1} F_{2} q^{4} - F_{2}^{2}\, \mathbf{q}^{2} \bigr) +
    (2\,F_{1} + 1) \bigl( F_{1}^{2} + F_{1} F_{2} q^{4} - F_{2}^{2}\, \mathbf{q}^{2} \bigr)\; - \\
  - &
    \displaystyle\frac{2\,F_{1} }{mc^{2}}\, \bigl( ze\, A_{0} + V(\mathbf{r}) \bigr) \bigl( F_{1}^{2} + F_{1} F_{2} q^{4} - F_{2}^{2}\, \mathbf{q}^{2} \bigr)
  \Bigr]\, \varepsilon_{mjl}\, q^{j}\, \sigma_{l}.
\end{array}
\label{eq.app.2.2.16}
\end{equation}
We consider summation of two terms:
\begin{equation}
\begin{array}{lll}
  (1 - 2\,F_{1})\, \bigl( F_{1}^{2} - F_{1} F_{2} q^{4} \bigr) + (2\,F_{1} + 1) \bigl( F_{1}^{2} + F_{1} F_{2} q^{4} \bigr)\; = 
  2\, F_{1}^{2} + 4\,F_{1} \cdot F_{1} F_{2} q^{4} =
  2\, F_{1}^{2} (1 + 2\,F_{2} q^{4}).
\end{array}
\label{eq.app.2.2.17}
\end{equation}
Then expression (\ref{eq.app.2.2.16}) can be rewritten as
\begin{equation}
\begin{array}{lll}
\vspace{0.1mm}
  &
  \Bigl[
    2\, F_{1}^{2} (1 + 2\,F_{2} q^{4}) -
    \displaystyle\frac{2\,F_{1} }{mc^{2}}\, \bigl( ze\, A_{0} + V(\mathbf{r}) \bigr) \bigl( F_{1}^{2} + F_{1} F_{2} q^{4} \bigr)
 \Bigr]\, q^{m}\; + \\
\vspace{2.8mm}
  + &
  i\, \Bigl[
    2\, F_{1}^{2} (1 + 2\,F_{2} q^{4}) -
    (1 - 2\,F_{1})\, F_{2}^{2}\, \mathbf{q}^{2} - (2\,F_{1} + 1)\, F_{2}^{2}\, \mathbf{q}^{2} -
    \displaystyle\frac{2\,F_{1} }{mc^{2}}\, \bigl( ze\, A_{0} + V(\mathbf{r}) \bigr) \bigl( F_{1}^{2} + F_{1} F_{2} q^{4} - F_{2}^{2}\, \mathbf{q}^{2} \bigr)
  \Bigr]\, \varepsilon_{mjl}\, q^{j}\, \sigma_{l}\; = \\

\vspace{0.1mm}
  = &
  \Bigl[
    2\, F_{1}^{2} (1 + 2\,F_{2} q^{4}) -
    \displaystyle\frac{2\,F_{1} }{mc^{2}}\, \bigl( ze\, A_{0} + V(\mathbf{r}) \bigr) \bigl( F_{1}^{2} + F_{1} F_{2} q^{4} \bigr)
 \Bigr]\, q^{m}\; + \\
  + &
  i\, \Bigl[
    2\, F_{1}^{2} (1 + 2\,F_{2} q^{4}) - 2\, F_{2}^{2}\, \mathbf{q}^{2} -
    \displaystyle\frac{2\,F_{1} }{mc^{2}}\, \bigl( ze\, A_{0} + V(\mathbf{r}) \bigr) \bigl( F_{1}^{2} + F_{1} F_{2} q^{4} - F_{2}^{2}\, \mathbf{q}^{2} \bigr)
  \Bigr]\, \varepsilon_{mjl}\, q^{j}\, \sigma_{l}.
\end{array}
\label{eq.app.2.2.18}
\end{equation}
Taking this expression into account, we rewrite solution for $B$ in Eq.~(\ref{eq.app.2.2.15}):
%
%
%
%
%
\begin{equation}
  B =
  B_{1} + B_{2}\, \Bigl( \mathbf{p} - \displaystyle\frac{ze}{c} \mathbf{A} \Bigr)_{m} -
  m c^{2}\, (1 - 2\,F_{1})\, F_{2}^{2}\, \mathbf{q}^{2} +
  2\,(1 - 2\,F_{1})\, F_{1}F_{2}^{2}\, \bigl[ ze\, A_{0} + V(\mathbf{r}) - m c^{2} \bigr]\, \mathbf{q}^{2},
\label{eq.app.2.2.19}
\end{equation}
where %
\begin{equation}
\begin{array}{lll}
\vspace{0.3mm}
  B_{2} & = &
  icF_{2}\;
  \biggl\{
  \Bigl[
    2\, F_{1}^{2} (1 + 2\,F_{2} q^{4}) -
    \displaystyle\frac{2\,F_{1} }{mc^{2}}\, \bigl( ze\, A_{0} + V(\mathbf{r}) \bigr) \bigl( F_{1}^{2} + F_{1} F_{2} q^{4} \bigr)
 \Bigr]\, q^{m}\; + \\
  & + &
    i\, \Bigl[
      2\, F_{1}^{2} (1 + 2\,F_{2} q^{4}) - 2\, F_{2}^{2}\, \mathbf{q}^{2} -
      \displaystyle\frac{2\,F_{1} }{mc^{2}}\, \bigl( ze\, A_{0} + V(\mathbf{r}) \bigr) \bigl( F_{1}^{2} + F_{1} F_{2} q^{4} - F_{2}^{2}\, \mathbf{q}^{2} \bigr)
    \Bigr]\, \varepsilon_{mjl}\, q^{j}\, \sigma_{l}
  \biggr\}.
\end{array}
\label{eq.app.2.2.20}
\end{equation}

As next step, we calculate such a term:
\begin{equation}
\begin{array}{lll}
\vspace{1.7mm}
  & B_{2}\, +
  c\, F_{2}\,
  \Bigl[ - i \bigl( F_{1} + F_{2} q^{4} \bigr)\, q^{m} + F_{2} q^{4} q^{k}\, \varepsilon_{kmj} \sigma_{j} \Bigr]
  \cdot f(|\mathbf{q}|)\; = \\


\vspace{0.1mm}
  = &
  icF_{2}\;
  \Bigl[
    2\, F_{1}^{2} (1 + 2\,F_{2} q^{4}) -
    \displaystyle\frac{2\,F_{1} }{mc^{2}}\, \bigl( ze\, A_{0} + V(\mathbf{r}) \bigr) \bigl( F_{1}^{2} + F_{1} F_{2} q^{4} \bigr) -
    \bigl( F_{1} + F_{2} q^{4} \bigr) f(|\mathbf{q}|)
 \Bigr]\, q^{m}\; - \\
  - &
    cF_{2}\, \Bigl[
      2\, F_{1}^{2} (1 + 2\,F_{2} q^{4}) - 2\, F_{2}^{2}\, \mathbf{q}^{2} -
      \displaystyle\frac{2\,F_{1} }{mc^{2}}\, \bigl( ze\, A_{0} + V(\mathbf{r}) \bigr) \bigl( F_{1}^{2} + F_{1} F_{2} q^{4} - F_{2}^{2}\, \mathbf{q}^{2} \bigr) -
      F_{2} q^{4} f(|\mathbf{q}|)\,
    \Bigr]\, \varepsilon_{mjl}\, q^{j}\, \sigma_{l}.
\end{array}
\label{eq.app.2.2.21}
\end{equation}
We obtain:
\begin{equation}
\begin{array}{lll}
  B - B_{1} + A \cdot f(|\mathbf{q}|) =
  i\, cF_{2}\,
  \Bigl\{ b_{1}\, q^{m} + b_{2}\, \varepsilon_{mjl}\, q^{j}\, \sigma_{l} \Bigr\}
  \Bigl( \mathbf{p} - \displaystyle\frac{ze}{c} \mathbf{A} \Bigr)_{m}\; +
  mc^{2}\,b_{3},
\end{array}
\label{eq.app.2.2.22}
\end{equation}
where
\begin{equation}
\begin{array}{lll}
\vspace{1.1mm}
  b_{1} & = &
    2\, F_{1}^{2} (1 + 2\,F_{2} q^{4}) -
    \displaystyle\frac{2\,F_{1} }{mc^{2}}\, \bigl( ze\, A_{0} + V(\mathbf{r}) \bigr) \bigl( F_{1}^{2} + F_{1} F_{2} q^{4} \bigr) -
    \bigl( F_{1} + F_{2} q^{4} \bigr) f(|\mathbf{q}|), \\

\vspace{1.1mm}
  b_{2} & = &
    i\, \Bigl[
      2\, F_{1}^{2} (1 + 2\,F_{2} q^{4}) - 2\, F_{2}^{2}\, \mathbf{q}^{2} -
      \displaystyle\frac{2\,F_{1} }{mc^{2}}\, \bigl( ze\, A_{0} + V(\mathbf{r}) \bigr) \bigl( F_{1}^{2} + F_{1} F_{2} q^{4} - F_{2}^{2}\, \mathbf{q}^{2} \bigr) -
      F_{2} q^{4} f(|\mathbf{q}|)\,
    \Bigr], \\

\vspace{0.9mm}
  b_{3} & = &
  \displaystyle\frac{1}{mc^{2}}
  \Bigl\{
  - m c^{2}\, (1 - 2\,F_{1})\, F_{2}^{2}\, \mathbf{q}^{2} +
  2\,(1 - 2\,F_{1})\, F_{1}F_{2}^{2}\, \bigl[ ze\, A_{0} + V(\mathbf{r}) - m c^{2} \bigr]\, \mathbf{q}^{2}\; + \\
  & + &
  \Bigl[
    F_{1}\, m c^{2} +
    \bigl(F_{1}^{2} + F_{2}^{2}\, \mathbf{q}^{2} \bigr)\,
    \bigl( ze\, A_{0} + V(\mathbf{r}) - m c^{2} \bigr)
  \Bigr]\, f(|\mathbf{q}|)
  \Bigr\}.
\end{array}
\label{eq.app.2.2.23}
\end{equation}

\subsection{Calculations of $b_{1}$, $b_{2}$, $b_{3}$
\label{sec.2.2.2}}

Taking into account Eq.~(\ref{eq.2.37})
\[
  f(|\mathbf{q}|) = F_{1} + F_{1}^{2} + F_{2}^{2}\, \mathbf{q}^{2},
\]
we have
%
\begin{equation}
\begin{array}{lll}
\vspace{1.1mm}
  b_{1} & = &
    2\, F_{1}^{2} (1 + 2\,F_{2} q^{4}) -
    \displaystyle\frac{2\,F_{1} }{mc^{2}}\, \bigl( ze\, A_{0} + V(\mathbf{r}) \bigr) \bigl( F_{1}^{2} + F_{1} F_{2} q^{4} \bigr) -
    \bigl( F_{1} + F_{2} q^{4} \bigr)\, \bigl[ F_{1} + F_{1}^{2} + F_{2}^{2}\, \mathbf{q}^{2} \bigr]\; = \\

%
%
  & = &
  F_{1}^{2} (1 - F_{1}) + F_{1} F_{2} (3F_{1} - 1)\, q^{4} -
  F_{2}^{2} \bigl( F_{1} + F_{2} q^{4} \bigr)\, \mathbf{q}^{2} -
  \displaystyle\frac{2\,F_{1}^{2} }{mc^{2}} \bigl( F_{1} + F_{2} q^{4} \bigr) \bigl( ze\, A_{0} + V(\mathbf{r}) \bigr).
\end{array}
\label{eq.app.2.2.2.1}
\end{equation}
Now we simplify solution for $b_{2}$. Taking Eq.~(\ref{eq.2.37}) into account, we obtain:
%
\begin{equation}
\begin{array}{lll}
\vspace{1.1mm}
  b_{2} & = &
    i\, \Bigl[
      2\, F_{1}^{2} (1 + 2\,F_{2} q^{4}) - 2\, F_{2}^{2}\, \mathbf{q}^{2} -
      \displaystyle\frac{2\,F_{1} }{mc^{2}}\, \bigl( ze\, A_{0} + V(\mathbf{r}) \bigr) \bigl( F_{1}^{2} + F_{1} F_{2} q^{4} - F_{2}^{2}\, \mathbf{q}^{2} \bigr) -
      F_{2} q^{4}\, \bigl( F_{1} + F_{1}^{2} + F_{2}^{2}\, \mathbf{q}^{2} \bigr)
    \Bigr]\; = \\


  & = &
    i\, \Bigl[
      2\, F_{1}^{2} +
      F_{2} \bigl( 3 F_{1}^{2} - F_{1}\bigr)\, q^{4} -
      F_{2}^{2} \bigl( 2 + F_{2} q^{4} \bigr)\, \mathbf{q}^{2} -
      \displaystyle\frac{2\,F_{1}}{mc^{2}} \bigl( F_{1}^{2} + F_{1} F_{2} q^{4} - F_{2}^{2}\, \mathbf{q}^{2} \bigr) \bigl( ze\, A_{0} + V(\mathbf{r}) \bigr)
    \Bigr].
\end{array}
\label{eq.app.2.2.2.2}
\end{equation}
We simplify solution for $b_{3}$, taking (\ref{eq.2.37}) into account:
\begin{equation}
\begin{array}{lll}
\vspace{0.2mm}
  mc^{2}\, b_{3} & = &
  m c^{2}\,
  \Bigl\{
    F_{1}^{2}\, (1 - F_{1}^{2}) -
    (1 - 2\,F_{1}^{2})\, F_{2}^{2}\, \mathbf{q}^{2} -
    F_{2}^{4}\, \mathbf{q}^{4}
  \Bigr\} +
  \Bigl\{
    F_{1}^{3}\, (1 + F_{1}) +
    F_{1} F_{2}^{2}\, (3 - 2 F_{1})\, \mathbf{q}^{2} +
    F_{2}^{4}\, \mathbf{q}^{4}
  \Bigr\}\, \bigl[ ze\, A_{0} + V(\mathbf{r}) \bigr].
\end{array}
\label{eq.app.2.2.2.3}
\end{equation}
We summarize the found solution:
\begin{equation}
\begin{array}{lll}
\vspace{1.1mm}
  b_{1} & = &
  F_{1}^{2} (1 - F_{1}) + F_{1} F_{2} (3F_{1} - 1)\, q^{4} -
  F_{2}^{2} \bigl( F_{1} + F_{2} q^{4} \bigr)\, \mathbf{q}^{2} -
  \displaystyle\frac{2\,F_{1}^{2}}{mc^{2}} \bigl( F_{1} + F_{2} q^{4} \bigr) \bigl[ ze\, A_{0} + V(\mathbf{r}) \bigr], \\

\vspace{1.1mm}
  b_{2} & = &
    i\, \Bigl[
      2\, F_{1}^{2} +
      F_{1} F_{2} \bigl( 3 F_{1} - 1 \bigr)\, q^{4} -
      F_{2}^{2} \bigl( 2 + F_{2} q^{4} \bigr)\, \mathbf{q}^{2} -
      \displaystyle\frac{2\,F_{1}}{mc^{2}} \bigl( F_{1}^{2} + F_{1} F_{2} q^{4} - F_{2}^{2}\, \mathbf{q}^{2} \bigr) \bigl( ze\, A_{0} + V(\mathbf{r}) \bigr)
    \Bigr], \\

  b_{3} & = &
  \Bigl\{
    F_{1}^{2}\, (1 - F_{1}^{2}) -
    (1 - 2\,F_{1}^{2})\, F_{2}^{2}\, \mathbf{q}^{2} -
    F_{2}^{4}\, \mathbf{q}^{4}
  \Bigr\} +
  \displaystyle\frac{1}{mc^{2}}\,
  \Bigl\{
    F_{1}^{3}\, (1 + F_{1}) +
    F_{1} F_{2}^{2}\, (3 - 2 F_{1})\, \mathbf{q}^{2} +
    F_{2}^{4}\, \mathbf{q}^{4}
  \Bigr\} \bigl[ ze\, A_{0} + V(\mathbf{r}) \bigr].
\end{array}
\label{eq.app.2.2.2.4}
\end{equation}

\subsection{Calculation of $B_{1}$
\label{sec.2.3}}

Let us calculate $B_{1}$ in Eq.~(\ref{eq.app.2.2.5}):
\begin{equation}
\begin{array}{lll}
\vspace{0.5mm}
  m B_{1} & = &
%
  \Bigl[ \bigl[ F_{1}^{2} - F_{1} F_{2} q^{4} - F_{2}^{2}\, \mathbf{q}^{2} \bigr]\, \sigma_{m} \Bigr]
    \Bigl( p_{m} - \displaystyle\frac{ze}{c} A_{m} \Bigr) \cdot
  \Bigl[ \bigl[ F_{1}^{2} + F_{1} F_{2} q^{4} - F_{2}^{2}\, \mathbf{q}^{2} \bigr]\, \sigma_{m^{\prime}} \Bigr]
    \Bigl( p_{m'} - \displaystyle\frac{ze}{c} A_{m'} \Bigr)\; + \\

\vspace{0.7mm}
  & + &
    F_{2}^{2}\, q^{m} q^{l}\, \sigma_{l}\; \Bigl( p_{m} - \displaystyle\frac{ze}{c} A_{m} \Bigr) \cdot
    \Bigl[ \bigl[ F_{1}^{2} + F_{1} F_{2} q^{4} - F_{2}^{2}\, \mathbf{q}^{2} \bigr]\, \sigma_{m^{\prime}} \Bigr]
      \Bigl( p_{m'} - \displaystyle\frac{ze}{c} A_{m'} \Bigr)\; + \\

\vspace{0.7mm}
  & + &
    \Bigl[ \bigl[ F_{1}^{2} - F_{1} F_{2} q^{4} - F_{2}^{2}\, \mathbf{q}^{2} \bigr]\, \sigma_{m} \Bigr]
      \Bigl( p_{m} - \displaystyle\frac{ze}{c} A_{m} \Bigr) \cdot
    F_{2}^{2}\: q^{m^{\prime}} q^{l}\, \sigma_{l}\; \Bigl( p_{m'} - \displaystyle\frac{ze}{c} A_{m'} \Bigr)\; + \\

  & + &
    F_{2}^{2}\, q^{m} q^{l}\, \sigma_{l}\; \Bigl( p_{m} - \displaystyle\frac{ze}{c} A_{m} \Bigr) \cdot
    F_{2}^{2}\: q^{m^{\prime}} q^{l}\, \sigma_{l}\; \Bigl( p_{m'} - \displaystyle\frac{ze}{c} A_{m'} \Bigr).
\end{array}
\label{eq.app.2.3.1}
\end{equation}
We simplify the first term in the obtained expression:
\begin{equation}
\begin{array}{lll}
\vspace{0.8mm}
  &  &
  \Bigl[ \bigl[ F_{1}^{2} - F_{1} F_{2} q^{4} - F_{2}^{2}\, \mathbf{q}^{2} \bigr]\, \sigma_{m} \Bigr]
    \Bigl( p_{m} - \displaystyle\frac{ze}{c} A_{m} \Bigr) \cdot
  \Bigl[ \bigl[ F_{1}^{2} + F_{1} F_{2} q^{4} - F_{2}^{2}\, \mathbf{q}^{2} \bigr]\, \sigma_{m^{\prime}} \Bigr]
    \Bigl( p_{m'} - \displaystyle\frac{ze}{c} A_{m'} \Bigr)\; = \\


  & = &
  \bigl[ (F_{1}^{2} - F_{2}^{2}\, \mathbf{q}^{2})^{2} - F_{1}^{2} F_{2}^{2} (q^{4})^{2} \bigr] \cdot
  \Bigl[ \sigmabf\, \Bigl( \mathbf{p} - \displaystyle\frac{z e}{c} \mathbf{A} \Bigr) \Bigr]^{2} =

  a_{1} \cdot \Bigl[ \sigmabf\, \Bigl( \mathbf{p} - \displaystyle\frac{z e}{c} \mathbf{A} \Bigr) \Bigr]^{2},
\end{array}
\label{eq.app.2.3.2}
\end{equation}
where
\begin{equation}
  a_{1} = (F_{1}^{2} - F_{2}^{2}\, \mathbf{q}^{2})^{2} - F_{1}^{2} F_{2}^{2} (q^{4})^{2}.
\label{eq.app.2.3.3}
\end{equation}
Using properties of Dirac's matrices, we have
\begin{equation}
\begin{array}{lcl}
  \Bigl[ \sigmabf \Bigl( \mathbf{p} - \displaystyle\frac{ze}{c} \mathbf{A} \Bigr) \Bigr]^{2} =
  \Bigl( \mathbf{p} - \displaystyle\frac{ze}{c} \mathbf{A} \Bigr)^{2} -
    \displaystyle\frac{ze}{c}\: \sigmabf \mathbf{H},
\end{array}
\label{eq.app.2.3.4}
\end{equation}
where $\mathbf{H} = \mathbf{rot\, A}$ is magnetic field.
Substituting this equation to Eq.~(\ref{eq.app.2.3.2}), we obtain:
\begin{equation}
\begin{array}{lll}
\vspace{0.7mm}
  &
  \Bigl[ \bigl[ F_{1}^{2} - F_{1} F_{2} q^{4} - F_{2}^{2}\, \mathbf{q}^{2} \bigr]\, \sigma_{m} \Bigr]
    \Bigl( p_{m} - \displaystyle\frac{ze}{c} A_{m} \Bigr) \cdot
  \Bigl[ \bigl[ F_{1}^{2} + F_{1} F_{2} q^{4} - F_{2}^{2}\, \mathbf{q}^{2} \bigr]\, \sigma_{m^{\prime}} \Bigr]
    \Bigl( p_{m'} - \displaystyle\frac{ze}{c} A_{m'} \Bigr)\; = \\

  = &
  a_{1}\;
  \Bigl[
    \Bigl( \mathbf{p} - \displaystyle\frac{ze}{c} \mathbf{A} \Bigr)^{2} -
      \displaystyle\frac{ze}{c}\, \sigmabf \mathbf{H}
  \Bigr].
\end{array}
\label{eq.app.2.3.5}
\end{equation}

We simplify fourth term in the obtained Eq.~(\ref{eq.app.2.3.1}) [$m,m^{\prime}=1,2,3$]:
%
\begin{equation}
\begin{array}{lll}
    F_{2}^{2}\, q^{m} q^{l}\, \sigma_{l}\; \Bigl( p_{m} - \displaystyle\frac{ze}{c} A_{m} \Bigr) \cdot
    F_{2}^{2}\: q^{m^{\prime}} q^{l}\, \sigma_{l}\; \Bigl( p_{m'} - \displaystyle\frac{ze}{c} A_{m'} \Bigr)\; = 


    F_{2}^{4} \mathbf{q}^{2}\,
    \Bigl( q^{m} p_{m} - \displaystyle\frac{ze}{c}\, q^{m} A_{m} \Bigr)^{2} =
    F_{2}^{4} \mathbf{q}^{2}\,
    \Bigl( \mathbf{qp} - \displaystyle\frac{ze}{c}\, \mathbf{qA} \Bigr)^{2}.
\end{array}
\label{eq.app.2.3.6}
\end{equation}
Now we calculate summation of the second and third terms in Eq.~(\ref{eq.app.2.3.1}):
%
\begin{equation}
\begin{array}{lll}
\vspace{0.7mm}
  & &
    F_{2}^{2}\, q^{m} q^{l}\, \sigma_{l}\; \Bigl( p_{m} - \displaystyle\frac{ze}{c} A_{m} \Bigr)\; \cdot
    \Bigl[ \bigl[ F_{1}^{2} + F_{1} F_{2} q^{4} - F_{2}^{2}\, \mathbf{q}^{2} \bigr]\, \sigma_{m^{\prime}} \Bigr]
      \Bigl( p_{m'} - \displaystyle\frac{ze}{c} A_{m'} \Bigr)\; + \\
\vspace{2.5mm}
  & + &
    \Bigl[ \bigl[ F_{1}^{2} - F_{1} F_{2} q^{4} - F_{2}^{2}\, \mathbf{q}^{2} \bigr]\, \sigma_{m} \Bigr]
      \Bigl( p_{m} - \displaystyle\frac{ze}{c} A_{m} \Bigr) \cdot
    F_{2}^{2}\: q^{m^{\prime}} q^{l}\, \sigma_{l}\; \Bigl( p_{m'} - \displaystyle\frac{ze}{c} A_{m'} \Bigr)\; = \\

\vspace{0.7mm}
  & = &
    F_{2}^{2} \bigl[ F_{1}^{2} + F_{1} F_{2} q^{4} - F_{2}^{2}\, \mathbf{q}^{2} \bigr]\,
    \mathbf{q \sigmabf}\;
    q^{m} \Bigl( p_{m} - \displaystyle\frac{ze}{c} A_{m} \Bigr)\; \cdot
    \sigma_{m^{\prime}} \Bigl( p_{m'} - \displaystyle\frac{ze}{c} A_{m'} \Bigr)\; + \\
  & + &
    F_{2}^{2}\, \bigl[ F_{1}^{2} - F_{1} F_{2} q^{4} - F_{2}^{2}\, \mathbf{q}^{2} \bigr]\, \sigma_{m} \mathbf{q \sigmabf}\;
      \Bigl( p_{m} - \displaystyle\frac{ze}{c} A_{m} \Bigr) \cdot
    q^{m^{\prime}} \Bigl( p_{m'} - \displaystyle\frac{ze}{c} A_{m'} \Bigr).
\end{array}
\label{eq.app.2.3.7}
\end{equation}
Introducing new functions:
%
\begin{equation}
\begin{array}{ll}
  a_{2} = F_{2}^{2} \bigl[ F_{1}^{2} + F_{1} F_{2} q^{4} - F_{2}^{2}\, \mathbf{q}^{2} \bigr], &
  a_{3} = F_{2}^{2}\, \bigl[ F_{1}^{2} - F_{1} F_{2} q^{4} - F_{2}^{2}\, \mathbf{q}^{2} \bigr],
\end{array}
\label{eq.app.2.3.8}
\end{equation}
we rewrite this summation as
%
\begin{equation}
\begin{array}{lll}
\vspace{0.7mm}
  & &
    F_{2}^{2}\, q^{m} q^{l}\, \sigma_{l}\; \Bigl( p_{m} - \displaystyle\frac{ze}{c} A_{m} \Bigr)\; \cdot
    \Bigl[ \bigl[ F_{1}^{2} + F_{1} F_{2} q^{4} - F_{2}^{2}\, \mathbf{q}^{2} \bigr]\, \sigma_{m^{\prime}} \Bigr]
      \Bigl( p_{m'} - \displaystyle\frac{ze}{c} A_{m'} \Bigr)\; + \\
\vspace{2.5mm}
  & + &
    \Bigl[ \bigl[ F_{1}^{2} - F_{1} F_{2} q^{4} - F_{2}^{2}\, \mathbf{q}^{2} \bigr]\, \sigma_{m} \Bigr]
      \Bigl( p_{m} - \displaystyle\frac{ze}{c} A_{m} \Bigr) \cdot
    F_{2}^{2}\: q^{m^{\prime}} q^{l}\, \sigma_{l}\; \Bigl( p_{m'} - \displaystyle\frac{ze}{c} A_{m'} \Bigr)\; = \\

  & = &
    a_{2}\,
      \mathbf{q \sigmabf}\;
      q^{m} \Bigl( p_{m} - \displaystyle\frac{ze}{c} A_{m} \Bigr)\; \cdot
      \sigma_{m^{\prime}} \Bigl( p_{m'} - \displaystyle\frac{ze}{c} A_{m'} \Bigr) +

    a_{3}\, \sigma_{m} \mathbf{q \sigmabf}\;
      \Bigl( p_{m} - \displaystyle\frac{ze}{c} A_{m} \Bigr) \cdot
      q^{m^{\prime}} \Bigl( p_{m'} - \displaystyle\frac{ze}{c} A_{m'} \Bigr).
\end{array}
\label{eq.app.2.3.9}
\end{equation}
So, we obtain the following expression for $B_{1}$:
%
\begin{equation}
\begin{array}{lll}
\vspace{0.8mm}
  m B_{1} & = &
    a_{1}\;
    \Bigl[
      \Bigl( \mathbf{p}_{i} - \displaystyle\frac{z_{i}e}{c} \mathbf{A}_{i} \Bigr)^{2} -
      \displaystyle\frac{z_{i}e}{c}\, \sigmabf \mathbf{H}
    \Bigr]\; +
    F_{2}^{4} \mathbf{q}^{2}\, \Bigl( \mathbf{qp} - \displaystyle\frac{ze}{c}\, \mathbf{qA} \Bigr)^{2}\; + \\

\vspace{0.8mm}
  & + &
    a_{2}\,
      \mathbf{q \sigmabf}\;
      q^{m} \Bigl( p_{m} - \displaystyle\frac{ze}{c} A_{m} \Bigr)\; \cdot
      \sigma_{m^{\prime}} \Bigl( p_{m'} - \displaystyle\frac{ze}{c} A_{m'} \Bigr) +

    a_{3}\, \sigma_{m} \mathbf{q \sigmabf}\;
      \Bigl( p_{m} - \displaystyle\frac{ze}{c} A_{m} \Bigr) \cdot
      q^{m^{\prime}} \Bigl( p_{m'} - \displaystyle\frac{ze}{c} A_{m'} \Bigr)\; = \\

  & = &
    a_{1}\;
    \Bigl[
      \Bigl( \mathbf{p}_{i} - \displaystyle\frac{z_{i}e}{c} \mathbf{A}_{i} \Bigr)^{2} -
      \displaystyle\frac{z_{i}e}{c}\, \sigmabf \mathbf{H}
    \Bigr]\; +
    F_{2}^{4} \mathbf{q}^{2}\, \Bigl( \mathbf{qp} - \displaystyle\frac{ze}{c}\, \mathbf{qA} \Bigr)^{2}\; + m\, B_{10},
\end{array}
\label{eq.app.2.3.10}
\end{equation}
where
\begin{equation}
\begin{array}{lll}
  m B_{10} & = &
    a_{2}\,
      \mathbf{q \sigmabf}\;
      q^{m} \Bigl( p_{m} - \displaystyle\frac{ze}{c} A_{m} \Bigr)\; \cdot
      \sigma_{m^{\prime}} \Bigl( p_{m'} - \displaystyle\frac{ze}{c} A_{m'} \Bigr) +

    a_{3}\, \sigma_{m} \mathbf{q \sigmabf}\;
      \Bigl( p_{m} - \displaystyle\frac{ze}{c} A_{m} \Bigr) \cdot
      q^{m^{\prime}} \Bigl( p_{m'} - \displaystyle\frac{ze}{c} A_{m'} \Bigr).
\end{array}
\label{eq.app.2.3.11}
\end{equation}

Taking properties (\ref{eq.app.2.2.1}) into account, we simplify the first term in Eq.~(\ref{eq.app.2.3.11}):
\begin{equation}
\begin{array}{lll}
\vspace{0.8mm}
  & &
    a_{2}\,
      \mathbf{q \sigmabf}\;
      q^{m} \Bigl( p_{m} - \displaystyle\frac{ze}{c} A_{m} \Bigr) \cdot
      \sigma_{m^{\prime}} \Bigl( p_{m'} - \displaystyle\frac{ze}{c} A_{m'} \Bigr) =

    a_{2}\,
      q^{l} \sigma_{l}\; \sigma_{m^{\prime}}\;
      q^{m} \Bigl( p_{m} - \displaystyle\frac{ze}{c} A_{m} \Bigr) \cdot
      \Bigl( p_{m'} - \displaystyle\frac{ze}{c} A_{m'} \Bigr)\; = \\


  & = &
    a_{2}\, q^{m^{\prime}} q^{m}\,
      \Bigl( p_{m} - \displaystyle\frac{ze}{c} A_{m} \Bigr)\,
      \Bigl( p_{m'} - \displaystyle\frac{ze}{c} A_{m'} \Bigr) +

    a_{2}\,i\, \varepsilon_{lm^{\prime}k}\: q^{l} q^{m}\, \sigma_{k}
      \Bigl( p_{m} - \displaystyle\frac{ze}{c} A_{m} \Bigr)\,
      \Bigl( p_{m'} - \displaystyle\frac{ze}{c} A_{m'} \Bigr)
\end{array}
\label{eq.app.2.3.12}
\end{equation}
and the second term in Eq.~(\ref{eq.app.2.3.11}):
%
\begin{equation}
\begin{array}{lll}
\vspace{0.8mm}
  & &
  a_{3}\, \sigma_{m} \mathbf{q \sigmabf}\;
      \Bigl( p_{m} - \displaystyle\frac{ze}{c} A_{m} \Bigr) \cdot
      q^{m^{\prime}} \Bigl( p_{m'} - \displaystyle\frac{ze}{c} A_{m'} \Bigr) =
  a_{3}\, q^{l} q^{m^{\prime}}\: \sigma_{m} \sigma_{l}\;
      \Bigl( p_{m} - \displaystyle\frac{ze}{c} A_{m} \Bigr) \cdot
      \Bigl( p_{m'} - \displaystyle\frac{ze}{c} A_{m'} \Bigr)\; = \\


  & = &
  a_{3}\, q^{m} q^{m^{\prime}}\,
    \Bigl( p_{m} - \displaystyle\frac{ze}{c} A_{m} \Bigr)
    \Bigl( p_{m'} - \displaystyle\frac{ze}{c} A_{m'} \Bigr) +
  a_{3}\, i\, \varepsilon_{mlk}\; q^{l} q^{m^{\prime}} \sigma_{k}\:
    \Bigl( p_{m} - \displaystyle\frac{ze}{c} A_{m} \Bigr)
    \Bigl( p_{m'} - \displaystyle\frac{ze}{c} A_{m'} \Bigr).
\end{array}
\label{eq.app.2.3.13}
\end{equation}
We find summation of these two terms:
%
\begin{equation}
\begin{array}{lll}
\vspace{0.8mm}
  m B_{10} & = &
%
%
%
  (a_{2} + a_{3})\, \Bigl( \mathbf{qp} - \displaystyle\frac{ze}{c}\, \mathbf{qA} \Bigr)^{2}\; + m \bar{B}_{10},
\end{array}
\label{eq.app.2.3.14}
\end{equation}
where
\begin{equation}
\begin{array}{lll}
  m \bar{B}_{10} & = &
  a_{2}\,i\, \varepsilon_{lm^{\prime}k}\, q^{l}\, q^{m} \sigma_{k}\:
      \Bigl( p_{m} - \displaystyle\frac{ze}{c} A_{m} \Bigr)\;
      \Bigl( p_{m'} - \displaystyle\frac{ze}{c} A_{m'} \Bigr) +

  a_{3}\, i\, \varepsilon_{mlk}\; q^{l} q^{m^{\prime}} \sigma_{k}\:
    \Bigl( p_{m} - \displaystyle\frac{ze}{c} A_{m} \Bigr)
    \Bigl( p_{m'} - \displaystyle\frac{ze}{c} A_{m'} \Bigr).
\end{array}
\label{eq.app.2.3.15}
\end{equation}
Now we rewrite the found solution (\ref{eq.app.2.3.10}) as
\begin{equation}
\begin{array}{lll}
  m B_{1} & = &
    a_{1}\;
    \Bigl[
      \Bigl( \mathbf{p}_{i} - \displaystyle\frac{z_{i}e}{c} \mathbf{A}_{i} \Bigr)^{2} -
      \displaystyle\frac{z_{i}e}{c}\, \sigmabf \mathbf{H}
    \Bigr]\; +
    \bigl(F_{2}^{4} \mathbf{q}^{2} + a_{2} + a_{3}\bigr)\, \Bigl( \mathbf{qp} - \displaystyle\frac{ze}{c}\, \mathbf{qA} \Bigr)^{2}\; + m\, \bar{B}_{10}.
\end{array}
\label{eq.app.2.3.16}
\end{equation}

\subsection{Calculation of $\bar{B}_{10}$
\label{sec.2.4}}

Let us rewrite Eq.~(\ref{eq.app.2.3.15}):
%
\[
\begin{array}{lll}
  m \bar{B}_{10} & = &
  a_{2}\,i\, \varepsilon_{lm^{\prime}k}\, q^{l}\, q^{m} \sigma_{k}\:
      \Bigl( p_{m} - \displaystyle\frac{ze}{c} A_{m} \Bigr) \Bigl( p_{m'} - \displaystyle\frac{ze}{c} A_{m'} \Bigr) +
  a_{3}\, i\, \varepsilon_{mlk}\; q^{l} q^{m^{\prime}} \sigma_{k}\:
    \Bigl( p_{m} - \displaystyle\frac{ze}{c} A_{m} \Bigr) \Bigl( p_{m'} - \displaystyle\frac{ze}{c} A_{m'} \Bigr).
\end{array}
\]
We find summations:
%
\begin{equation}
\begin{array}{lll}
\vspace{0.5mm}
  &
  i\,a_{2}\, \varepsilon_{lm^{\prime}k}\, q^{l}\, q^{m} \sigma_{k}\: p_{m} p_{m'} +
  i\,a_{3}\, \varepsilon_{mlk}\, q^{l} q^{m^{\prime}} \sigma_{k}\: p_{m} p_{m'} =

%
  i\,(a_{2}-a_{3})\, \varepsilon_{lm^{\prime}k}\, \sigma_{k}\: q^{l} q^{m}\, p_{m} p_{m'}, \\

  &
  i\,a_{2}\, \varepsilon_{lm^{\prime}k}\, q^{l}\, q^{m} \sigma_{k}\: \displaystyle\frac{ze}{c} A_{m} \displaystyle\frac{ze}{c} A_{m'} +
  i\,a_{3}\, \varepsilon_{mlk}\, q^{l} q^{m^{\prime}} \sigma_{k}\: \displaystyle\frac{ze}{c} A_{m} \displaystyle\frac{ze}{c} A_{m'} =
  i\,(a_{2}-a_{3})\, \displaystyle\frac{z^{2}e^{2}}{c^{2}}\, \varepsilon_{lm^{\prime}k}\, \sigma_{k}\: q^{l} q^{m}\, A_{m} A_{m'}.
\end{array}
\label{eq.app.2.4.1}
\end{equation}
Now let us consider term:
%
\begin{equation}
\begin{array}{lll}
  i\,a_{2}\, \varepsilon_{lm^{\prime}k}\, q^{l}\, q^{m} \sigma_{k}\:
    \Bigl[ p_{m} \displaystyle\frac{ze}{c} A_{m'}  + \displaystyle\frac{ze}{c} A_{m} p_{m'} \Bigr] =

  i\,\displaystyle\frac{ze}{c}\, a_{2}\, \varepsilon_{lm^{\prime}k}\, q^{l}\, q^{m} \sigma_{k}\:
    \bigl[ -i\hbar \displaystyle\frac{dA_{m'}}{dx_{m}} + A_{m'} p_{m} + A_{m} p_{m'} \bigr]
\end{array}
\label{eq.app.2.4.2}
\end{equation}
and we obtain:
%
\begin{equation}
\begin{array}{lll}
\vspace{0.5mm}
  &
  i\,a_{2}\, \varepsilon_{lm^{\prime}k}\, q^{l}\, q^{m} \sigma_{k}\:
    \Bigl[ p_{m} \displaystyle\frac{ze}{c} A_{m'}  + \displaystyle\frac{ze}{c} A_{m} p_{m'} \Bigr] +
  i\,a_{3}\, \varepsilon_{mlk}\, q^{l}\, q^{m^{\prime}} \sigma_{k}\:
    \Bigl[ p_{m} \displaystyle\frac{ze}{c} A_{m'}  + \displaystyle\frac{ze}{c} A_{m} p_{m'} \Bigr]\; = \\

%
%
%
  = &
  i\,\displaystyle\frac{ze}{c}\, \varepsilon_{lm^{\prime}k}\, q^{l}\, q^{m} \sigma_{k}\:
    \Bigl[
      -i\hbar \bigl( a_{2}\, \displaystyle\frac{dA_{m'}}{dx_{m}} - a_{3}\, \displaystyle\frac{dA_{m}}{dx_{m'}} \bigr) +
      (a_{2} - a_{3}) \, ( A_{m'} p_{m} + A_{m} p_{m'} \bigr)
    \Bigr].
\end{array}
\label{eq.app.2.4.3}
\end{equation}
Now we find the final solution for $\bar{B}_{10}$, performing summation in (\ref{eq.app.2.4.1})--(\ref{eq.app.2.4.3}):
\begin{equation}
\begin{array}{lll}
  m \bar{B}_{10} & = &
  i\, \varepsilon_{lm^{\prime}k}\, \sigma_{k}\: q^{l} q^{m}\,
  \Bigl\{
    (a_{2}-a_{3}) \Bigl(p_{m} p_{m'} - \displaystyle\frac{ze}{c} ( A_{m'} p_{m} + A_{m} p_{m'} ) + \displaystyle\frac{z^{2}e^{2}}{c^{2}}\, A_{m} A_{m'} \Bigr) +
    \displaystyle\frac{i\hbar ze}{c} \Bigl[ a_{2}\, \displaystyle\frac{dA_{m'}}{dx_{m}} - a_{3}\, \displaystyle\frac{dA_{m}}{dx_{m'}} \Bigr]
  \Bigr\}.
\end{array}
\label{eq.app.2.4.4}
\end{equation}
We write the found solutions for the coefficients $a_{1}$, $a_{2}$ and $a_{3}$:
%
\begin{equation}
\begin{array}{lll}
  a_{1} = (F_{1}^{2} - F_{2}^{2}\, \mathbf{q}^{2})^{2} - F_{1}^{2} F_{2}^{2} (q^{4})^{2}, &
  a_{2} = F_{2}^{2} \bigl[ F_{1}^{2} + F_{1} F_{2} q^{4} - F_{2}^{2}\, \mathbf{q}^{2} \bigr], &
  a_{3} = F_{2}^{2}\, \bigl[ F_{1}^{2} - F_{1} F_{2} q^{4} - F_{2}^{2}\, \mathbf{q}^{2} \bigr].
\end{array}
\label{eq.app.2.4.5}
\end{equation}

\section{Operator of emission of bremsstrahlung photons
\label{sec.app.4}}

In approximation of $A_{0} = 0$, and using (\ref{eq.2.37})
\[
  f(|\mathbf{q}|) = F_{1} + F_{1}^{2} + F_{2}^{2}\, \mathbf{q}^{2},
\]
we rewrite equation (\ref{eq.4.1.1}) as
\begin{equation}
  i\hbar\, \Bigl\{ F_{1}^{3} (1 + F_{1}) - F_{1} F_{2}^{2}\, \mathbf{q}^{2} - F_{4}^{2}\, \mathbf{q}^{4} \Bigr\}\, \displaystyle\frac{\partial \varphi}{\partial t} =
  h_{0} \cdot \varphi + h_{\gamma} \varphi,
\label{eq.app.4.1}
\end{equation}
where
\begin{equation}
\begin{array}{lll}
\vspace{1.1mm}
  h_{0} & = &
    i\, cF_{2}\,
    \bigl[ b_{1}\, q^{m} + b_{2}\, \varepsilon_{mjl}\, q^{j}\, \sigma_{l} \bigr]\, \mathbf{p}_{m} +
    mc^{2}\,b_{3} +
  \displaystyle\frac{1}{m}\,
    \Bigl\{
      a_{1}\, \mathbf{p}_{i}^{2} + \bigl(F_{2}^{4} \mathbf{q}^{2} + a_{2} + a_{3}\bigr)\, \bigl( \mathbf{qp} \bigr)^{2} +
      i\, \varepsilon_{lm^{\prime}k}\, \sigma_{k}\: q^{l} q^{m}\, (a_{2}-a_{3}) p_{m} p_{m'}
    \Bigr\}, \\

\vspace{0.5mm}
  h_{\gamma} & = &
  -\, i\, cF_{2}\,
    \Bigl\{ b_{1}\, q^{m} + b_{2}\, \varepsilon_{mjl}\, q^{j}\, \sigma_{l} \Bigr\}\,
    \displaystyle\frac{ze}{c} \mathbf{A}_{m}\; + \\

\vspace{0.5mm}
  & + &
  \displaystyle\frac{1}{m}\,
  \Bigl\{
    a_{1}\:
    \Bigl[
      \Bigl( - \displaystyle\frac{z_{i}e}{c} (\mathbf{pA} + \mathbf{Ap}) \Bigr) +
      \displaystyle\frac{z_{i}^{2}e^{2}}{c^{2}}\, \mathbf{A}^{2} -
      \displaystyle\frac{z_{i}e}{c}\, \sigmabf \mathbf{H}
    \Bigr] +
    \bigl(F_{2}^{4} \mathbf{q}^{2} + a_{2} + a_{3}\bigr)\,
      \Bigl[ -2 \mathbf{(qp)} \displaystyle\frac{ze}{c}\, \mathbf{(qA)} + \displaystyle\frac{z^{2}e^{2}}{c^{2}}\, \mathbf{(qA)}^{2} \Bigr] + \\

  & + &
  i\, \varepsilon_{lm^{\prime}k}\, \sigma_{k}\: q^{l} q^{m}\,
  \Bigl[
    (a_{2}-a_{3}) \Bigl(- \displaystyle\frac{ze}{c} ( A_{m'} p_{m} + A_{m} p_{m'} ) + \displaystyle\frac{z^{2}e^{2}}{c^{2}}\, A_{m} A_{m'} \Bigr) +
    \displaystyle\frac{i\hbar ze}{c} \Bigl( a_{2}\, \displaystyle\frac{dA_{m'}}{dx_{m}} - a_{3}\, \displaystyle\frac{dA_{m}}{dx_{m'}} \Bigr) \Bigr]
  \Bigr\}.
\end{array}
\label{eq.app.4.2}
\end{equation}


One can separate explicitly terms with interacting potential in hamiltonian $h_{0}$.
We calculate:
\begin{equation}
\begin{array}{lll}
\vspace{1.5mm}
  & i\, cF_{2}\, \bigl[ b_{1}\, q^{m} + b_{2}\, \varepsilon_{mjl}\, q^{j}\, \sigma_{l} \bigr]\, \mathbf{p}_{m} + mc^{2}\,b_{3}\; = \\

\vspace{0.1mm}
  = &
  i\, cF_{2}\,
  \Bigl\{
  \Bigl[
    F_{1}^{2} (1 - F_{1}) + F_{1} F_{2} (3F_{1} - 1)\, q^{4} -
    F_{2}^{2} \bigl( F_{1} + F_{2} q^{4} \bigr)\, \mathbf{q}^{2}
  \Bigr]\, q^{m}\; + \\
\vspace{2.1mm}
  + &
  i\, \Bigl[
    2\, F_{1}^{2} +
    F_{1} F_{2} \bigl( 3 F_{1} - 1 \bigr)\, q^{4} -
    F_{2}^{2} \bigl( 2 + F_{2} q^{4} \bigr)\, \mathbf{q}^{2}
  \Bigr]\, \varepsilon_{mjl}\, q^{j}\, \sigma_{l} \Bigr\}\, \mathbf{p}_{m} +

  mc^{2}
  \Bigl[
    F_{1}^{2}\, (1 - F_{1}^{2}) -
    (1 - 2\,F_{1}^{2})\, F_{2}^{2}\, \mathbf{q}^{2} -
    F_{2}^{4}\, \mathbf{q}^{4}
  \Bigr]\; + \\

\vspace{0.1mm}
  - &
  i\, \displaystyle\frac{2F_{1}F_{2}}{mc}\,
  \Bigl\{
    F_{1} \bigl( F_{1} + F_{2} q^{4} \bigr)\, q^{m} +
    i\, \bigl( F_{1}^{2} + F_{1} F_{2} q^{4} - F_{2}^{2}\, \mathbf{q}^{2} \bigr) \varepsilon_{mjl}\, q^{j}\, \sigma_{l}
  \Bigr\}
  \bigl[ ze\, A_{0} + V(\mathbf{r}) \bigr] \mathbf{p}_{m}\; + \\
  + &
  \Bigl(
    F_{1}^{3}\, (1 + F_{1}) +
    F_{1} F_{2}^{2}\, (3 - 2 F_{1})\, \mathbf{q}^{2} +
    F_{2}^{4}\, \mathbf{q}^{4}
  \Bigr) \bigl[ ze\, A_{0} + V(\mathbf{r}) \bigr]
\end{array}
\label{eq.app.4.3}
\end{equation}
and obtain:
\begin{equation}
\begin{array}{lll}
\vspace{0.1mm}
  h_{0} & = &
  \displaystyle\frac{a_{1}\, \mathbf{p}_{i}^{2}}{m} +
  \Bigl(
    F_{1}^{3}\, (1 + F_{1}) +
    F_{1} F_{2}^{2}\, (3 - 2 F_{1})\, \mathbf{q}^{2} +
    F_{2}^{4}\, \mathbf{q}^{4}
  \Bigr) \bigl[ ze\, A_{0} + V(\mathbf{r}) \bigr]\; + \\
\vspace{1.5mm}
  & + &
  mc^{2}
  \Bigl[
    F_{1}^{2}\, (1 - F_{1}^{2}) -
    (1 - 2\,F_{1}^{2})\, F_{2}^{2}\, \mathbf{q}^{2} -
    F_{2}^{4}\, \mathbf{q}^{4}
  \Bigr]\; - \\

\vspace{0.1mm}
  & - &
  i\, \displaystyle\frac{2F_{1}F_{2}}{mc}\,
  \Bigl\{
    F_{1} \bigl( F_{1} + F_{2} q^{4} \bigr)\, q^{m} +
    i\, \bigl( F_{1}^{2} + F_{1} F_{2} q^{4} - F_{2}^{2}\, \mathbf{q}^{2} \bigr) \varepsilon_{mjl}\, q^{j}\, \sigma_{l}
  \Bigr\}
  \bigl[ ze\, A_{0} + V(\mathbf{r}) \bigr] \mathbf{p}_{m}\; + \\
\vspace{0.1mm}
  & + &
  \displaystyle\frac{1}{m}\,
    \Bigl\{
      \bigl(F_{2}^{4} \mathbf{q}^{2} + a_{2} + a_{3}\bigr)\, \bigl( \mathbf{qp} \bigr)^{2} +
      i\, \varepsilon_{lm^{\prime}k}\, \sigma_{k}\: q^{l} q^{m}\, (a_{2}-a_{3}) p_{m} p_{m'}
    \Bigr\}\; + \\
\vspace{0.1mm}
  & + &
  i\, cF_{2}\,
  \Bigl\{
  \Bigl[
    F_{1}^{2} (1 - F_{1}) + F_{1} F_{2} (3F_{1} - 1)\, q^{4} -
    F_{2}^{2} \bigl( F_{1} + F_{2} q^{4} \bigr)\, \mathbf{q}^{2}
  \Bigr]\, q^{m}\; + \\
  & + &
  i\, \Bigl[
    2\, F_{1}^{2} +
    F_{1} F_{2} \bigl( 3 F_{1} - 1 \bigr)\, q^{4} -
    F_{2}^{2} \bigl( 2 + F_{2} q^{4} \bigr)\, \mathbf{q}^{2}
  \Bigr]\, \varepsilon_{mjl}\, q^{j}\, \sigma_{l} \Bigr\}\, \mathbf{p}_{m}.
\end{array}
\label{eq.app.4.4}
\end{equation}


In operator of emission we separate term, corresponding to old our formalism in Ref.~\cite{Maydanyuk.2012.PRC}, where virtual photons were not included into analysis.
From Eq.~(\ref{eq.app.4.2}) we obtain:
\begin{equation}
  h_{\gamma} = h_{\gamma 0} + h_{\gamma 1},
\label{eq.app.4.5}
\end{equation}
where
\begin{equation}
\begin{array}{lll}
\vspace{1.1mm}
  h_{\gamma 0} & = &
  \displaystyle\frac{a_{1}}{m}\,
    \Bigl[
      - \displaystyle\frac{z_{i}e}{c} (-i\hbar\, \mathbf{div A} + 2\,\mathbf{Ap}) +
      \displaystyle\frac{z_{i}^{2}e^{2}}{c^{2}}\, \mathbf{A}^{2} -
      \displaystyle\frac{z_{i}e}{c}\, \sigmabf \mathbf{H}
    \Bigr], \\

\vspace{0.5mm}
  h_{\gamma 1} & = &
  -\, i\, cF_{2}\,
    \Bigl\{ b_{1}\, q^{m} + b_{2}\, \varepsilon_{mjl}\, q^{j}\, \sigma_{l} \Bigr\}\,
    \displaystyle\frac{ze}{c} \mathbf{A}_{m} +
  \displaystyle\frac{1}{m}\,
  \Bigl\{
    \bigl(F_{2}^{4} \mathbf{q}^{2} + a_{2} + a_{3}\bigr)\,
      \Bigl[ -2 \mathbf{(qp)} \displaystyle\frac{ze}{c}\, \mathbf{(qA)} + \displaystyle\frac{z^{2}e^{2}}{c^{2}}\, \mathbf{(qA)}^{2} \Bigr]\; + \\
  & + &
  i\, \varepsilon_{lm^{\prime}k}\, \sigma_{k}\: q^{l} q^{m}\,
  \Bigl[
    (a_{2}-a_{3}) \Bigl(- \displaystyle\frac{ze}{c} ( A_{m'} p_{m} + A_{m} p_{m'} ) + \displaystyle\frac{z^{2}e^{2}}{c^{2}}\, A_{m} A_{m'} \Bigr) +
    \displaystyle\frac{i\hbar ze}{c} \Bigl( a_{2}\, \displaystyle\frac{dA_{m'}}{dx_{m}} - a_{3}\, \displaystyle\frac{dA_{m}}{dx_{m'}} \Bigr) \Bigr]
  \Bigr\}.
\end{array}
\label{eq.app.4.6}
\end{equation}
The first term $h_{\gamma 0}$ is operator of emission in old formalism in Ref.~\cite{Maydanyuk.2012.PRC}, without inclusion of the virtual photons and possibility to consider internal ctructure of the scattered proton.
The second term $h_{\gamma 1}$ is correction of old operator of emission $h_{\gamma 0}$, which is appeared after inclusion of virtual photons to formalism.
%

\section{Elastic scattering of virtual photon on proton (scattered off nucleus)
\label{sec.app.5}}

Let us calculate hamiltonians $h_{0}$ and $h_{\gamma}$ for the elastic scattering of the virtual photon off proton.
From Eq.~(\ref{eq.4.2.2}) we obtain:
\begin{equation}
\begin{array}{lll}
\vspace{0.1mm}
  h_{0} & = &
  \displaystyle\frac{a_{1}\, \mathbf{p}^{2}}{m} +
  \Bigl(
    F_{1}^{3}\, (1 + F_{1}) +
    F_{1} F_{2}^{2}\, (3 - 2 F_{1})\, Q^{2} +
    F_{2}^{4}\, Q^{4}
  \Bigr)\, V(\mathbf{r})\; + \\
\vspace{1.5mm}
  & + &
  mc^{2}
  \Bigl[
    F_{1}^{2}\, (1 - F_{1}^{2}) -
    (1 - 2\,F_{1}^{2})\, F_{2}^{2}\, Q^{2} -
    F_{2}^{4}\, Q^{4}
  \Bigr]\; - \\

\vspace{0.1mm}
  & - &
  i\, \displaystyle\frac{2F_{1}F_{2}}{mc}\,
  \Bigl\{
    F_{1}^{2}\, q^{m} +
    i\, \bigl( F_{1}^{2} - F_{2}^{2}\, Q^{2} \bigr) \varepsilon_{mjl}\, q^{j}\, \sigma_{l}
  \Bigr\}\, V(\mathbf{r})\, \mathbf{p}_{m}\; + \\
\vspace{0.1mm}
  & + &
  \displaystyle\frac{1}{m}\, \bigl(F_{2}^{4} Q^{2} + 2a_{2}\bigr)\, \bigl( \mathbf{qp} \bigr)^{2}\; + 

  i\, cF_{2}\,
  \Bigl\{
    \Bigl[ F_{1}^{2} (1 - F_{1}) - F_{2}^{2} F_{1} \, Q^{2} \Bigr]\, q^{m} +
    i\, \Bigl[ 2\, F_{1}^{2} - 2 F_{2}^{2}\, Q^{2} \Bigr]\,
    \varepsilon_{mjl}\, q^{j}\, \sigma_{l}
  \Bigr\}\, \mathbf{p}_{m}.
\end{array}
\label{eq.app.5.1.1}
\end{equation}
$h_{\gamma 0}$ is not changed. From Eq.~(\ref{eq.4.3.2}) for $h_{\gamma 1}$ we obtain:
\begin{equation}
\begin{array}{lll}
 \vspace{0.5mm}
  h_{\gamma 1} & = &
  -\, i\, ze\, F_{2}\,
    \Bigl\{ b_{1}\, \mathbf{qA} + b_{2}\, \varepsilon_{mjl}\, q^{j}\, \sigma_{l} \mathbf{A}_{m} \Bigr\}\; + \\
\vspace{1.1mm}
  & + &
  \displaystyle\frac{1}{m}\,
  \Bigl\{
    \bigl(F_{2}^{4} Q^{2} + 2 a_{2} \bigr)\,
      \Bigl[ -2 \mathbf{(qp)} \displaystyle\frac{ze}{c}\, \mathbf{(qA)} + \displaystyle\frac{z^{2}e^{2}}{c^{2}}\, \mathbf{(qA)}^{2} \Bigr] +
  i\, \varepsilon_{lm^{\prime}k}\, \sigma_{k}\: q^{l} q^{m}\,
    \displaystyle\frac{i\hbar ze}{c} a_{2}\, \Bigl( \displaystyle\frac{dA_{m'}}{dx_{m}} - \displaystyle\frac{dA_{m}}{dx_{m'}} \Bigr)
  \Bigr\}\; = \\

\vspace{0.5mm}
  & = &
  -\, i\, ze\, F_{2}\,
    \Bigl\{ b_{1}\, \mathbf{qA} + b_{2}\, \varepsilon_{mjl}\, q^{j}\, \sigma_{l} \mathbf{A}_{m} \Bigr\}\; + \\
\vspace{1.1mm}
  & + &
  \displaystyle\frac{ze}{mc}\,
  \Bigl\{
    F_{2}^{2} \bigl( 2F_{1}^{2} - F_{2}^{2}\, Q^{2} \bigr)
      \Bigl[ -2 \mathbf{(qp)}\, \mathbf{(qA)} + \displaystyle\frac{ze}{c}\, \mathbf{(qA)}^{2} \Bigr] -

  \hbar a_{2}\, \varepsilon_{lm^{\prime}k}\, \sigma_{k}\: q^{l} q^{m}\,
    \Bigl( \displaystyle\frac{dA_{m'}}{dx_{m}} - \displaystyle\frac{dA_{m}}{dx_{m'}} \Bigr)
  \Bigr\}.
\end{array}
\label{eq.app.5.1.2}
\end{equation}


For the elastic scattering we use kinematic relation~(\ref{eq.5.2.1}),
and unperturbed hamiltonian $h_{0}$ from Eq.~(\ref{eq.app.5.1.1}) is simplified as
\begin{equation}
\begin{array}{lll}
\vspace{0.1mm}
  h_{0} & = &
  \displaystyle\frac{a_{1}\, \mathbf{p}^{2}}{m} +
  \Bigl( F_{1}^{3}\, (1 + F_{1}) + F_{1} F_{2}^{2}\, (3 - 2 F_{1})\, Q^{2} + F_{2}^{4}\, Q^{4} + i\, \displaystyle\frac{F_{1}^{3}F_{2}\, Q^{2}}{mc} \Bigr) \bigl[ ze\, A_{0} + V(\mathbf{r}) \bigr]\; + \\
\vspace{1.1mm}
  & + &
  mc^{2} \Bigl[  F_{1}^{2}\, (1 - F_{1}^{2}) - (1 - 2\,F_{1}^{2})\, F_{2}^{2}\, Q^{2} - F_{2}^{4}\, Q^{4} \Bigr] +
  \displaystyle\frac{Q^{4}}{4m}\, \bigl(F_{2}^{4} Q^{2} + 2a_{2}\bigr) -
  \displaystyle\frac{i\, cF_{2}\, Q^{2}}{2}\, \Bigl[ F_{1}^{2} (1 - F_{1}) - F_{2}^{2} F_{1} \, Q^{2} \Bigr]\; + \\
  & + &
  \Bigl\{ \displaystyle\frac{2F_{1}F_{2}}{mc}\, \bigl[ ze\, A_{0} + V(\mathbf{r}) \bigr]\, - 2cF_{2} \Bigr\}
    \bigl( F_{1}^{2} - F_{2}^{2}\, Q^{2} \bigr) \varepsilon_{mjl}\, q^{j}\, \sigma_{l}\, \mathbf{p}_{m}.
\end{array}
\label{eq.app.5.2.1}
\end{equation}
Operator of emission $h_{\gamma 1}$ from Eq.~(\ref{eq.app.5.1.2}) is transformed as
\begin{equation}
\begin{array}{lll}
 \vspace{0.5mm}
  h_{\gamma 1} & = &
  -\, i\, ze\, F_{2}\, \Bigl\{ b_{1}\, \mathbf{qA} + b_{2}\, \varepsilon_{mjl}\, q^{j}\, \sigma_{l} \mathbf{A}_{m} \Bigr\}\; + \\
  & + &
  \displaystyle\frac{ze}{mc}\,
  \Bigl\{
    F_{2}^{2} \bigl( 2F_{1}^{2} - F_{2}^{2}\, Q^{2} \bigr) \Bigl[ Q^{2} \mathbf{(qA)} + \displaystyle\frac{ze}{c}\, \mathbf{(qA)}^{2} \Bigr] -
    \hbar a_{2}\, \varepsilon_{lm^{\prime}k}\, \sigma_{k}\: q^{l} q^{m}\, \Bigl( \displaystyle\frac{dA_{m'}}{dx_{m}} - \displaystyle\frac{dA_{m}}{dx_{m'}} \Bigr)
  \Bigr\}.
\end{array}
\label{eq.app.5.2.2}
\end{equation}
We assume that last term, having Plank constant, is smaller essentially in comparison with other terms.
In such a case, we neglect by such a term and obtain:
\begin{equation}
\begin{array}{lll}
  h_{\gamma 1} & = &
  -\, i\, ze\, F_{2}\, \Bigl\{ b_{1}\, \mathbf{qA} + b_{2}\, \varepsilon_{mjl}\, q^{j}\, \sigma_{l} \mathbf{A}_{m} \Bigr\} +
  \displaystyle\frac{ze}{mc}\,
    F_{2}^{2} \bigl( 2F_{1}^{2} - F_{2}^{2}\, Q^{2} \bigr) \Bigl[ Q^{2} \mathbf{(qA)} + \displaystyle\frac{ze}{c}\, \mathbf{(qA)}^{2} \Bigr].
\end{array}
\label{eq.app.5.2.3}
\end{equation}


For the QED representation (\ref{eq.5.3.1}) for the vector potential of the electromagnetic field,
we introduce new angle $\varphi_{ph}$ between vectors $\mathbf{q}$ and $\mathbf{A}$ for determination of the scalar multiplication of them:
\begin{equation}
\begin{array}{lcl}
  \mathbf{qA} & = & qA \cdot \sin \varphi_{ph}.
\end{array}
\label{eq.app.5.3.1}
\end{equation}
We find properties:
\begin{equation}
\begin{array}{lcl}
  \mathbf{qA} = 2Q\, \sqrt{\displaystyle\frac{\pi\hbar c^{2}}{w_{\rm ph}}}\; e^{-i\, \mathbf{k_{\rm ph}r}}\, \sin \varphi_{ph}, &
  (\mathbf{qA})^{2} = 4Q^{2}\, \displaystyle\frac{\pi\hbar c^{2}}{w_{\rm ph}}\; e^{-2i\, \mathbf{k_{\rm ph}r}}\, \sin^{2} \varphi_{ph},
\end{array}
\label{eq.app.5.3.2}
\end{equation}
and we calculate operator of bremsstrahlung emission, related with the virtual photons:
\begin{equation}
\begin{array}{lll}
  h_{\gamma 1} & = &
  ze\, F_{2}\, \sqrt{\displaystyle\frac{\pi\hbar c^{2}}{w_{\rm ph}}}\; e^{-i\, \mathbf{k_{\rm ph}r}}\;
  \Bigl\{
    2 Q\, \sin \varphi_{ph}\,
      \Bigl[ -\, i\, b_{1} + \displaystyle\frac{F_{2}\, Q^{2}}{mc}\, \bigl( 2F_{1}^{2} - F_{2}^{2}\, Q^{2} \bigr) \Bigr]\; - \\
  & - &
    i\, \sqrt{2}\, b_{2}\, \varepsilon_{mjl}\, q^{j}\, \sigma_{l} \sum\limits_{\alpha=1,2} \mathbf{e}_{m}^{(\alpha),\,*} +
    \displaystyle\frac{4ze}{mc}\, \sqrt{\displaystyle\frac{\pi\hbar}{w_{\rm ph}}}\;
      F_{2}Q^{2}\, \bigl( 2F_{1}^{2} - F_{2}^{2}\, Q^{2} \bigr)\;
      e^{-i\, \mathbf{k_{\rm ph}r}}\, \sin^{2} \varphi_{ph}
  \Bigr\}.
\end{array}
\label{eq.app.5.3.3}
\end{equation}

\section{Matrix element of emission of bremsstrahlung photons
\label{sec.app.6}}

\subsection{Calculations of $p_{\rm eq}$, $p_{\rm mag,2}$ and $p_{\rm mag,2}$
\label{sec.app.6.1}}

Taking into account Eqs.~(\ref{eq.5.3.4}) and (\ref{eq.5.3.5}) for the operator of emission, we obtain:
\[
\begin{array}{lcl}
  \vspace{1mm}
  F_{fi,0} & = &
  \bigl< k_{f} \bigl|\,  h_{\gamma 0}\, \bigr| \,k_{i} \bigr> = \\
  \vspace{2mm}
  & = &
  \biggl< k_{f} \biggl|\,
  Z_{\rm eff}\, \displaystyle\frac{e}{mc}\,
  \sqrt{\displaystyle\frac{2\pi\hbar c^{2}}{w}}\;
    \displaystyle\sum\limits_{\alpha=1,2}
  e^{-i\,\mathbf{kr}}\;
  \Big(
    i\, \mathbf{e}^{(\alpha)}\, \nabla -
    \displaystyle\frac{1}{2}\: \sigmabf\cdot
      \Bigl[ \nabla \times \mathbf{e}^{(\alpha)} \Bigr] +
    i\,\displaystyle\frac{1}{2}\: \sigmabf\cdot
      \Bigl[ \mathbf{k} \times \mathbf{e}^{(\alpha)} \Bigr]
  \Bigr)\,
  \biggr| \,k_{i} \biggr> = \\

  \vspace{2mm}
  & = &
  Z_{\rm eff}\, \displaystyle\frac{e}{mc}\,
  \sqrt{\displaystyle\frac{2\pi\hbar c^{2}}{w}}\;
  \displaystyle\sum\limits_{\alpha=1,2}
    \Bigl< k_{f} \Bigl|\,
    e^{-i\,\mathbf{kr}}\;
    \Big(
      i\, \mathbf{e}^{(\alpha)}\, \nabla -
      \displaystyle\frac{1}{2}\: \sigmabf\cdot
      \Bigl[ \mathbf{e}^{(\alpha)} \times \nabla \Bigr] +
      i\,\displaystyle\frac{1}{2}\: \sigmabf\cdot
      \Bigl[ \mathbf{k} \times \mathbf{e}^{(\alpha)} \Bigr]
    \Bigr)\, \Bigr| \,k_{i} \Bigr> = \\

  \vspace{2mm}
  & = &
  Z_{\rm eff}\, \displaystyle\frac{e}{mc}\,
  \sqrt{\displaystyle\frac{2\pi\hbar c^{2}}{w}}\;
  \biggl\{\,
  i \displaystyle\sum\limits_{\alpha=1,2}
    \mathbf{e}^{(\alpha)}\,
    \Bigl< k_{f} \Bigl|\, e^{-i\,\mathbf{kr}}\: \nabla\, \Bigr| \,k_{i} \Bigr>\, -

  \displaystyle\frac{1}{2}\:
  \displaystyle\sum\limits_{\alpha=1,2}
  \Bigl< k_{f} \Bigl|\,
    e^{-i\,\mathbf{kr}}\;
    \sigmabf\cdot \Bigl[ \mathbf{e}^{(\alpha)} \times \nabla \Bigr]\,
  \Bigr| \,k_{i} \Bigr>\quad + \\

  & + &
  i\,\displaystyle\frac{1}{2}\:
  \displaystyle\sum\limits_{\alpha=1,2}
  \Bigl[ \mathbf{k} \times \mathbf{e}^{(\alpha)} \Bigr]\:
  \Bigl< k_{f} \Bigl|\, e^{-i\,\mathbf{kr}}\; \sigmabf\, \Bigr| \,k_{i} \Bigr>
  \biggr\},
\end{array}
\]
\[
\begin{array}{lcl}
  \vspace{1mm}
  F_{fi,1} & = &
  \bigl< k_{f} \bigl|\,  h_{\gamma 1}\, \bigr| \,k_{i} \bigr> = \\

  \vspace{0.1mm}
  & = &
  \biggl< k_{f} \biggl|\,
  ze\, F_{2}\, \sqrt{\displaystyle\frac{\pi\hbar c^{2}}{w_{\rm ph}}}\; e^{-i\, \mathbf{k_{\rm ph}r}}\;
  \Bigl\{
    2 Q\, \sin \varphi_{ph}\,
      \Bigl[ -\, i\, b_{1} + \displaystyle\frac{F_{2}\, Q^{2}}{mc}\, \bigl( 2F_{1}^{2} - F_{2}^{2}\, Q^{2} \bigr) \Bigr]\; - \\
  \vspace{2.5mm}
  & - &
    i\, \sqrt{2}\, b_{2}\, \varepsilon_{mjl}\, q^{j}\, \sigma_{l} \sum\limits_{\alpha=1,2} \mathbf{e}_{m}^{(\alpha),\,*} +
    \displaystyle\frac{4ze}{mc}\, \sqrt{\displaystyle\frac{\pi\hbar}{w_{\rm ph}}}\;
      F_{2}Q^{2}\, \bigl( 2F_{1}^{2} - F_{2}^{2}\, Q^{2} \bigr)\;
      e^{-i\, \mathbf{k_{\rm ph}r}}\, \sin^{2} \varphi_{ph}
  \Bigr\}
  \biggr| \,k_{i} \biggr>\; = \\

  \vspace{0.1mm}
  & = &
  ze\, F_{2}\, \sqrt{\displaystyle\frac{\pi\hbar c^{2}}{w_{\rm ph}}}\;
  \biggl\{

  2 Q\, \sin \varphi_{ph}\,
  \Bigl< k_{f} \Bigl|\,
    e^{-i\, \mathbf{k_{\rm ph}r}}\; \Bigl[ -\, i\, b_{1} + \displaystyle\frac{F_{2}\, Q^{2}}{mc}\, \bigl( 2F_{1}^{2} - F_{2}^{2}\, Q^{2} \bigr) \Bigr]
  \Bigr| \,k_{i} \Bigr>\; - \\
  \vspace{2.5mm}
  & - &

  i\, \sqrt{2}\, \varepsilon_{mjl}\, q^{j}\, \sigma_{l} \sum\limits_{\alpha=1,2} \mathbf{e}_{m}^{(\alpha),\,*}
  \Bigl< k_{f} \Bigl|\,
    e^{-i\, \mathbf{k_{\rm ph}r}}\; b_{2}\,
  \Bigr| \,k_{i} \Bigr> +

  \displaystyle\frac{4ze}{mc}\, \sqrt{\displaystyle\frac{\pi\hbar}{w_{\rm ph}}}\;
  F_{2}Q^{2}\, \bigl( 2F_{1}^{2} - F_{2}^{2}\, Q^{2} \bigr)\;
  \sin^{2} \varphi_{ph}
  \Bigl< k_{f} \Bigl|\,
    e^{-2i\, \mathbf{k_{\rm ph}r}}\,
  \Bigr\}
  \Bigr| \,k_{i} \Bigr>
  \biggr\}
\end{array}
\]
or %
\begin{equation}
\begin{array}{lll}
\vspace{0.5mm}
  F_{fi,0} & = &
  \bigl< k_{f} \bigl|\,  h_{\gamma 0}\, \bigr| \,k_{i} \bigr> \quad = \quad
  Z_{\rm eff}\, \displaystyle\frac{e}{mc}\,
    \sqrt{\displaystyle\frac{2\pi\hbar c^{2}}{w}}\;
    \Bigl\{ p_{\rm el} + p_{\rm mag, 1} + p_{\rm mag, 2} \Bigr\}, \\

  F_{fi,1} & = &
  \bigl< k_{f} \bigl|\,  h_{\gamma 0}\, \bigr| \,k_{i} \bigr> \quad = \quad
    Z_{\rm eff}\; e\, F_{2}\, \sqrt{\displaystyle\frac{\pi\hbar c^{2}}{w_{\rm ph}}}\;
    \Bigl\{ p_{\rm q,1} + p_{\rm q, 2} + p_{\rm q, 3} \Bigr\},
\end{array}
\label{eq.app.6.1.1}
\end{equation}
where
\begin{equation}
\begin{array}{lcl}
  \vspace{2mm}
  p_{\rm el} & = &
  i \displaystyle\sum\limits_{\alpha=1,2}
    \mathbf{e}^{(\alpha)}\,
    \Bigl< k_{f} \Bigl|\, e^{-i\,\mathbf{kr}}\: \nabla\, \Bigr| \,k_{i} \Bigr>, \\

  \vspace{2mm}
  p_{\rm mag, 1} & = &
  \displaystyle\frac{1}{2}\:
  \displaystyle\sum\limits_{\alpha=1,2}
  \Bigl< k_{f} \Bigl|\,
    e^{-i\,\mathbf{kr}}\;
    \sigmabf\cdot \Bigl[ \mathbf{e}^{(\alpha)} \times \nabla \Bigr]\,
  \Bigr| \,k_{i} \Bigr>, \\

  p_{\rm mag, 2} & = & -
  i\,\displaystyle\frac{1}{2}\:
  \displaystyle\sum\limits_{\alpha=1,2}
  \Bigl[ \mathbf{k} \times \mathbf{e}^{(\alpha)} \Bigr]\:
  \Bigl< k_{f} \Bigl|\, e^{-i\,\mathbf{kr}}\; \sigmabf\, \Bigr| \,k_{i} \Bigr>,
\end{array}
\label{eq.app.6.1.2}
\end{equation}
\begin{equation}
\begin{array}{lcl}
\vspace{1.7mm}
  p_{\rm q,1} & = &
    2 Q\, \sin \varphi_{ph}\,
    \Bigl< k_{f} \Bigl|\,
      e^{-i\, \mathbf{k_{\rm ph}r}}\; \Bigl[ -\, i\, b_{1} + \displaystyle\frac{F_{2}\, Q^{2}}{mc}\, \bigl( 2F_{1}^{2} - F_{2}^{2}\, Q^{2} \bigr) \Bigr]
    \Bigr| \,k_{i} \Bigr>, \\

\vspace{1.0mm}
  p_{\rm q,2} & = &
  -\, i\, \sqrt{2}\, \varepsilon_{mjl}\, q^{j}\, \sigma_{l} \sum\limits_{\alpha=1,2} \mathbf{e}_{m}^{(\alpha),\,*}
  \Bigl< k_{f} \Bigl|\,
    e^{-i\, \mathbf{k_{\rm ph}r}}\; b_{2}\,
  \Bigr| \,k_{i} \Bigr>, \\

  p_{\rm q,3} & = &
  \displaystyle\frac{4ze}{mc}\, \sqrt{\displaystyle\frac{\pi\hbar}{w_{\rm ph}}}\;
  F_{2}Q^{2}\, \bigl( 2F_{1}^{2} - F_{2}^{2}\, Q^{2} \bigr)\;
  \sin^{2} \varphi_{ph}
  \Bigl< k_{f} \Bigl|\,
    e^{-2i\, \mathbf{k_{\rm ph}r}}\,
  \Bigr\}
  \Bigr| \,k_{i} \Bigr>.
\end{array}
\label{eq.app.6.1.3}
\end{equation}

\subsection{Calculations of $p_{\rm q,1}$, $p_{\rm q,2}$ and $p_{\rm q,3}$, and averaging over polarizations of virtual photons
\label{sec.app.6.2}}

At $A_{0} = 0$
\begin{equation}
\begin{array}{lll}
\vspace{0.9mm}
  b_{1} & = &
  F_{1}^{2} (1 - F_{1}) -
  F_{1} F_{2}^{2}\, Q^{2} -
  \displaystyle\frac{2\,F_{1}^{3}}{mc^{2}}\, V(\mathbf{r}), \\

  b_{2} & = &
    i\, \Bigl[
      2\, F_{1}^{2} -
      2 F_{2}^{2}\, Q^{2} -
      \displaystyle\frac{2\,F_{1}}{mc^{2}} \bigl( F_{1}^{2} - F_{2}^{2} Q^{2} \bigr)\, V(\mathbf{r})
    \Bigr].
\end{array}
\label{eq.app.6.2.1}
\end{equation}
Substitute such solutions to Eqs.~(\ref{eq.app.6.1.3}) and obtain:
\begin{equation}
\begin{array}{lcl}
\vspace{1.0mm}
  p_{\rm q,1} & = &
  2 Q\, \sin \varphi_{ph}\,
  \Bigl\{ - i\, \Bigl[ F_{1}^{2} (1 - F_{1}) - F_{1} F_{2}^{2}\, Q^{2} \Bigr] +
         \displaystyle\frac{F_{2}\, Q^{2}}{mc}\, \bigl( 2F_{1}^{2} - F_{2}^{2}\, Q^{2} \bigr) \Bigr\}\,
    \Bigl< k_{f} \Bigl|\, e^{-i\, \mathbf{k_{\rm ph}r}}\; \Bigr| \,k_{i} \Bigr>\; + \\
\vspace{1.5mm}
  & + &
    2 Q\, \sin \varphi_{ph}\,
    i\, \displaystyle\frac{2\,F_{1}^{3}}{mc^{2}}\,
    \Bigl< k_{f} \Bigl|\, e^{-i\, \mathbf{k_{\rm ph}r}}\; V(\mathbf{r}) \Bigr| \,k_{i} \Bigr>, \\

\vspace{0.5mm}
  p_{\rm q,2} & = &
  2\, \bigl( F_{1}^{2} - F_{2}^{2}\, Q^{2} \bigr)\,
  \sqrt{2}\, \varepsilon_{mjl}\, q^{j}\, \sigma_{l} \sum\limits_{\alpha=1,2} \mathbf{e}_{m}^{(\alpha),\,*}
  \Bigl< k_{f} \Bigl|\, e^{-i\, \mathbf{k_{\rm ph}r}}\; \Bigr| \,k_{i} \Bigr>\; - \\
  & - &
  2\, \bigl( F_{1}^{2} - F_{2}^{2} Q^{2} \bigr)\,
  \displaystyle\frac{\sqrt{2}\, F_{1}}{mc^{2}}\,
  \varepsilon_{mjl}\, q^{j}\, \sigma_{l} \sum\limits_{\alpha=1,2} \mathbf{e}_{m}^{(\alpha),\,*}
  \Bigl< k_{f} \Bigl|\, e^{-i\, \mathbf{k_{\rm ph}r}}\; V(\mathbf{r}) \Bigr| \,k_{i} \Bigr>.
\end{array}
\label{eq.app.6.2.2}
\end{equation}


We assume that there is no way to fix direction of polarization of the virtual photons (concerning to vectors of polarization of the bremsstrahlung photons) experimentally.
So, we have to integrate the matrix elements $p_{\rm q,i}$ over all such a possible directions (i.e. we integrate over angle $\varphi_{\rm ph}$, $i = 1,2,3$):
\begin{equation}
\begin{array}{lcl}
  \tilde{p}_{\rm q,i} = N \cdot \displaystyle\int\limits_{0}^{\pi} p_{\rm q,i}\; d \varphi_{\rm ph}, &
  N = \displaystyle\frac{1}{\pi}.
\end{array}
\label{eq.app.6.3.1}
\end{equation}
Taking into account that
\begin{equation}
\begin{array}{lcl}
  \displaystyle\int\limits_{0}^{\pi} \sin\varphi_{\rm ph}\; d \varphi_{\rm ph} =
  2, &

  \displaystyle\int\limits_{0}^{\pi} d \varphi_{\rm ph} =
  \pi, &

  \displaystyle\int\limits_{0}^{\pi} \sin^{2}\varphi_{\rm ph}\; d \varphi_{\rm ph} =
  \displaystyle\frac{\pi}{2},
\end{array}
\label{eq.app.6.3.2}
\end{equation}
from (\ref{eq.app.6.2.2}) and (\ref{eq.app.6.1.3}) we obtain:
\begin{equation}
\begin{array}{lcl}
\vspace{1.3mm}
  \tilde{p}_{\rm q,1} & = &
    i\, A_{1} (Q, F_{1}, F_{2})\, \Bigl< k_{f} \Bigl|\, e^{-i\, \mathbf{k_{\rm ph}r}}\; \Bigr| \,k_{i} \Bigr> +
    i\, B_{1} (Q, F_{1}, F_{2})\, \Bigl< k_{f} \Bigl|\, e^{-i\, \mathbf{k_{\rm ph}r}}\; V(\mathbf{r}) \Bigr| \,k_{i} \Bigr>, \\

\vspace{1.3mm}
  \tilde{p}_{\rm q,2} & = &
    i\, A_{2} (Q, F_{1}, F_{2})\, \Bigl< k_{f} \Bigl|\, e^{-i\, \mathbf{k_{\rm ph}r}}\; \Bigr| \,k_{i} \Bigr> +
    i\, B_{2} (Q, F_{1}, F_{2})\, \Bigl< k_{f} \Bigl|\, e^{-i\, \mathbf{k_{\rm ph}r}}\; V(\mathbf{r}) \Bigr| \,k_{i} \Bigr>, \\

  \tilde{p}_{\rm q,3} & = &
    i\, A_{3} (Q, F_{1}, F_{2})\, \Bigl< k_{f} \Bigl|\, e^{-i\, \mathbf{2k_{\rm ph}r}}\; \Bigr| \,k_{i} \Bigr>,
\end{array}
\label{eq.app.6.3.3}
\end{equation}
where
\begin{equation}
\begin{array}{lcl}
\vspace{1.5mm}
  A_{1} (Q, F_{1}, F_{2}) & = &
    -\, \displaystyle\frac{4Q}{\pi}
    \Bigl\{ \Bigl[ F_{1}^{2} (1 - F_{1}) - F_{1} F_{2}^{2}\, Q^{2} \Bigr] +
       i\, \displaystyle\frac{F_{2}\, Q^{2}}{\pi mc}\, \bigl( 2F_{1}^{2} - F_{2}^{2}\, Q^{2} \bigr) \Bigr\}, \\
\vspace{1.5mm}
  B_{1} (Q, F_{1}, F_{2}) & = & 8 Q\, \displaystyle\frac{F_{1}^{3}}{\pi mc^{2}}, \\

\vspace{1.5mm}
  A_{2} (Q, F_{1}, F_{2}) & = &
    - i\, 2\, \bigl( F_{1}^{2} - F_{2}^{2}\, Q^{2} \bigr)\,
    \sqrt{2}\, \varepsilon_{mjl}\, q^{j}\, \sigma_{l} \sum\limits_{\alpha=1,2} \mathbf{e}_{m}^{(\alpha),\,*}, \\

\vspace{0.9mm}
  B_{2} (Q, F_{1}, F_{2}) & = &
    i\, 2\, \bigl( F_{1}^{2} - F_{2}^{2} Q^{2} \bigr)\,
    \displaystyle\frac{\sqrt{2}\, F_{1}}{mc^{2}}\,
    \varepsilon_{mjl}\, q^{j}\, \sigma_{l} \sum\limits_{\alpha=1,2} \mathbf{e}_{m}^{(\alpha),\,*}, \\

  A_{3} (Q, F_{1}, F_{2}) & = &
  - i\, \displaystyle\frac{2ze}{mc}\, \sqrt{\displaystyle\frac{\pi\hbar}{w_{\rm ph}}}\;
  F_{2}Q^{2}\, \bigl( 2F_{1}^{2} - F_{2}^{2}\, Q^{2} \bigr).
\end{array}
\label{eq.app.6.3.4}
\end{equation}

\section{Calculations of matrix elements of emission in multipolar expansion
\label{sec.app.9}}


We shall calculate the following matrix elements:
\begin{equation}
\begin{array}{ll}
  \Bigl< k_{f} \Bigl| \,  e^{-i\mathbf{kr}} \, \Bigr| \,k_{i} \Bigr>_\mathbf{r} =
  \displaystyle\int
    \varphi^{*}_{f}(\mathbf{r}) \:
    e^{-i\mathbf{kr}}\:
    \varphi_{i}(\mathbf{r}) \;
    \mathbf{dr}, &

  \hspace{5mm}
  \biggl< k_{f} \biggl| \,  e^{-i\mathbf{kr}} \displaystyle\frac{\partial}{\partial \mathbf{r}} \,
  \biggr| \,k_{i} \biggr>_\mathbf{r} =
  \displaystyle\int
    \varphi^{*}_{f}(\mathbf{r}) \:
    e^{-i\mathbf{kr}} \displaystyle\frac{\partial}{\partial \mathbf{r}}\:
    \varphi_{i}(\mathbf{r}) \;
    \mathbf{dr}.
\end{array}
\label{eq.app.9.1.1}
\end{equation}

\subsection{Expansion of the vector potential $\mathbf{A}$ by multipoles
\label{sec.9.2}}

Let us expand the vectorial potential $\mathbf{A}$ of electromagnetic field by multipolar terms.
According to Ref.~\cite{Eisenberg.1973} [see~(2.106), p.~58], in the spherical symmetric approximation we have:
\begin{equation}
  \mathbf{\xi}_{\mu}\, e^{i \mathbf{kr}} =
    \mu\, \sqrt{2\pi}\, \sum_{l=1}\,
    (2l+1)^{1/2}\, i^{l}\,  \cdot
    \Bigl[ \mathbf{A}_{l\mu} (\mathbf{r}, M) +
    i\mu\, \mathbf{A}_{l\mu} (\mathbf{r}, E) \Bigr],
\label{eq.app.9.2.1}
\end{equation}
where (see~\cite{Eisenberg.1973}, (2.73) in p.~49, (2.80) in p.~51)
\begin{equation}
\begin{array}{lcl}
  \vspace{2mm}
  \mathbf{A}_{l\mu}(\mathbf{r}, M) & = &
        j_{l}(kr) \: \mathbf{T}_{ll,\mu} (\mathbf{n}_{\rm r}), \\
  \vspace{2mm}
  \mathbf{A}_{l\mu}(\mathbf{r}, E) & = &
        \sqrt{\displaystyle\frac{l+1}{2l+1}}\,
        j_{l-1}(kr) \: \mathbf{T}_{ll-1,\mu}(\mathbf{n}_{\rm r})\; -
        \sqrt{\displaystyle\frac{l}{2l+1}}\,
        j_{l+1}(kr) \: \mathbf{T}_{ll+1,\mu}(\mathbf{n}_{\rm r}).
\end{array}
\label{eq.app.9.2.2}
\end{equation}
Here, $\mathbf{A}_{l\mu}(\textbf{r}, M)$ and $\mathbf{A}_{l\mu}(\textbf{r}, E)$ are \emph{magnetic} and \emph{electric multipoles},
$j_{l}(kr)$ is \emph{spherical Bessel function of order $l$},
$\mathbf{T}_{ll',\mu}(\mathbf{n}_{\rm r})$ are \emph{vector spherical harmonics}.
We orient the frame so that axis $z$ be directed along the vector $\mathbf{k}$ (see~\cite{Eisenberg.1973}, (2.105) in p.~57). According to \cite{Eisenberg.1973} (see p.~45), the functions $\mathbf{T}_{ll',\mu}(\mathbf{n}_{\rm r})$ have the following form
(${\mathbf \xi}_{0} = 0$):
\begin{equation}
  \mathbf{T}_{jl,m} (\mathbf{n}_{\rm r}) =
  \sum\limits_{\mu = \pm 1} (l, 1, j \,\big| \,m-\mu, \mu, m) \; Y_{l,m-\mu}(\mathbf{n}_{\rm r})\;
  \mathbf{\xi}_{\mu},
\label{eq.app.9.2.3}
\end{equation}
where $(l, 1, j \,\bigl| \, m-\mu, \mu, m)$ are \emph{Clebsh-Gordon coefficients},
$Y_{lm}(\theta, \varphi)$ are \emph{spherical functions} defined, according to~\cite{Landau.v3.1989} (see p.~119, (28,7)--(28,8)).
From eq.app.~(\ref{eq.app.9.2.1}) one can obtain such a formula (at $\mathbf{e}^{(3)}=0$):
\begin{equation}
  e^{-i \mathbf{kr}} =
  \displaystyle\frac{1}{2}\,
  \displaystyle\sum\limits_{\mu = \pm 1}
    \mathbf{\xi}_{\mu}\, \mu\, \sqrt{2\pi}\, \sum_{l=1}\,
    (2l+1)^{1/2}\, (-i)^{l}\,  \cdot
    \Bigl[ \mathbf{A}_{l\mu}^{*} (\mathbf{r}, M) -
    i\mu\, \mathbf{A}_{l\mu}^{*} (\mathbf{r}, E) \Bigr].
\label{eq.app.9.2.4}
\end{equation}

\subsection{Spherically symmetric nucleus
\label{sec.9.3}}

Using (\ref{eq.app.9.2.4}), for (\ref{eq.app.9.2.1}) we find:
\begin{equation}
\begin{array}{ll}
  \vspace{1mm}
  \Bigl< k_{f} \Bigl| \,  e^{-i\mathbf{kr}} \, \Bigr| \,k_{i} \Bigr>_\mathbf{r} =
  \sqrt{\displaystyle\frac{\pi}{2}}\:
  \displaystyle\sum\limits_{l_{\rm ph}=1}\,
    (-i)^{l_{\rm ph}}\, \sqrt{2l_{\rm ph}+1}\;
  \displaystyle\sum\limits_{\mu = \pm 1}
    \Bigl[ \mu\,\tilde{p}_{l_{\rm ph}\mu}^{M} - i\, \tilde{p}_{l_{\rm ph}\mu}^{E} \Bigr], \\

  \biggl< k_{f} \biggl| \,  e^{-i\mathbf{kr}} \displaystyle\frac{\partial}{\partial \mathbf{r}}\,
  \biggr| \,k_{i} \biggr>_\mathbf{r} =
  \sqrt{\displaystyle\frac{\pi}{2}}\:
  \displaystyle\sum\limits_{l_{\rm ph}=1}\,
    (-i)^{l_{\rm ph}}\, \sqrt{2l_{\rm ph}+1}\;
  \displaystyle\sum\limits_{\mu = \pm 1}
    \xibf_{\mu}\, \mu\, \times
    \Bigl[ p_{l_{\rm ph}\mu}^{M} - i\mu\: p_{l_{\rm ph}\mu}^{E} \Bigr],
\end{array}
\label{eq.app.9.3.1}
\end{equation}
where
\begin{equation}
\begin{array}{lcllcl}
  p_{l_{\rm ph}\mu}^{M} & = &
    \displaystyle\int
        \varphi^{*}_{f}(\mathbf{r}) \,
        \biggl( \displaystyle\frac{\partial}{\partial \mathbf{r}}\, \varphi_{i}(\mathbf{r}) \biggr) \,
        \mathbf{A}_{l_{\rm ph}\mu}^{*} (\mathbf{r}, M) \;
        \mathbf{dr}, &

  \hspace{7mm}
  p_{l_{\rm ph}\mu}^{E} & = &
    \displaystyle\int
        \varphi^{*}_{f}(\mathbf{r}) \,
        \biggl( \displaystyle\frac{\partial}{\partial \mathbf{r}}\, \varphi_{i}(\mathbf{r}) \biggr)\,
        \mathbf{A}_{l_{\rm ph}\mu}^{*} (\mathbf{r}, E) \;
        \mathbf{dr},
\end{array}
\label{eq.app.9.3.2}
\end{equation}
and
\begin{equation}
\begin{array}{lcllcl}
  \tilde{p}_{l_{\rm ph}\mu}^{M} & = &
    \xibf_{\mu}\,
    \displaystyle\int
      \varphi^{*}_{f}(\mathbf{r})\,
      \varphi_{i}(\mathbf{r})\;
      \mathbf{A}_{l_{\rm ph}\mu}^{*} (\mathbf{r}, M) \;
      \mathbf{dr}, &
  \hspace{7mm}
  \tilde{p}_{l_{\rm ph}\mu}^{E} & = &
    \xibf_{\mu}\,
    \displaystyle\int
      \varphi^{*}_{f}(\mathbf{r})\,
      \varphi_{i}(\mathbf{r})\;
      \mathbf{A}_{l_{\rm ph}\mu}^{*} (\mathbf{r}, E)\;
      \mathbf{dr}.
\end{array}
\label{eq.app.9.3.3}
\end{equation}

Now we shall calculate components in Eqs.~(\ref{eq.8.1.12}). For the first and third items we obtain:
\begin{equation}
\begin{array}{lcl}
  \vspace{1mm}
  p_{\rm el} & = &
  i\, \sqrt{\displaystyle\frac{\pi}{2}}
  \displaystyle\sum\limits_{m_{i}, m_{f}}
  \displaystyle\sum\limits_{\mu_{i},\, \mu_{f} = \pm 1/2}
    C_{l_{f}m_{f} 1/2 \mu_{f}}^{j_{f}M_{f},\,*}\,
    C_{l_{i}m_{i} 1/2 \mu_{i}}^{j_{i}M_{i}} \cdot
    \displaystyle\sum\limits_{l_{\rm ph}=1}\,
      (-i)^{l_{\rm ph}}\, \sqrt{2l_{\rm ph}+1}\; \cdot
    \Bigl[
      p_{l_{\rm ph}}^{M} -
      i\,p_{l_{\rm ph}}^{E}
    \Bigr], \\

  \vspace{1mm}
  p_{\rm mag,\,2} & = &
  \displaystyle\frac{-i\,k}{2}\,
  \sqrt{\displaystyle\frac{\pi}{2}}\;
  \displaystyle\sum\limits_{m_{i}, m_{f}}
  \displaystyle\sum\limits_{\mu_{i},\, \mu_{f} = \pm 1/2}
    C_{l_{f}m_{f} 1/2 \mu_{f}}^{j_{f}M_{f},\,*}\,
    C_{l_{i}m_{i} 1/2 \mu_{i}}^{j_{i}M_{i}}\quad \times \\
  \vspace{1mm}
  & \times &
  \Bigl[ -1 + i\, \Bigl\{ \delta_{\mu_{i}, +1/2}\; -\; \delta_{\mu_{i}, -1/2} \Bigr\} \Bigr] \cdot
    \displaystyle\sum\limits_{l_{\rm ph}=1}\,
      (-i)^{l_{\rm ph}}\, \sqrt{2l_{\rm ph}+1} \cdot
      \Bigl[ \tilde{p}_{l_{\rm ph}}^{M} - i\, \tilde{p}_{l_{\rm ph}}^{E} \Bigr],
\end{array}
\label{eq.app.9.3.4}
\end{equation}
where
\begin{equation}
\begin{array}{cccc}
  p_{l_{\rm ph}}^{M} = \displaystyle\sum\limits_{\mu = \pm 1} h_{\mu}\, \mu\, p_{l_{\rm ph}\mu}^{M}, &
  \hspace{7mm}
  p_{l_{\rm ph}}^{E} = \displaystyle\sum\limits_{\mu = \pm 1} h_{\mu}\, p_{l_{\rm ph}\mu}^{E}, &
  \hspace{7mm}
  \tilde{p}_{l_{\rm ph}}^{M} =
    \displaystyle\sum\limits_{\mu = \pm 1} \mu\; \tilde{p}_{l_{\rm ph}\mu}^{M}, &
  \hspace{7mm}
  \tilde{p}_{l_{\rm ph}}^{M} =
    \displaystyle\sum\limits_{\mu = \pm 1} \tilde{p}_{l_{\rm ph}\mu}^{E}.
\end{array}
\label{eq.app.9.3.5}
\end{equation}
Now we shall analyze the second item in Eqs.~(\ref{eq.8.1.12}) and find:
\begin{equation}
\begin{array}{ll}
  \vspace{1mm}
  p_{\rm mag,\,1} =
    \displaystyle\frac{1}{2}
  \displaystyle\sum\limits_{m_{i}, m_{f}}
  \displaystyle\sum\limits_{\mu_{i},\, \mu_{f} = \pm 1/2}
    C_{l_{f}m_{f} 1/2 \mu_{f}}^{j_{f}M_{f},\,*}\,
    C_{l_{i}m_{i} 1/2 \mu_{i}}^{j_{i}M_{i}} \cdot
    \Bigl[
      \displaystyle\frac{1}{\sqrt{2}}\, \bigl(\xibf_{-1} - \xibf_{+1}\bigr) +
      \displaystyle\frac{i}{\sqrt{2}}\, \bigl(\xibf_{-1} + \xibf_{+1}\bigr)\,
      i\, \Bigl\{ \delta_{\mu_{i}, +1/2}\; - \\

  - \quad
    \delta_{\mu_{i}, -1/2} \Bigr\} +  \mathbf{e}_{\rm z} \Bigr]\quad \times
    \biggl[
      \displaystyle\sum\limits_{\mu=\pm 1} h_{\mu} \xibf_{\mu}^{*} \times
    \sqrt{\displaystyle\frac{\pi}{2}}\:
    \displaystyle\sum\limits_{l}\,
      (-i)^{l}\, \sqrt{2l+1}\;
    \displaystyle\sum\limits_{\mu^{\prime} = \pm 1}
      \xibf_{\mu^{\prime}}\, \mu^{\prime}\, \times
      \Bigl[ p_{l\mu^{\prime}}^{M} - i\mu^{\prime}\: p_{l\mu^{\prime}}^{E} \Bigr]\, \biggr].
\end{array}
\label{eq.app.9.3.6}
\end{equation}
Taking properties (\ref{eq.5.3.2}) into account, we calculate Eq.~(\ref{eq.app.9.3.6}):
\begin{equation}
\begin{array}{ll}
  p_{\rm mag,\,1} =
    - \displaystyle\frac{1}{2}
  \sqrt{\displaystyle\frac{\pi}{2}}\:
  \displaystyle\sum\limits_{m_{i}, m_{f}}
  \displaystyle\sum\limits_{\mu_{i},\, \mu_{f} = \pm 1/2}
    C_{l_{f}m_{f} 1/2 \mu_{f}}^{j_{f}M_{f},\,*}\,
    C_{l_{i}m_{i} 1/2 \mu_{i}}^{j_{i}M_{i}}\;
    \displaystyle\sum\limits_{l_{\rm ph}}\,
      (-i)^{l_{\rm ph}}\, \sqrt{2l_{\rm ph}+1}\; \cdot
    \displaystyle\sum\limits_{\mu=\pm 1}
      i\, h_{\mu}\, \mu \Bigl[\mu\, p_{l_{\rm ph}\mu}^{M} - i\, p_{l_{\rm ph}\mu}^{E} \Bigr].
\end{array}
\label{eq.app.9.3.7}
\end{equation}
So, we have found all components in Eqs.~(\ref{eq.8.1.12}):
\begin{equation}
\begin{array}{lcl}
  \vspace{3mm}
  p_{\rm el} & = &
  \sqrt{\displaystyle\frac{\pi}{2}}\:
  \displaystyle\sum\limits_{l_{\rm ph}=1}\,
    (-i)^{l_{\rm ph}}\, \sqrt{2l_{\rm ph}+1}\; \cdot
  \displaystyle\sum\limits_{\mu=\pm 1}
    h_{\mu} \cdot
  \displaystyle\sum\limits_{m_{i}, m_{f}}
  \displaystyle\sum\limits_{\mu_{i},\, \mu_{f} = \pm 1/2}
    C_{l_{f}m_{f} 1/2 \mu_{f}}^{j_{f}M_{f},\,*}\,
    C_{l_{i}m_{i} 1/2 \mu_{i}}^{j_{i}M_{i}}\,
    \Bigl[
      i\,\mu\, p_{l_{\rm ph}\mu}^{M m_{i} m_{f}} +
      p_{l_{\rm ph}\mu}^{E m_{i} m_{f}}
    \Bigr], \\

  \vspace{3mm}
  p_{\rm mag,1} & = &
  \displaystyle\frac{1}{2}\:
  \sqrt{\displaystyle\frac{\pi}{2}}\:
  \displaystyle\sum\limits_{l_{\rm ph}=1}\,
    (-i)^{l_{\rm ph}}\, \sqrt{2l_{\rm ph}+1}\; \cdot
  \displaystyle\sum\limits_{\mu=\pm 1}
    h_{\mu}\, \mu
  \displaystyle\sum\limits_{m_{i}, m_{f}}
  \displaystyle\sum\limits_{\mu_{i},\, \mu_{f} = \pm 1/2}
    C_{l_{f}m_{f} 1/2 \mu_{f}}^{j_{f}M_{f},\,*}\,
    C_{l_{i}m_{i} 1/2 \mu_{i}}^{j_{i}M_{i}}\,
    \Bigl[i\,\mu\, p_{l_{\rm ph}\mu}^{M m_{i} m_{f}} + p_{l_{\rm ph}\mu}^{E m_{i} m_{f}} \Bigr], \\

  \vspace{1mm}
  p_{\rm mag, 2} & = &
  \sqrt{\displaystyle\frac{\pi}{8}}\: k\;
  \displaystyle\sum\limits_{l_{\rm ph}=1}\,
    (-i)^{l_{\rm ph}}\, \sqrt{2l_{\rm ph}+1} \cdot
  \displaystyle\sum\limits_{\mu=\pm 1}
  \displaystyle\sum\limits_{m_{i}, m_{f}}
  \displaystyle\sum\limits_{\mu_{i},\, \mu_{f} = \pm 1/2}
    C_{l_{f}m_{f} 1/2 \mu_{f}}^{j_{f}M_{f},\,*}\,
    C_{l_{i}m_{i} 1/2 \mu_{i}}^{j_{i}M_{i}}\quad \times \\
  & \times &
  \Bigl[ -1 + i\, \Bigl\{ \delta_{\mu_{i}, +1/2}\; -\; \delta_{\mu_{i}, -1/2} \Bigr\} \Bigr] \cdot
    \Bigl[ i\,\mu\,\tilde{p}_{l_{\rm ph} \mu}^{M m_{i} m_{f}} + \tilde{p}_{l_{\rm ph} \mu}^{E m_{i} m_{f}} \Bigr].
\end{array}
\label{eq.app.9.3.8}
\end{equation}

\subsection{Matrix element $\tilde{p}_{\rm q, 1}$ at spherically symmetric description of nucleus
\label{sec.9.4}}

On the basis of the obtained formalism for the matrix elements above, we write matrix element $\tilde{p}_{\rm q, 1}$:
%
\begin{equation}
\begin{array}{lcl}
\vspace{1.3mm}
  \tilde{p}_{\rm q,1} & = &
  \sqrt{\displaystyle\frac{\pi}{2}}
  \displaystyle\sum\limits_{l_{\rm ph}=1}\, (-i)^{l_{\rm ph}}\, \sqrt{2l_{\rm ph}+1} \cdot
  \displaystyle\sum\limits_{\mu \pm 1}
  \displaystyle\sum\limits_{m_{f}, m_{i}}
  \displaystyle\sum\limits_{\mu_{i},\, \mu_{f} = \pm 1/2}
    C_{l_{f}m_{f} 1/2 \mu_{f}}^{j_{f}M_{f},\,*}\,
    C_{l_{i}m_{i} 1/2 \mu_{i}}^{j_{i}M_{i}}\; \times \\
  & \times &
  \Bigl\{
    A_{1} (Q, F_{1}, F_{2})\, \Bigl[ i\mu\, \tilde{p}_{l_{\rm ph} \mu}^{M m_{i} m_{f}} + \tilde{p}_{l_{\rm ph} \mu}^{E m_{i} m_{f}} \Bigr] +
    B_{1} (Q, F_{1}, F_{2})\, \Bigl[ i\mu\, \breve{p}_{l_{\rm ph} \mu}^{M m_{i} m_{f}} + \breve{p}_{l_{\rm ph} \mu}^{E m_{i} m_{f}} \Bigr]
  \Bigr\},
\end{array}
\label{eq.app.9.4.1}
\end{equation}
where $\tilde{p}_{l_{\rm ph} \mu}^{M}$ and $\tilde{p}_{l_{\rm ph} \mu}^{E}$ are defined in Eqs.~(\ref{eq.app.9.3.5}) and
\begin{equation}
\begin{array}{lcllcl}
  \breve{p}_{l_{\rm ph}\mu}^{M m_{i} m_{f}} & = &
    \xibf_{\mu}\,
    \displaystyle\int
      \varphi^{*}_{f}(\mathbf{r})\,
      \varphi_{i}(\mathbf{r})\,
      V(\mathbf{r})\;
      \mathbf{A}_{l_{\rm ph}\mu}^{*} (\mathbf{r}, M) \;
      \mathbf{dr}, &
  \hspace{7mm}
  \breve{p}_{l_{\rm ph}\mu}^{E m_{i} m_{f}} & = &
    \xibf_{\mu}\,
    \displaystyle\int
      \varphi^{*}_{f}(\mathbf{r})\,
      \varphi_{i}(\mathbf{r})\,
      V(\mathbf{r})\;
      \mathbf{A}_{l_{\rm ph}\mu}^{*} (\mathbf{r}, E)\;
      \mathbf{dr}.
\end{array}
\label{eq.app.9.4.2}
\end{equation}

\subsection{Calculations of components $p_{l_{\rm ph}\mu}^{M}$, $p_{l_{\rm ph}\mu}^{E}$ and\, $\tilde{p}_{l_{\rm ph}\mu}^{M}$, $\tilde{p}_{l_{\rm ph}\mu}^{E}$: case of $l_{i} \ne 0$
\label{sec.9.7}}

Let us consider a case, when the full nuclear system has $l_{i} \ne 0$ in the initial state.
Using gradient formula, we obtain:
%
\begin{equation}
\begin{array}{ll}
\vspace{1mm}
  &
  \displaystyle\frac{\partial}{\partial \mathbf{r}}\: \varphi_{i}(\mathbf{r}) =
  \displaystyle\frac{\partial}{\partial \mathbf{r}}\: \Bigl\{ R_{i} (r)\: Y_{l_{i}m_{i}}({\mathbf n}_{\rm r}^{i}) \Bigr\} = \\

  = &
    \sqrt{\displaystyle\frac{l_{i}}{2l_{i}+1}}\:
    \biggl( \displaystyle\frac{dR_{i}(r)}{dr} + \displaystyle\frac{l_{i}+1}{r}\, R_{i}(r) \biggr)\,
      \mathbf{T}_{l_{i} l_{i}-1, m_{i}}({\mathbf n}_{\rm r}^{i}) - 
  \sqrt{\displaystyle\frac{l_{i}+1}{2l_{i}+1}}\:
    \biggl( \displaystyle\frac{dR_{i}(r)}{dr} - \displaystyle\frac{l_{i}}{r}\, R_{i}(r) \biggr)\,
      \mathbf{T}_{l_{i} l_{i}+1, m_{i}}({\mathbf n}_{\rm r}^{i}).
\end{array}
\label{eq.app.9.7.1}
\end{equation}
Using such a formula, for the magnetic component $p_{l\mu}^{M}$ при $l_{i} \ne 0$ we have:
%
\[
\begin{array}{lcl}
  p_{l\mu}^{M} & = &
  \displaystyle\int\limits^{+\infty}_{0} dr
    \displaystyle\int d\Omega \:
    r^{2} \varphi^{*}_{f}(\mathbf{r}) \:
    \biggl( \displaystyle\frac{\partial}{\partial \mathbf{r}}\,
            \varphi_{i}(\mathbf{r}) \biggr) \:
    \mathbf{A}_{l\mu}^{*} (\mathbf{r}, M) = \\

  & = &
  \displaystyle\int\limits^{+\infty}_{0} dr
    \displaystyle\int d\Omega \:
    r^{2} R^{*}_{f}(r)\:
    Y_{l_{f}m}^{*}({\mathbf n}_{\rm r}^{f}) \:
    \biggl\{
      \sqrt{\displaystyle\frac{l_{i}}{2l_{i}+1}}\:
      \biggl( \displaystyle\frac{dR_{i}(r)}{dr} + \displaystyle\frac{l_{i}+1}{r}\, R_{i}(r) \biggr)\,
      \mathbf{T}_{l_{i} l_{i}-1, m_{i}}({\mathbf n}_{\rm r}^{i})\; - \\
    & - &
    \sqrt{\displaystyle\frac{l_{i}+1}{2l_{i}+1}}\:
      \biggl( \displaystyle\frac{dR_{i}(r)}{dr} - \displaystyle\frac{l_{i}}{r}\, R_{i}(r) \biggr)\,
      \mathbf{T}_{l_{i} l_{i}+1, m_{i}}({\mathbf n}_{\rm r}^{i})
    \biggr\} \:
    j_{l_{\rm ph}}(kr) \: \mathbf{T}_{l_{\rm ph}l_{\rm ph},\mu}^{*} ({\mathbf n}_{\rm ph}) = \\

  & = &
  \sqrt{\displaystyle\frac{l_{i}}{2l_{i}+1}}\:
  \displaystyle\int\limits^{+\infty}_{0}
      R^{*}_{f}(r)\,
      \biggl( \displaystyle\frac{dR_{i}(r)}{dr} + \displaystyle\frac{l_{i}+1}{r}\, R_{i}(r) \biggr)\,
      j_{l_{\rm ph}}(kr)\;
      r^{2} dr \cdot
    \displaystyle\int
      Y_{l_{f}m}^{*}({\mathbf n}_{\rm r}^{f})\:
      \mathbf{T}_{l_{i} l_{i}-1, m_{i}}({\mathbf n}_{\rm r}^{i})\,
      \mathbf{T}_{l_{\rm ph}l_{\rm ph},\mu}^{*} ({\mathbf n}_{\rm ph})\; d\Omega\; - \\
    & - &
  \sqrt{\displaystyle\frac{l_{i}+1}{2l_{i}+1}}\:
  \displaystyle\int\limits^{+\infty}_{0}
      R^{*}_{f}(r)\,
      \biggl( \displaystyle\frac{dR_{i}(r)}{dr} - \displaystyle\frac{l_{i}}{r}\, R_{i}(r) \biggr)\, j_{l_{\rm ph}}(kr)\;
      r^{2} dr \cdot
    \displaystyle\int
      Y_{l_{f}m}^{*}({\mathbf n}_{\rm r}^{f}) \:
      \mathbf{T}_{l_{i} l_{i}+1, m_{i}}({\mathbf n}_{\rm r}^{i})\,
      \mathbf{T}_{l_{\rm ph}l_{\rm ph},\mu}^{*} ({\mathbf n}_{\rm ph})\; d\Omega.
\end{array}
\]
For electric component $p^{E}_{l_{\rm ph}\mu}$ we have:
%
\[
\begin{array}{cl}
  & p_{l_{\rm ph}\mu}^{E} =
  \displaystyle\int\limits^{+\infty}_{0} dr
    \displaystyle\int d\Omega \:
    r^{2} \varphi^{*}_{f}(\mathbf{r})
    \biggl( \displaystyle\frac{\partial}{\partial \mathbf{r}}
            \varphi_{i}(\mathbf{r}) \biggr)
    \mathbf{A}_{l_{\rm ph} \mu}^{*} (\mathbf{r}, E) = \\

  = &
  \displaystyle\int\limits^{+\infty}_{0} dr
    \displaystyle\int d\Omega \:
    r^{2} R^{*}_{f}(r)\,
    Y_{l_{f}m_{f}}^{*}({\mathbf n}_{\rm r}^{f}) \cdot
    \biggl\{
    \sqrt{\displaystyle\frac{l_{i}}{2l_{i}+1}}\:
    \biggl( \displaystyle\frac{dR_{i}(r)}{dr} + \displaystyle\frac{l_{i}+1}{r}\, R_{i}(r) \biggr)\,
      \mathbf{T}_{l_{i} l_{i}-1, m_{i}}({\mathbf n}_{\rm r}^{i}) - \\
  - &
    \sqrt{\displaystyle\frac{l_{i}+1}{2l_{i}+1}}\:
      \biggl( \displaystyle\frac{dR_{i}(r)}{dr} - \displaystyle\frac{l_{i}}{r}\, R_{i}(r) \biggr)\,
        \mathbf{T}_{l_{i} l_{i}+1, m_{i}}({\mathbf n}_{\rm r}^{i})
    \biggr\} \times \\
  \vspace{4mm}
  \times &
    \Biggl\{
      \sqrt{\displaystyle\frac{l_{\rm ph}+1}{2l_{\rm ph}+1}}
      j_{l_{\rm ph}-1}(kr) \: \mathbf{T}_{l_{\rm ph}l_{\rm ph}-1,\mu}^{*}({\mathbf n}_{\rm ph}) -
      \sqrt{\displaystyle\frac{l_{\rm ph}}{2l_{\rm ph}+1}}
      j_{l_{\rm ph}+1}(kr) \: \mathbf{T}_{l_{\rm ph}l_{\rm ph}+1,\mu}^{*}({\mathbf n}_{\rm ph})
    \Biggr\} = \\

  = &
  \sqrt{\displaystyle\frac{l_{i}}{2l_{i}+1}}
  \sqrt{\displaystyle\frac{l_{\rm ph}+1}{2l_{\rm ph}+1}}
    \displaystyle\int\limits^{+\infty}_{0}
      R^{*}_{f}(r)\,
      \biggl( \displaystyle\frac{dR_{i}(r)}{dr} + \displaystyle\frac{l_{i}+1}{r}\, R_{i}(r) \biggr)\,
      j_{l_{\rm ph}-1}(kr)\:
      r^{2} dr
    \displaystyle\int
      Y_{l_{f}m_{f}}^{*}({\mathbf n}_{\rm r}^{f})\,
      \mathbf{T}_{l_{i}l_{i}-1,m_{i}}(\mathbf{n}^{i}_{\rm r})\,
      \mathbf{T}_{l_{\rm ph}l_{\rm ph}-1,\mu}^{*}({\mathbf n}_{\rm ph})\; d\Omega - \\

  - &
  \sqrt{\displaystyle\frac{l_{i}}{2l_{i}+1}}
  \sqrt{\displaystyle\frac{l_{\rm ph}}{2l_{\rm ph}+1}}
    \displaystyle\int\limits^{+\infty}_{0}
      R^{*}_{f}(r)\,
      \biggl( \displaystyle\frac{dR_{i}(r)}{dr} + \displaystyle\frac{l_{i}+1}{r}\, R_{i}(r) \biggr)\,
      j_{l_{\rm ph}+1}(kr)\:
      r^{2} dr
    \displaystyle\int
      Y_{l_{f}m_{f}}^{*}({\mathbf n}_{\rm r}^{f})\,
      \mathbf{T}_{l_{i}l_{i}-1,m_{i}}(\mathbf{n}^{i}_{\rm r})\,
      \mathbf{T}_{l_{\rm ph}l_{\rm ph}+1,\mu}^{*}({\mathbf n}_{\rm ph})\; d\Omega + \\

  + &
  \sqrt{\displaystyle\frac{l_{i}+1}{2l_{i}+1}}
  \sqrt{\displaystyle\frac{l_{\rm ph}+1}{2l_{\rm ph}+1}}
    \displaystyle\int\limits^{+\infty}_{0}
      R^{*}_{f}(r)\,
      \biggl( \displaystyle\frac{dR_{i}(r)}{dr} - \displaystyle\frac{l_{i}}{r}\, R_{i}(r) \biggr)\,
      j_{l_{\rm ph}-1}(kr)\:
      r^{2} dr
    \displaystyle\int
      Y_{l_{f}m_{f}}^{*}({\mathbf n}_{\rm r}^{f})\,
      \mathbf{T}_{l_{i}l_{i}+1,m_{i}}(\mathbf{n}^{i}_{\rm r})\,
      \mathbf{T}_{l_{\rm ph}l_{\rm ph}-1,\mu}^{*}({\mathbf n}_{\rm ph})\; d\Omega - \\

  - &
  \sqrt{\displaystyle\frac{l_{i}+1}{2l_{i}+1}}
  \sqrt{\displaystyle\frac{l_{\rm ph}}{2l_{\rm ph}+1}}
    \displaystyle\int\limits^{+\infty}_{0}
      R^{*}_{f}(r)\,
      \biggl( \displaystyle\frac{dR_{i}(r)}{dr} - \displaystyle\frac{l_{i}}{r}\, R_{i}(r) \biggr)\,
      j_{l_{\rm ph}+1}(kr)\:
      r^{2} dr
    \displaystyle\int
      Y_{l_{f}m_{f}}^{*}({\mathbf n}_{\rm r}^{f})\,
      \mathbf{T}_{l_{i}l_{i}+1,m_{i}}(\mathbf{n}^{i}_{\rm r})\,
      \mathbf{T}_{l_{\rm ph}l_{\rm ph}+1,\mu}^{*}({\mathbf n}_{\rm ph})\; d\Omega.
\end{array}
\]
So, components $p_{l_{\rm ph}\mu}^{M}$ and $p_{l_{\rm ph}\mu}^{E}$ have form:
%
\begin{equation}
\begin{array}{cl}
  & p_{l_{\rm ph}\mu}^{M} =
  \sqrt{\displaystyle\frac{l_{i}}{2l_{i}+1}}\:
  \displaystyle\int\limits^{+\infty}_{0}
      R^{*}_{f}(r)\,
      \biggl( \displaystyle\frac{dR_{i}(r)}{dr} + \displaystyle\frac{l_{i}+1}{r}\, R_{i}(r) \biggr)\,
      j_{l_{\rm ph}}(kr)\;
      r^{2} dr \cdot
    \displaystyle\int
      Y_{l_{f} m_{f}}^{*}({\mathbf n}_{\rm r}^{f})\:
      \mathbf{T}_{l_{i} l_{i}-1, m_{i}}({\mathbf n}_{\rm r}^{i})\,
      \mathbf{T}_{l_{\rm ph}l_{\rm ph},\mu}^{*} ({\mathbf n}_{\rm ph})\; d\Omega\; - \\
  \vspace{2mm}
  - &
  \sqrt{\displaystyle\frac{l_{i}+1}{2l_{i}+1}}\:
  \displaystyle\int\limits^{+\infty}_{0}
      R^{*}_{f}(r)\,
      \biggl( \displaystyle\frac{dR_{i}(r)}{dr} - \displaystyle\frac{l_{i}}{r}\, R_{i}(r) \biggr)\, j_{l_{\rm ph}}(kr)\;
      r^{2} dr \cdot
    \displaystyle\int
      Y_{l_{f} m_{f}}^{*}({\mathbf n}_{\rm r}^{f}) \:
      \mathbf{T}_{l_{i} l_{i}+1, m_{i}}({\mathbf n}_{\rm r}^{i})\,
      \mathbf{T}_{l_{\rm ph}l_{\rm ph},\mu}^{*} ({\mathbf n}_{\rm ph})\; d\Omega, \\

  & p_{l_{\rm ph}\mu}^{E} =
  \sqrt{\displaystyle\frac{l_{i}}{2l_{i}+1}}
  \sqrt{\displaystyle\frac{l_{\rm ph}+1}{2l_{\rm ph}+1}}
    \displaystyle\int\limits^{+\infty}_{0}
      R^{*}_{f}(r)\,
      \biggl( \displaystyle\frac{dR_{i}(r)}{dr} + \displaystyle\frac{l_{i}+1}{r}\, R_{i}(r) \biggr)\,
      j_{l_{\rm ph}-1}(kr)\:
      r^{2} dr \times \\
  \times &
    \displaystyle\int
      Y_{l_{f} m_{f}}^{*}({\mathbf n}_{\rm r}^{f})\,
      \mathbf{T}_{l_{i}l_{i}-1,m_{i}}(\mathbf{n}^{i}_{\rm r})\,
      \mathbf{T}_{l_{\rm ph}l_{\rm ph}-1,\mu}^{*}({\mathbf n}_{\rm ph})\; d\Omega - \\

  - &
  \sqrt{\displaystyle\frac{l_{i}}{2l_{i}+1}}
  \sqrt{\displaystyle\frac{l_{\rm ph}}{2l_{\rm ph}+1}}
    \displaystyle\int\limits^{+\infty}_{0}
      R^{*}_{f}(r)\,
      \biggl( \displaystyle\frac{dR_{i}(r)}{dr} + \displaystyle\frac{l_{i}+1}{r}\, R_{i}(r) \biggr)\,
      j_{l_{\rm ph}+1}(kr)\:
      r^{2} dr \times \\
  \times &
    \displaystyle\int
      Y_{l_{f} m_{f}}^{*}({\mathbf n}_{\rm r}^{f})\,
      \mathbf{T}_{l_{i}l_{i}-1,m_{i}}(\mathbf{n}^{i}_{\rm r})\,
      \mathbf{T}_{l_{\rm ph}l_{\rm ph}+1,\mu}^{*}({\mathbf n}_{\rm ph})\; d\Omega + \\

  + &
  \sqrt{\displaystyle\frac{l_{i}+1}{2l_{i}+1}}
  \sqrt{\displaystyle\frac{l_{\rm ph}+1}{2l_{\rm ph}+1}}
    \displaystyle\int\limits^{+\infty}_{0}
      R^{*}_{f}(r)\,
      \biggl( \displaystyle\frac{dR_{i}(r)}{dr} - \displaystyle\frac{l_{i}}{r}\, R_{i}(r) \biggr)\,
      j_{l_{\rm ph}-1}(kr)\:
      r^{2} dr
    \displaystyle\int
      Y_{l_{f} m_{f}}^{*}({\mathbf n}_{\rm r}^{f})\,
      \mathbf{T}_{l_{i}l_{i}+1,m_{i}}(\mathbf{n}^{i}_{\rm r})\,
      \mathbf{T}_{l_{\rm ph}l_{\rm ph}-1,\mu}^{*}({\mathbf n}_{\rm ph})\; d\Omega - \\

  - &
  \sqrt{\displaystyle\frac{l_{i}+1}{2l_{i}+1}}
  \sqrt{\displaystyle\frac{l_{\rm ph}}{2l_{\rm ph}+1}}
    \displaystyle\int\limits^{+\infty}_{0}
      R^{*}_{f}(r)\,
      \biggl( \displaystyle\frac{dR_{i}(r)}{dr} - \displaystyle\frac{l_{i}}{r}\, R_{i}(r) \biggr)\,
      j_{l_{\rm ph}+1}(kr)\:
      r^{2} dr
    \displaystyle\int
      Y_{l_{f} m_{f}}^{*}({\mathbf n}_{\rm r}^{f})\,
      \mathbf{T}_{l_{i}l_{i}+1,m_{i}}(\mathbf{n}^{i}_{\rm r})\,
      \mathbf{T}_{l_{\rm ph}l_{\rm ph}+1,\mu}^{*}({\mathbf n}_{\rm ph})\; d\Omega.
\end{array}
\label{eq.app.9.7.2}
\end{equation}
Let us introduce the following notions:
%
\begin{equation}
\begin{array}{ccl}
  J_{1}(l_{i},l_{f},n) & = &
  \displaystyle\int\limits^{+\infty}_{0}
    \displaystyle\frac{dR_{i}(r, l_{i})}{dr}\: R^{*}_{f}(l_{f},r)\,
    j_{n}(kr)\; r^{2} dr, \\

  J_{2}(l_{i},l_{f},n) & = &
  \displaystyle\int\limits^{+\infty}_{0}
    R_{i}(r, l_{i})\, R^{*}_{f}(l_{f},r)\: j_{n}(kr)\; r\, dr, \\

  I_{M}\, (l_{i}, l_{f}, l_{\rm ph}, l_{1}, \mu) & = &
    \displaystyle\int
      Y_{l_{f}m_{f}}^{*}({\mathbf n}_{\rm r}^{f})\,
      \mathbf{T}_{l_{i}\, l_{1},\, m_{i}}(\mathbf{n}^{i}_{\rm r})\,
      \mathbf{T}_{l_{\rm ph}\,l_{\rm ph},\, \mu}^{*}({\mathbf n}_{\rm ph})\; d\Omega, \\

  I_{E}\, (l_{i}, l_{f}, l_{\rm ph}, l_{1}, l_{2}, \mu) & = &
    \displaystyle\int
      Y_{l_{f}m_{f}}^{*}({\mathbf n}_{\rm r}^{f})\,
      \mathbf{T}_{l_{i} l_{1},\, m_{i}}(\mathbf{n}^{i}_{\rm r})\,
      \mathbf{T}_{l_{\rm ph} l_{2},\, \mu}^{*}({\mathbf n}_{\rm ph})\; d\Omega.
\end{array}
\label{eq.app.9.7.3}
\end{equation}
Then, one can rewrite Eqs.~(\ref{eq.app.9.7.2}) as
%
\begin{equation}
\begin{array}{lcl}
\vspace{1mm}
  p_{l_{\rm ph,\mu}}^{M} & = &
    \sqrt{\displaystyle\frac{l_{i}}{2l_{i}+1}}\:
      I_{M}(l_{i},l_{f}, l_{\rm ph}, l_{i}-1, \mu) \cdot
      \Bigl\{
        J_{1}(l_{i},l_{f},l_{\rm ph}) + (l_{i}+1) \cdot J_{2}(l_{i},l_{f},l_{\rm ph})
      \Bigr\}\; - \\
\vspace{3mm}
  & - &
    \sqrt{\displaystyle\frac{l_{i}+1}{2l_{i}+1}}\:
      I_{M}(l_{i},l_{f}, l_{\rm ph}, l_{i}+1, \mu) \cdot
      \Bigl\{
        J_{1}(l_{i},l_{f},l_{\rm ph}) - l_{i} \cdot J_{2}(l_{i},l_{f},l_{\rm ph})
      \Bigr\}, \\

\vspace{1mm}
  p_{l_{\rm ph,\mu}}^{E} & = &
    \sqrt{\displaystyle\frac{l_{i}\,(l_{\rm ph}+1)}{(2l_{i}+1)(2l_{\rm ph}+1)}} \cdot
      I_{E}(l_{i},l_{f}, l_{\rm ph}, l_{i}-1, l_{\rm ph}-1, \mu) \cdot
      \Bigl\{
        J_{1}(l_{i},l_{f},l_{\rm ph}-1)\; +
        (l_{i}+1) \cdot J_{2}(l_{i},l_{f},l_{\rm ph}-1)
      \Bigr\}\; - \\
\vspace{1mm}
    & - &
    \sqrt{\displaystyle\frac{l_{i}\,l_{\rm ph}}{(2l_{i}+1)(2l_{\rm ph}+1)}} \cdot
      I_{E} (l_{i},l_{f}, l_{\rm ph}, l_{i}-1, l_{\rm ph}+1, \mu) \cdot
      \Bigl\{
        J_{1}(l_{i},l_{f},l_{\rm ph}+1)\; +
        (l_{i}+1) \cdot J_{2}(l_{i},l_{f},l_{\rm ph}+1)
      \Bigr\}\; + \\
\vspace{1mm}
  & + &
    \sqrt{\displaystyle\frac{(l_{i}+1)(l_{\rm ph}+1)}{(2l_{i}+1)(2l_{\rm ph}+1)}} \cdot
      I_{E} (l_{i},l_{f},l_{\rm ph}, l_{i}+1, l_{\rm ph}-1, \mu) \cdot
      \Bigl\{
        J_{1}(l_{i},l_{f},l_{\rm ph}-1)\; -
        l_{i} \cdot J_{2}(l_{i},l_{f},l_{\rm ph}-1)
      \Bigr\}\; - \\
  & - &
    \sqrt{\displaystyle\frac{(l_{i}+1)\,l_{\rm ph}}{(2l_{i}+1)(2l_{\rm ph}+1)}} \cdot
      I_{E} (l_{i},l_{f}, l_{\rm ph}, l_{i}+1, l_{\rm ph}+1, \mu) \cdot
      \Bigl\{
        J_{1}(l_{i},l_{f},l_{\rm ph}+1)\; -
        l_{i} \cdot J_{2}(l_{i},l_{f},l_{\rm ph}+1)
      \Bigr\}.
\end{array}
\label{eq.app.9.7.4}
\end{equation}

By the same way, we find components $\tilde{p}_{l_{\rm ph}\mu}^{M}$ and $\tilde{p}_{l_{\rm ph}\mu}^{E}$:
%
\begin{equation}
\begin{array}{lcl}
  \tilde{p}_{l_{\rm ph}\mu}^{M} & = &
  \displaystyle\int\limits^{+\infty}_{0}
    R^{*}_{f}(r)\,
    R_{i}(r)\,
    j_{l_{\rm ph}}(kr)\:
    r^{2} dr \cdot
  \xibf_{\mu} \displaystyle\int
    Y_{l_{f}m_{f}}^{*}({\mathbf n}_{\rm r}^{f})\,
    Y_{l_{i}m_{i}}({\mathbf n}_{\rm r}^{i})\:
    \mathbf{T}_{l_{\rm ph}l_{\rm ph},\mu}^{*} ({\mathbf n}_{\rm ph}) \: d\Omega, \\

  \tilde{p}_{l_{\rm ph}\mu}^{E} & = &
  \sqrt{\displaystyle\frac{l_{\rm ph}+1}{2l_{\rm ph}+1}}
    \displaystyle\int\limits^{+\infty}_{0}
      R^{*}_{f}(r)\, R_{i}(r)\, j_{l_{\rm ph}-1}(kr)\;
      r^{2} dr \cdot
  \xibf_{\mu} \displaystyle\int
    Y_{l_{f}m_{f}}^{*}({\mathbf n}_{\rm r}^{f})\,
    Y_{l_{i}m_{i}}({\mathbf n}_{\rm r}^{i})\:
    \mathbf{T}_{l_{\rm ph}l_{\rm ph}-1,\mu}^{*}({\mathbf n}_{\rm ph})\: d\Omega\; - \\

  & & - \;
  \sqrt{\displaystyle\frac{l_{\rm ph}}{2l_{\rm ph}+1}}
    \displaystyle\int\limits^{+\infty}_{0}
      R^{*}_{f}(r)\, R_{i}(r)\, j_{l_{\rm ph}+1}(kr)\;
      r^{2} dr \cdot
  \xibf_{\mu} \displaystyle\int
    Y_{l_{f}m_{f}}^{*}({\mathbf n}_{\rm r}^{f})\,
    Y_{l_{i}m_{i}}({\mathbf n}_{\rm r}^{i})\:
    \mathbf{T}_{l_{\rm ph}l_{\rm ph}+1,\mu}^{*}({\mathbf n}_{\rm ph})\; d\Omega.
\end{array}
\label{eq.app.9.7.5}
\end{equation}
Introducing new integrals:
%
\begin{equation}
\begin{array}{lcl}
  \tilde{J}\,(l_{i},l_{f},n) & = &
  \displaystyle\int\limits^{+\infty}_{0}
    R_{i}(r)\, R^{*}_{f}(l,r)\, j_{n}(kr)\; r^{2} dr, \\

  \tilde{I}\,(l_{i}, l_{f}, l_{\rm ph}, n, \mu) & = &
  \xibf_{\mu} \displaystyle\int
    Y_{l_{i}m_{i}}({\mathbf n}_{\rm r}^{i})\:
    Y_{l_{f}m_{f}}^{*}({\mathbf n}_{\rm r}^{f})\:
    \mathbf{T}_{l_{\rm ph} n,\mu}^{*}({\mathbf n}_{\rm ph}) \: d\Omega,
\end{array}
\label{eq.app.9.7.6}
\end{equation}
we rewrite (\ref{eq.app.9.7.5}) as
\begin{equation}
\begin{array}{lcl}
  \tilde{p}_{l_{\rm ph}\mu}^{M} & = &
    \tilde{I}\,(l_{i},l_{f},l_{\rm ph}, l_{\rm ph}, \mu) \cdot \tilde{J}\, (l_{i},l_{f},l_{\rm ph}), \\
  \tilde{p}_{l_{\rm ph}\mu}^{E} & = &
    \sqrt{\displaystyle\frac{l_{\rm ph}+1}{2l_{\rm ph}+1}}
      \tilde{I}\,(l_{i},l_{f},l_{\rm ph},l_{\rm ph}-1,\mu) \cdot \tilde{J}\,(l_{i},l_{f},l_{\rm ph}-1) -
    \sqrt{\displaystyle\frac{l_{\rm ph}}{2l_{\rm ph}+1}}
      \tilde{I}\,(l_{i},l_{f},l_{\rm ph},l_{\rm ph}+1,\mu) \cdot \tilde{J}\,(l_{i},l_{f},l_{\rm ph}+1).
\end{array}
\label{eq.app.9.7.7}
\end{equation}

\subsection{Calculation of terms $\breve{p}_{l_{\rm ph}\mu}^{M}$ and $\breve{p}_{l_{\rm ph}\mu}^{E}$: case of $l_{i} \ne 0$
\label{sec.9.8}}

In this Section we calculate components $\breve{p}_{l_{\rm ph}\mu}^{M}$ and $\breve{p}_{l_{\rm ph}\mu}^{E}$ at $l_{i} \ne 0$.
From (\ref{eq.app.9.4.2}) we have:
\begin{equation}
\begin{array}{lcllcl}
  \breve{p}_{l_{\rm ph}\mu}^{M} & = &
    \xibf_{\mu}\,
    \displaystyle\int
      \varphi^{*}_{f}(\mathbf{r})\,
      \varphi_{i}(\mathbf{r})\,
      V(\mathbf{r})\;
      \mathbf{A}_{l_{\rm ph}\mu}^{*} (\mathbf{r}, M) \;
      \mathbf{dr}, &
  \hspace{7mm}
  \breve{p}_{l_{\rm ph}\mu}^{E} & = &
    \xibf_{\mu}\,
    \displaystyle\int
      \varphi^{*}_{f}(\mathbf{r})\, \varphi_{i}(\mathbf{r})\, V(\mathbf{r})\;
      \mathbf{A}_{l_{\rm ph}\mu}^{*} (\mathbf{r}, E)\;
      \mathbf{dr},
\end{array}
\label{eq.app.9.8.1}
\end{equation}
\begin{equation}
\begin{array}{lcllcl}
  \varphi_{i} (\mathbf{r}) = R_{i} (r)\: Y_{l_{i}m_{i}}({\mathbf n}_{\rm r}^{i}), &
  \varphi_{f} (\mathbf{r}) = R_{f} (r)\: Y_{l_{f}m_{f}}({\mathbf n}_{\rm r}^{f}).
\end{array}
\label{eq.app.9.8.2}
\end{equation}
For magnetic component $\breve{p}_{l\mu}^{M}$ при $l_{i} \ne 0$ we obtain:
\[
\begin{array}{lcl}
  \breve{p}_{l\mu}^{M} & = &
  \xibf_{\mu}
  \displaystyle\int\limits^{+\infty}_{0} dr
    \displaystyle\int d\Omega \:
    r^{2} \varphi^{*}_{f}(\mathbf{r})\, \varphi_{i}(\mathbf{r})\, V(\mathbf{r})\;
    \mathbf{A}_{l_{\rm ph}\mu}^{*} (\mathbf{r}, M) = \\

  & = &
  \xibf_{\mu}
  \displaystyle\int\limits^{+\infty}_{0} dr
    \displaystyle\int d\Omega \:
    r^{2} R^{*}_{f}(r)\: Y_{l_{f}m}^{*}({\mathbf n}_{\rm r}^{f})\:
    R_{i}(r)\: Y_{l_{i}m}({\mathbf n}_{\rm r}^{i})\,  V(\mathbf{r})\, j_{l_{\rm ph}}(kr)\:
    \mathbf{T}_{l_{\rm ph}l_{\rm ph},\mu}^{*} ({\mathbf n}_{\rm ph})\; = \\

  & = &
  \displaystyle\int\limits^{+\infty}_{0}
      R^{*}_{f}(r)\, R_{i}(r)\, V(\mathbf{r})\, j_{l_{\rm ph}}(kr)\; r^{2} dr \cdot
  \xibf_{\mu}
  \displaystyle\int
    Y_{l_{f}m}^{*}({\mathbf n}_{\rm r}^{f})\, Y_{l_{i}m}({\mathbf n}_{\rm r}^{i})\,
    \mathbf{T}_{l_{\rm ph}l_{\rm ph},\mu}^{*} ({\mathbf n}_{\rm ph})\; d\Omega.
\end{array}
\]
For electric component $\breve{p}^{E}_{l_{\rm ph}\mu}$ we have:
\[
\begin{array}{cl}
  & \breve{p}_{l_{\rm ph}\mu}^{E} =
  \xibf_{\mu}
  \displaystyle\int\limits^{+\infty}_{0} dr
  \displaystyle\int d\Omega\:
    r^{2} \varphi^{*}_{f}(\mathbf{r})\, \varphi_{i}(\mathbf{r})\, V(\mathbf{r})\;
    \mathbf{A}_{l_{\rm ph} \mu}^{*} (\mathbf{r}, E)\; = \\

  = &
  \xibf_{\mu}
  \displaystyle\int\limits^{+\infty}_{0} dr
  \displaystyle\int d\Omega \:
    r^{2} R^{*}_{f}(r)\, Y_{l_{f}m_{f}}^{*}({\mathbf n}_{\rm r}^{f})\:
    R_{i}(r)\, Y_{l_{i}m_{i}}({\mathbf n}_{\rm r}^{i})\, V(\mathbf{r})\; \times \\
  \vspace{4mm}
  \times &
    \Biggl\{
      \sqrt{\displaystyle\frac{l_{\rm ph}+1}{2l_{\rm ph}+1}}
      j_{l_{\rm ph}-1}(kr) \: \mathbf{T}_{l_{\rm ph}l_{\rm ph}-1,\mu}^{*}({\mathbf n}_{\rm ph}) -
      \sqrt{\displaystyle\frac{l_{\rm ph}}{2l_{\rm ph}+1}}
      j_{l_{\rm ph}+1}(kr) \: \mathbf{T}_{l_{\rm ph}l_{\rm ph}+1,\mu}^{*}({\mathbf n}_{\rm ph})
    \Biggr\} = \\

  = &
  \sqrt{\displaystyle\frac{l_{\rm ph}+1}{2l_{\rm ph}+1}}
  \displaystyle\int\limits^{+\infty}_{0}
    R^{*}_{f}(r)\, R_{i}(r)\, V(\mathbf{r})\, j_{l_{\rm ph}-1}(kr)\: r^{2} dr \cdot
  \xibf_{\mu}
  \displaystyle\int
    Y_{l_{f}m_{f}}^{*}({\mathbf n}_{\rm r}^{f})\,
    Y_{l_{i}m_{i}}({\mathbf n}_{\rm r}^{i})\,
    \mathbf{T}_{l_{\rm ph}l_{\rm ph}-1,\mu}^{*}({\mathbf n}_{\rm ph})\; d\Omega\; - \\

  - &
  \sqrt{\displaystyle\frac{l_{\rm ph}}{2l_{\rm ph}+1}}
  \displaystyle\int\limits^{+\infty}_{0}
    R^{*}_{f}(r)\, R_{i}(r)\, V(\mathbf{r})\, j_{l_{\rm ph}+1}(kr)\: r^{2} dr \cdot
  \xibf_{\mu}
  \displaystyle\int
    Y_{l_{f}m_{f}}^{*}({\mathbf n}_{\rm r}^{f})\,
    Y_{l_{i}m_{i}}({\mathbf n}_{\rm r}^{i})\,
    \mathbf{T}_{l_{\rm ph}l_{\rm ph}+1,\mu}^{*}({\mathbf n}_{\rm ph})\; d\Omega.
\end{array}
\]
So, components $\breve{p}_{l_{\rm ph}\mu}^{M}$ and $\breve{p}_{l_{\rm ph}\mu}^{E}$ have form:
%
\begin{equation}
\begin{array}{lcl}
  \breve{p}_{l_{\rm ph}\mu}^{M} & = &
  \displaystyle\int\limits^{+\infty}_{0}
      R^{*}_{f}(r)\, R_{i}(r)\, V(\mathbf{r})\, j_{l_{\rm ph}}(kr)\; r^{2} dr \cdot
  \xibf_{\mu}
  \displaystyle\int
    Y_{l_{f}m}^{*}({\mathbf n}_{\rm r}^{f})\, Y_{l_{i}m}({\mathbf n}_{\rm r}^{i})\,
    \mathbf{T}_{l_{\rm ph}l_{\rm ph},\mu}^{*} ({\mathbf n}_{\rm ph})\; d\Omega, \\

  \breve{p}_{l_{\rm ph}\mu}^{E} & = &
  \sqrt{\displaystyle\frac{l_{\rm ph}+1}{2l_{\rm ph}+1}}
  \displaystyle\int\limits^{+\infty}_{0}
    R^{*}_{f}(r)\, R_{i}(r)\, V(\mathbf{r})\, j_{l_{\rm ph}-1}(kr)\: r^{2} dr \cdot
  \xibf_{\mu}
  \displaystyle\int
    Y_{l_{f}m_{f}}^{*}({\mathbf n}_{\rm r}^{f})\,
    Y_{l_{i}m_{i}}({\mathbf n}_{\rm r}^{i})\,
    \mathbf{T}_{l_{\rm ph}l_{\rm ph}-1,\mu}^{*}({\mathbf n}_{\rm ph})\; d\Omega\; - \\

  & - &
  \sqrt{\displaystyle\frac{l_{\rm ph}}{2l_{\rm ph}+1}}
  \displaystyle\int\limits^{+\infty}_{0}
    R^{*}_{f}(r)\, R_{i}(r)\, V(\mathbf{r})\, j_{l_{\rm ph}+1}(kr)\: r^{2} dr \cdot
  \xibf_{\mu}
  \displaystyle\int
    Y_{l_{f}m_{f}}^{*}({\mathbf n}_{\rm r}^{f})\,
    Y_{l_{i}m_{i}}({\mathbf n}_{\rm r}^{i})\,
    \mathbf{T}_{l_{\rm ph}l_{\rm ph}+1,\mu}^{*}({\mathbf n}_{\rm ph})\; d\Omega.
\end{array}
\label{eq.app.9.8.3}
\end{equation}
Using notation (\ref{eq.app.9.7.6}) for the angular integral $\tilde{I}\,(l_{i}, l_{f}, l_{\rm ph}, n, \mu)$,
we rewrite Eqs.~(\ref{eq.app.9.8.3}) as
%
%
\begin{equation}
\begin{array}{lcl}
  \breve{p}_{l_{\rm ph}\mu}^{M} & = &
    \tilde{I}\,(l_{i},l_{f},l_{\rm ph}, l_{\rm ph}, \mu) \cdot \breve{J}\, (l_{i},l_{f},l_{\rm ph}), \\
  \breve{p}_{l_{\rm ph}\mu}^{E} & = &
    \sqrt{\displaystyle\frac{l_{\rm ph}+1}{2l_{\rm ph}+1}}
      \tilde{I}\,(l_{i},l_{f},l_{\rm ph},l_{\rm ph}-1,\mu) \cdot \breve{J}\,(l_{i},l_{f},l_{\rm ph}-1) -
    \sqrt{\displaystyle\frac{l_{\rm ph}}{2l_{\rm ph}+1}}
      \tilde{I}\,(l_{i},l_{f},l_{\rm ph},l_{\rm ph}+1,\mu) \cdot \breve{J}\,(l_{i},l_{f},l_{\rm ph}+1),
\end{array}
\label{eq.app.9.8.4}
\end{equation}
where we introduce the following new denotation:
\begin{equation}
\begin{array}{lcl}
  \breve{J}\,(l_{i}, l_{f},n) & = &
  \displaystyle\int\limits^{+\infty}_{0}
    R_{i}(r)\, R^{*}_{l,f}(r)\, V(\mathbf{r})\, j_{n}(kr)\; r^{2} dr.
\end{array}
\label{eq.app.9.8.5}
\end{equation}
%


\subsection{Calculation of components $p_{l_{\rm ph}\mu}^{M}$ and $p_{l_{\rm ph}\mu}^{E}$ and\, $\tilde{p}_{l_{\rm ph}\mu}^{M}$, $\tilde{p}_{l_{\rm ph}\mu}^{E}$: case of $l_{i} = 0$
\label{sec.9.5}}

In this section let su analyze case of $l_{i}=0$.
We have:
%
\begin{equation}
\begin{array}{lcl}
  \varphi_{i} (\mathbf{r}) & = & R_{i} (r) \: Y_{00}({\mathbf n}_{\rm r}^{i}).
\end{array}
\label{eq.app.9.5.1}
\end{equation}
Using gradient formula (see~\cite{Eisenberg.1973}, (2.56), p.~46):
%
\begin{equation}
  \displaystyle\frac{\partial}{\partial \mathbf{r}}\:
    f(r)\, Y_{lm}({\mathbf n}_{\rm r}) =
  \sqrt{\displaystyle\frac{l}{2l+1}}\:
    \biggl( \displaystyle\frac{df}{dr} + \displaystyle\frac{l+1}{r} f \biggr)\:
    \mathbf{T}_{l l-1, m}({\mathbf n}_{\rm r}) -
  \sqrt{\displaystyle\frac{l+1}{2l+1}}\:
    \biggl( \displaystyle\frac{df}{dr} - \displaystyle\frac{l}{r} f \biggr)\:
    \mathbf{T}_{l l+1, m}({\mathbf n}_{\rm r})
\label{eq.app.9.5.2}
\end{equation}
and taking into account Eq.~(\ref{eq.app.9.5.1}), we obtain:
%
\begin{equation}
  \displaystyle\frac{\partial}{\partial \mathbf{r}}\: \varphi_{i}(\mathbf{r}) =
    -\displaystyle\frac{d R_{i}(r)}{dr}\;
    \mathbf{T}_{01,0}(\mathbf{n}^{i}_{\rm r}).
\label{eq.app.9.5.3}
\end{equation}
Using this relation and Eq.~(\ref{eq.app.9.2.2}), we transform expressions (\ref{eq.app.9.4.2}).
For the magnetic component $p_{l\mu}^{M}$ we obtain:
%
%
\[
\begin{array}{lcl}
  p_{l_{\rm ph}\mu}^{M 0\, m_{f}} & = &
  \displaystyle\int\limits^{+\infty}_{0} dr
    \displaystyle\int d\Omega \:
    r^{2} \varphi^{*}_{f}(\mathbf{r}) \:
    \biggl( \displaystyle\frac{\partial}{\partial \mathbf{r}}
            \varphi_{i}(\mathbf{r}) \biggr) \:
    \mathbf{A}_{l_{\rm ph}\mu}^{*} (\mathbf{r}, M) = \\

  & = &
    \displaystyle\int\limits^{+\infty}_{0} dr
    \displaystyle\int d\Omega\:
    r^{2} R^{*}_{f}(r)\:
    Y_{l_{f}m_{f}}^{*}({\mathbf n}_{\rm r}^{f})\:
    \biggl(
      -\displaystyle\frac{dR_{i}(r)}{dr}
      \mathbf{T}_{01,0}(\mathbf{n}^{i}_{\rm r})
    \biggr) \:
    j_{l_{\rm ph}}(kr) \: \mathbf{T}_{l_{\rm ph}l_{\rm ph},\mu}^{*} ({\mathbf n}_{\rm ph}) = \\

  & = &
  -\displaystyle\int\limits^{+\infty}_{0}
    R^{*}_{f}(r)
    \displaystyle\frac{dR_{i}(r)}{dr}
    j_{l_{\rm ph}}(kr) \:
    r^{2} dr\; \cdot
    \displaystyle\int
    Y_{l_{f}m_{f}}^{*}({\mathbf n}_{\rm r}^{f}) \:
    \mathbf{T}_{01,0}(\mathbf{n}^{i}_{\rm r}) \:
    \mathbf{T}_{l_{\rm ph}l_{\rm ph},\mu}^{*} ({\mathbf n}_{\rm ph}) \: d\Omega.
\end{array}
\]
For the electric component $p^{E}_{l_{\rm ph} \mu}$ we obtain:
\[
\begin{array}{lcl}
  p_{l_{\rm ph}\mu}^{E 0\, m_{f}} & = &
  \displaystyle\int\limits^{+\infty}_{0} dr
    \displaystyle\int d\Omega \:
    r^{2} \varphi^{*}_{f}(\mathbf{r})
    \biggl( \displaystyle\frac{\partial}{\partial \mathbf{r}}
            \varphi_{i}(\mathbf{r}) \biggr)
    \mathbf{A}_{l_{\rm ph} \mu}^{*} (\mathbf{r}, E) = \\

  & = &
  \displaystyle\int\limits^{+\infty}_{0} dr
    \displaystyle\int d\Omega \:
    r^{2}\: R^{*}_{f}(r)\:
      Y_{l_{f}m_{f}}^{*}({\mathbf n}_{\rm r}^{f})\:
    \biggl(
      -\displaystyle\frac{dR_{i}(r)}{dr}
      \mathbf{T}_{01,0}(\mathbf{n}^{i}_{\rm r})
    \biggr) \times \\
    & & \times \;
    \Biggl\{
      \sqrt{\displaystyle\frac{l_{\rm ph}+1}{2l_{\rm ph}+1}}
      j_{l_{\rm ph}-1}(kr) \: \mathbf{T}_{l_{\rm ph}l_{\rm ph}-1,\mu}^{*}({\mathbf n}_{\rm ph}) -
      \sqrt{\displaystyle\frac{l_{\rm ph}}{2l_{\rm ph}+1}}
      j_{l_{\rm ph}+1}(kr) \: \mathbf{T}_{l_{\rm ph}l_{\rm ph}+1,\mu}^{*}({\mathbf n}_{\rm ph})
    \Biggr\} = \\

  & = &
  -\sqrt{\displaystyle\frac{l_{\rm ph}+1}{2l_{\rm ph}+1}}
    \displaystyle\int\limits^{+\infty}_{0}
    R^{*}_{f}(r)
    \displaystyle\frac{dR_{i}(r)}{dr}
    j_{l_{\rm ph}-1}(kr) \:
    r^{2} dr \cdot
    \displaystyle\int
    Y_{l_{f}m_{f}}^{*}({\mathbf n}_{\rm r}^{f})
    \mathbf{T}_{01,0}(\mathbf{n}^{i}_{\rm r})
    \: \mathbf{T}_{l_{\rm ph}l_{\rm ph}-1,\mu}^{*}({\mathbf n}_{\rm ph}) \: d\Omega - \\

  & & + \;
  \sqrt{\displaystyle\frac{l_{\rm ph}}{2l_{\rm ph}+1}}
    \displaystyle\int\limits^{+\infty}_{0}
    R^{*}_{f}(r)
    \displaystyle\frac{dR_{i}(r)}{dr} j_{l_{\rm ph}+1}(kr) \:
    r^{2} dr \cdot
    \displaystyle\int
    Y_{l_{f}m_{f}}^{*}({\mathbf n}_{\rm r}^{f})
    \mathbf{T}_{01,0}(\mathbf{n}^{i}_{\rm r})
    \: \mathbf{T}_{l_{\rm ph}l_{\rm ph}+1,\mu}^{*}({\mathbf n}_{\rm ph}) \: d\Omega.
\end{array}
\]
So, the components $p_{l_{\rm ph}\mu}^{M}$ and $p_{l_{\rm ph}\mu}^{E}$ have form:
\begin{equation}
\begin{array}{lcl}
  p_{l_{\rm ph}\mu}^{M 0\, m_{f}} & = &
  -\displaystyle\int\limits^{+\infty}_{0}
    R^{*}_{f}(r)
    \displaystyle\frac{dR_{i}(r)}{dr}
    j_{l_{\rm ph}}(kr)\:
    r^{2} dr \cdot
  \displaystyle\int
    Y_{l_{f}m_{f}}^{*}({\mathbf n}_{\rm r}^{f}) \:
    \mathbf{T}_{01,0}(\mathbf{n}^{i}_{\rm r}) \:
    \mathbf{T}_{l_{\rm ph}l_{\rm ph},\mu}^{*} ({\mathbf n}_{\rm ph}) \: d\Omega, \\

  p_{l_{\rm ph}\mu}^{E 0\, m_{f}} & = &
  -\sqrt{\displaystyle\frac{l_{\rm ph}+1}{2l_{\rm ph}+1}}
    \displaystyle\int\limits^{+\infty}_{0}
    R^{*}_{f}(r)
    \displaystyle\frac{dR_{i}(r)}{dr}
    j_{l_{\rm ph}-1}(kr)\;
    r^{2} dr \cdot
    \displaystyle\int
    Y_{l_{f}m_{f}}^{*}({\mathbf n}_{\rm r}^{f})
    \mathbf{T}_{01,0}(\mathbf{n}^{i}_{\rm r})
    \: \mathbf{T}_{l_{\rm ph}l_{\rm ph}-1,\mu}^{*}({\mathbf n}_{\rm ph}) \: d\Omega - \\

  & & + \;
  \sqrt{\displaystyle\frac{l_{\rm ph}}{2l_{\rm ph}+1}}
    \displaystyle\int\limits^{+\infty}_{0}
    R^{*}_{f}(r)
    \displaystyle\frac{dR_{i}(r)}{dr} j_{l_{\rm ph}+1}(kr)\;
    r^{2} dr \cdot
    \displaystyle\int
    Y_{l_{f}m_{f}}^{*}({\mathbf n}_{\rm r}^{f})
    \mathbf{T}_{01,0}(\mathbf{n}^{i}_{\rm r})\:
    \mathbf{T}_{l_{\rm ph}l_{\rm ph}+1,\mu}^{*}({\mathbf n}_{\rm ph})\; d\Omega.
\end{array}
\label{eq.app.9.5.4}
\end{equation}
We introduce the following definitions:
%
\begin{equation}
\begin{array}{lcl}
  J\,(l_{f},n) & = &
  \displaystyle\int\limits^{+\infty}_{0}
    \displaystyle\frac{dR_{i}(r)}{dr}\:
    R^{*}_{f}(l,r)\,
    j_{n}(kr)\; r^{2} dr, \\

  I\,(l_{f}, l_{\rm ph}, n, \mu) & = &
    \displaystyle\int
    Y_{l_{f}m_{f}}^{*}({\mathbf n}_{\rm r}^{f}) \:
    \mathbf{T}_{01,0}(\mathbf{n}^{i}_{\rm r})
    \: \mathbf{T}_{l_{\rm ph} n,\mu}^{*}({\mathbf n}_{\rm ph}) \: d\Omega.
\end{array}
\label{eq.app.9.5.5}
\end{equation}
Then one can write Eqs.~(\ref{eq.app.9.5.4}) as
%
\begin{equation}
\begin{array}{lcl}
  p_{l_{\rm ph}\mu}^{M 0 m_{f}} & = & - I\,(l_{f},l_{\rm ph}, l_{\rm ph}, \mu) \cdot J\,(l_{f},l_{\rm ph}), \\
  p_{l_{\rm ph}\mu}^{E 0 m_{f}} & = &
    -\sqrt{\displaystyle\frac{l_{\rm ph}+1}{2l_{\rm ph}+1}}
      I\,(l_{f},l_{\rm ph},l_{\rm ph}-1,\mu) \cdot J\,(l_{f},l_{\rm ph}-1) +
    \sqrt{\displaystyle\frac{l_{\rm ph}}{2l_{\rm ph}+1}}
      I\,(l_{f},l_{\rm ph},l_{\rm ph}+1,\mu) \cdot J\,(l_{f},l_{\rm ph}+1).
\end{array}
\label{eq.app.9.5.6}
\end{equation}

By the same way, we find the components $\tilde{p}_{l_{\rm ph}\mu}^{M 0 m_{f}}$ and $\tilde{p}_{l_{\rm ph}\mu}^{E0 m_{f}}$:
%
\begin{equation}
\begin{array}{lcl}
  \tilde{p}_{l_{\rm ph}\mu}^{M 0 m_{f}} & = &
  \displaystyle\int\limits^{+\infty}_{0}
    R^{*}_{f}(r)\,
    R_{i}(r)\,
    j_{l_{\rm ph}}(kr)\:
    r^{2} dr \cdot
  \xibf_{\mu}\,
  \displaystyle\int
    Y_{l_{f}m_{f}}^{*}({\mathbf n}_{\rm r}^{f}) \:
    \mathbf{T}_{l_{\rm ph}l_{\rm ph},\mu}^{*} ({\mathbf n}_{\rm ph}) \: d\Omega, \\

  \tilde{p}_{l_{\rm ph}\mu}^{E 0 m_{f}} & = &
  \sqrt{\displaystyle\frac{l_{\rm ph}+1}{2l_{\rm ph}+1}}
    \displaystyle\int\limits^{+\infty}_{0}
      R^{*}_{f}(r)\, R_{i}(r)\, j_{l_{\rm ph}-1}(kr)\;
      r^{2} dr \cdot
  \xibf_{\mu}
  \displaystyle\int
    Y_{l_{f}m_{f}}^{*}({\mathbf n}_{\rm r}^{f})\:
    \mathbf{T}_{l_{\rm ph}l_{\rm ph}-1,\mu}^{*}({\mathbf n}_{\rm ph})\: d\Omega\; - \\

  & & - \;
  \sqrt{\displaystyle\frac{l_{\rm ph}}{2l_{\rm ph}+1}}
    \displaystyle\int\limits^{+\infty}_{0}
      R^{*}_{f}(r)\, R_{i}(r)\, j_{l_{\rm ph}+1}(kr)\;
      r^{2} dr \cdot
    \xibf_{\mu}
    \displaystyle\int
    Y_{l_{f}m_{f}}^{*}({\mathbf n}_{\rm r}^{f})\:
    \mathbf{T}_{l_{\rm ph}l_{\rm ph}+1,\mu}^{*}({\mathbf n}_{\rm ph})\; d\Omega.
\end{array}
\label{eq.app.9.5.7}
\end{equation}
Introducing new integrals:
%
\begin{equation}
\begin{array}{lcl}
  \tilde{J}\,(l_{f},n) & = &
  \displaystyle\int\limits^{+\infty}_{0}
    R_{i}(r)\, R^{*}_{f}(l,r)\, j_{n}(kr)\; r^{2} dr, \\

  \tilde{I}\,(l_{f}, l_{\rm ph}, n, \mu) & = &
  \xibf_{\mu} \displaystyle\int
    Y_{l_{f}m_{f}}^{*}({\mathbf n}_{\rm r}^{f}) \:
    \mathbf{T}_{l_{\rm ph} n,\mu}^{*}({\mathbf n}_{\rm ph}) \: d\Omega,
\end{array}
\label{eq.app.9.5.8}
\end{equation}
we rewrite Eq.~(\ref{eq.app.9.5.7}) as
%
\begin{equation}
\begin{array}{lcl}
  \tilde{p}_{l_{\rm ph}\mu}^{M0 m_{f}} & = &
    \tilde{I}\,(l_{f},l_{\rm ph}, l_{\rm ph}, \mu) \cdot \tilde{J}\, (l_{f},l_{\rm ph}), \\
  \tilde{p}_{l_{\rm ph}\mu}^{E0 m_{f}} & = &
    \sqrt{\displaystyle\frac{l_{\rm ph}+1}{2l_{\rm ph}+1}}
      \tilde{I}\,(l_{f},l_{\rm ph},l_{\rm ph}-1,\mu) \cdot \tilde{J}\,(l_{f},l_{\rm ph}-1) -
    \sqrt{\displaystyle\frac{l_{\rm ph}}{2l_{\rm ph}+1}}
      \tilde{I}\,(l_{f},l_{\rm ph},l_{\rm ph}+1,\mu) \cdot \tilde{J}\,(l_{f},l_{\rm ph}+1).
\end{array}
\label{eq.app.9.5.9}
\end{equation}

\subsection{Calculation of components $\breve{p}_{l_{\rm ph}\mu}^{M}$ and $\breve{p}_{l_{\rm ph}\mu}^{E}$: case of $l_{i} = 0$
\label{sec.9.6}}

Now we calculate the components $\breve{p}_{l_{\rm ph}\mu}^{M}$ and $\breve{p}_{l_{\rm ph}\mu}^{E}$ at $l_{i}=0$.
From Eqs.~(\ref{eq.app.9.4.2}) and (\ref{eq.app.9.5.1}) we have:
%
\begin{equation}
\begin{array}{lcllcl}
  \breve{p}_{l_{\rm ph}\mu}^{M0 m_{f}} & = &
    \xibf_{\mu}\,
    \displaystyle\int
      \varphi^{*}_{f}(\mathbf{r})\,
      \varphi_{i}(\mathbf{r})\,
      V(\mathbf{r})\;
      \mathbf{A}_{l_{\rm ph}\mu}^{*} (\mathbf{r}, M) \;
      \mathbf{dr}, &
  \hspace{7mm}
  \breve{p}_{l_{\rm ph}\mu}^{E0 m_{f}} & = &
    \xibf_{\mu}\,
    \displaystyle\int
      \varphi^{*}_{f}(\mathbf{r})\,
      \varphi_{i}(\mathbf{r})\,
      V(\mathbf{r})\;
      \mathbf{A}_{l_{\rm ph}\mu}^{*} (\mathbf{r}, E)\;
      \mathbf{dr},
\end{array}
\label{eq.app.9.6.1}
\end{equation}
\begin{equation}
\begin{array}{lcllcl}
  \varphi_{i} (\mathbf{r}) = R_{i} (r)\: Y_{00}({\mathbf n}_{\rm r}^{i}), &
  \varphi_{f} (\mathbf{r}) = R_{f} (r)\: Y_{l_{f}m_{f}}({\mathbf n}_{\rm r}^{f}).
\end{array}
\label{eq.app.9.6.2}
\end{equation}
For the magnetic component $p_{l\mu}^{M0 m_{f}}$ we obtain:
%
\[
\begin{array}{lcl}
  \breve{p}_{l_{\rm ph}\mu}^{M0 m_{f}} & = &
  \xibf_{\mu}\,
  \displaystyle\int\limits^{+\infty}_{0} dr
  \displaystyle\int d\Omega \:
    r^{2} \varphi^{*}_{f}(\mathbf{r})\, \varphi_{i}(\mathbf{r})\, V(\mathbf{r})\; \mathbf{A}_{l_{\rm ph}\mu}^{*} (\mathbf{r}, M) = \\

  & = &
  \xibf_{\mu}\,
  \displaystyle\int\limits^{+\infty}_{0} dr
  \displaystyle\int d\Omega\:
    r^{2} R^{*}_{f}(r)\:
    Y_{l_{f}m_{f}}^{*}({\mathbf n}_{\rm r}^{f})\:
    R_{i}(r)\, V(\mathbf{r})\;
    j_{l_{\rm ph}}(kr) \: \mathbf{T}_{l_{\rm ph}l_{\rm ph},\mu}^{*} ({\mathbf n}_{\rm ph}) = \\

  & = &
  \xibf_{\mu}\,
  \displaystyle\int\limits^{+\infty}_{0}
    R^{*}_{f}(r)\, R_{i}(r)\, V(\mathbf{r})\; j_{l_{\rm ph}}(kr)\: r^{2} dr \cdot
    \displaystyle\int
      Y_{l_{f}m_{f}}^{*}({\mathbf n}_{\rm r}^{f})\, \mathbf{T}_{l_{\rm ph}l_{\rm ph},\mu}^{*} ({\mathbf n}_{\rm ph})\: d\Omega.
\end{array}
\]
For the electric component $p^{E0 m_{f}}_{l_{\rm ph} \mu}$ we obtain:
%
\[
\begin{array}{lcl}
  \breve{p}_{l_{\rm ph}\mu}^{E0 m_{f}} & = &
  \xibf_{\mu}\,
  \displaystyle\int\limits^{+\infty}_{0} dr
  \displaystyle\int d\Omega\:
    r^{2} \varphi^{*}_{f}(\mathbf{r})\, \varphi_{i}(\mathbf{r})\, V(\mathbf{r})\;
    \mathbf{A}_{l_{\rm ph} \mu}^{*} (\mathbf{r}, E) = \\

  & = &
  \xibf_{\mu}\,
  \displaystyle\int\limits^{+\infty}_{0} dr
  \displaystyle\int d\Omega \:
    r^{2}\: R^{*}_{f}(r)\:
    Y_{l_{f}m_{f}}^{*}({\mathbf n}_{\rm r}^{f})\:
    R_{i}(r)\, V(\mathbf{r})\; \times \\
    & & \times \;
    \Biggl\{
      \sqrt{\displaystyle\frac{l_{\rm ph}+1}{2l_{\rm ph}+1}}
      j_{l_{\rm ph}-1}(kr) \: \mathbf{T}_{l_{\rm ph}l_{\rm ph}-1,\mu}^{*}({\mathbf n}_{\rm ph}) -
      \sqrt{\displaystyle\frac{l_{\rm ph}}{2l_{\rm ph}+1}}
      j_{l_{\rm ph}+1}(kr) \: \mathbf{T}_{l_{\rm ph}l_{\rm ph}+1,\mu}^{*}({\mathbf n}_{\rm ph})
    \Biggr\}\; = \\

  & = &
  \xibf_{\mu}\,
  \sqrt{\displaystyle\frac{l_{\rm ph}+1}{2l_{\rm ph}+1}}
  \displaystyle\int\limits^{+\infty}_{0}
    R^{*}_{f}(r)\, R_{i}(r)\, V(\mathbf{r})\, j_{l_{\rm ph}-1}(kr)\: r^{2} dr \cdot
  \displaystyle\int
    Y_{l_{f}m_{f}}^{*}({\mathbf n}_{\rm r}^{f})\: \mathbf{T}_{l_{\rm ph}l_{\rm ph}-1,\mu}^{*}({\mathbf n}_{\rm ph}) \: d\Omega\; - \\

  & & - \;
  \xibf_{\mu}\,
  \sqrt{\displaystyle\frac{l_{\rm ph}}{2l_{\rm ph}+1}}
  \displaystyle\int\limits^{+\infty}_{0}
    R^{*}_{f}(r)\, R_{i}(r)\, V(\mathbf{r})\, j_{l_{\rm ph}+1}(kr)\: r^{2} dr \cdot
  \displaystyle\int
    Y_{l_{f}m_{f}}^{*}({\mathbf n}_{\rm r}^{f})\: \mathbf{T}_{l_{\rm ph}l_{\rm ph}+1,\mu}^{*}({\mathbf n}_{\rm ph}) \: d\Omega.
\end{array}
\]
So, the components $p_{l_{\rm ph}\mu}^{M0 m_{f}}$ and $p_{l_{\rm ph}\mu}^{E0 m_{f}}$ have form:
%
\begin{equation}
\begin{array}{lcl}
  \breve{p}_{l_{\rm ph}\mu}^{M0 m_{f}} & = &
  \xibf_{\mu}\,
  \displaystyle\int\limits^{+\infty}_{0}
    R^{*}_{f}(r)\, R_{i}(r)\, V(\mathbf{r})\; j_{l_{\rm ph}}(kr)\: r^{2} dr \cdot
    \displaystyle\int
      Y_{l_{f}m_{f}}^{*}({\mathbf n}_{\rm r}^{f})\, \mathbf{T}_{l_{\rm ph}l_{\rm ph},\mu}^{*} ({\mathbf n}_{\rm ph})\: d\Omega, \\

  \breve{p}_{l_{\rm ph}\mu}^{E0 m_{f}} & = &
  \xibf_{\mu}\,
  \sqrt{\displaystyle\frac{l_{\rm ph}+1}{2l_{\rm ph}+1}}
  \displaystyle\int\limits^{+\infty}_{0}
    R^{*}_{f}(r)\, R_{i}(r)\, V(\mathbf{r})\, j_{l_{\rm ph}-1}(kr)\: r^{2} dr \cdot
  \displaystyle\int
    Y_{l_{f}m_{f}}^{*}({\mathbf n}_{\rm r}^{f})\: \mathbf{T}_{l_{\rm ph}l_{\rm ph}-1,\mu}^{*}({\mathbf n}_{\rm ph}) \: d\Omega\; - \\

  & - &
  \xibf_{\mu}\,
  \sqrt{\displaystyle\frac{l_{\rm ph}}{2l_{\rm ph}+1}}
  \displaystyle\int\limits^{+\infty}_{0}
    R^{*}_{f}(r)\, R_{i}(r)\, V(\mathbf{r})\, j_{l_{\rm ph}+1}(kr)\: r^{2} dr \cdot
  \displaystyle\int
    Y_{l_{f}m_{f}}^{*}({\mathbf n}_{\rm r}^{f})\: \mathbf{T}_{l_{\rm ph}l_{\rm ph}+1,\mu}^{*}({\mathbf n}_{\rm ph}) \: d\Omega.
\end{array}
\label{eq.app.9.6.3}
\end{equation}
Using Eq.~(\ref{eq.app.9.5.8}) for the angular integral,
we rewrite (\ref{eq.app.9.6.3}) as
%
\begin{equation}
\begin{array}{lcl}
  \breve{p}_{l_{\rm ph}\mu}^{M0 m_{f}} & = &
    \tilde{I}\,(l_{f},l_{\rm ph}, l_{\rm ph}, \mu) \cdot \breve{J}\, (0, l_{f},l_{\rm ph}), \\
  \breve{p}_{l_{\rm ph}\mu}^{E0 m_{f}} & = &
    \sqrt{\displaystyle\frac{l_{\rm ph}+1}{2l_{\rm ph}+1}}
      \tilde{I}\,(l_{f},l_{\rm ph},l_{\rm ph}-1,\mu) \cdot \breve{J}\,(0, l_{f},l_{\rm ph}-1) -
    \sqrt{\displaystyle\frac{l_{\rm ph}}{2l_{\rm ph}+1}}
      \tilde{I}\,(l_{f},l_{\rm ph},l_{\rm ph}+1,\mu) \cdot \breve{J}\,(0, l_{f},l_{\rm ph}+1).
\end{array}
\label{eq.app.9.6.5}
\end{equation}


\end{document}